\documentclass[twocolumn,           % Format : preprint, twocolumn
               showpacs,            % Pacs : showpacs, noshowpacs
               nopreprintnumbers,   % Preprint: preprintnumbers, nopreprintnumbers
               aps,                 % Society: ...
               prd,                 % Journal Style : pra, prb, prc, prd, pre, prl, prstab, rmp
               letterpaper,         % Size : a4paper, ...
               groupeaddress,       % Affiliation (Title) : groupedaddress, superscriptaddress, unsortedaddress
               footinbib,           % Footnote: footinbib (activar notas al pie reales)
               tightenlines,        % Remove additional spaces in a line
               floats,floatfix,     % Floating pictures and tables
               showkeys
              ]{revtex4-2}
               
\usepackage[toc,page]{appendix}
\usepackage{graphicx}
\usepackage{dcolumn}
\usepackage{bm}
\usepackage{amsmath}
\usepackage{amsfonts,amssymb}
\usepackage{soul}
\usepackage{color}
\usepackage{xcolor}
\usepackage{enumerate}
\usepackage{float}
\usepackage{subfigure}
\usepackage{ulem}
\makeatletter
\newcommand{\vast}{\bBigg@{3}}
\newcommand{\Vast}{\bBigg@{4}}
\makeatother

\begin{document}

\title{Extended datasamples under the lens of Brane World Theory}
\author{Kyra Jacobo\textsuperscript{1}}
\author{Dorian Araya\textsuperscript{2}}
\email{Corresponding author: dorian.araya@alumnos.ucentral.cl}

\address{$^1$Depto. de F\'isica y Matem\'aticas, Universidad Iberoamericana Ciudad de  M\'exico, Prolongaci\'on Paseo de la Reforma 880, M\'exico D. F. 01219,  M\'exico.}
\address{$^1$School of Physics and Astronomy, University of Edinburgh, Peter Guthrie Tait Road, Edinburgh, EH9 3FD, Scotland, United Kingdom.}
\address{$^2$ Facultad de Ingenier\'ia y Arquitectura, Universidad Central de Chile, Avda. Fco. de Aguirre 0405, 1710164 La Serena, Coquimbo,  Chile.}

%-------------------------------------------------------------------------------------------------
%-------------------------------------------------------------------------------------------------
\begin{abstract}
\begin{center}
\textbf{\large Abstract}
\end{center}
This work revises the Brane World Theory known as Randall Sundrum with the modification of an exponential, redshift-dependent brane tension. This model is studied in a scenario assuming no dark energy, with the aim of determining whether it can reproduce the universe’s acceleration on its own, without the addition of a dark energy fluid. Bayesian Statistical analysis is performed in order to constrain the free parameters of each scenario for which the datasamples of SLS, SNIa, OHD and BAO are used, the last two considering newly added elements on the data. Both Planck and Riess priors for $h$ are used and compared. In both cases we are able to reproduce the late-time accelerated expansion in agreement with observational data. Interesting consistencies at transition redshift $z_t$ with $\Lambda$-CDM are found suggesting that this might be a suitable model for studying the evolution of the universe up to present date, however, some pathologies are detected in this model, namely a "Big Rip" divergence of $H(z)$ at $z=1$, as well as a strong relationship between the form of the functional for the brane tension and the future evolution of the universe in this model. 
\end{abstract}
%\draft
%\date{\today}
\maketitle

%%%%%%%%%%%%%%%%%%%%%%%%%%%%%%%%%%%%
\section{Introduction} \label{intro}
%%%%%%%%%%%%%%%%%%%%%%%%%%%%%%%%%%%%
%\JM{Escribire comentarios en este color\\}
%\textcolor{red}{Miguel en rojo}

In contemporary cosmology, $\Lambda$-CDM is the most accepted cosmological model. In this model, the Universe's structure is composed of baryonic and non-relativistic \textit{cold} dark matter, radiation density and a vacuum energy density that is called the cosmological constant, $\Lambda$. This last component is a type of dark energy fluid (DE) that causes the universe to expand in a accelerating form. $\Lambda$-CDM has gained the approval of most astrophysicists nowadays due to its success at predicting the evolution of the universe when compared with observational data \citep{Planck2020}. However, as both our understanding of the universe, as well as our technology to perform high-precision astronomical measurements has improved,$\Lambda$CDM has faced some challenges that it seems unable to overcome, or for which it cannot provide satisfactory explanations. Some of these problems are the Hubble tension \citep{di2021realm}, the fine-tuning problem \citep{friederich2017fine}, the coincidence problem or the mystery of the nature of dark matter (DM) and dark energy (DE) \citep{lima2004alternative}, just to name a few \citep{LCDMPROBLEMS}. These inabilities of the $\Lambda$-CDM  model have motivated the proposal of different kinds of dark matter and dark energy models \citep{motta2021taxonomy}. For the latest, instead of assuming a cosmological constant $\Lambda$ with an equation of state parameter (EoS) $w=-1$, alternative fields such as  \textit{quintessence} DE field with and EOS of $ -1/3<w<-1$\citep{caldwell2002phantom, quintessencemodelpaper}  or \textit{phantom} DE  with and EoS of $w<-1$ \citep{phantomDE} have been proposed. Furthermore, researchers have also turned to alternative theories of gravity to seek answers. These theories are called Modified Gravity theories and according to \cite{Kaluza-Kleinmodel,Kaluza-Kleinmodel2} can be classified in: theories that involve extra fields such as scalar-tensor theories \citep{scalar-thensortheories} or Einstein aether theories \citep{Einstein-aethertheory}, higher derivatives and non-local theories of gravity such as $f(R)$ theories \citep{f(r)gravitypaper1, f(r)gravitypaper2, Nojiri:2011, Nojiri:2017}, Horava-Lifschitz Gravity \citep{Horavapaper1,Horavapaper2, Nojiri:2011}, parameterized post-Friedmann approaches \citep{post-friedmannpaper} and higher-dimensional theories of gravity \citep{higherdimensionspaper1,higherdimensionalpaper2, Nojiri:2000brane, Nojiri:2002bulk} among others.

Among this last classification, another subdivision of higher-dimensional gravity theories arises called Brane World theories. These theories consist on assuming a 5 dimensional manifold in which 4 dimensional hypersurfaces are embedded. These hypersurfaces are called \textit{branes} and, in these models, the universe is considered to be one of these 4 dimensional sub-manifolds. Some theories that fall into Brane World theories are Kaluza-Klein models \citep{Kaluza:1921,Kaluzakleintheories,Kaluza-kleinpaper2}, Dvali-Gabadadze-Porrati Gravity \citep{DGPgravityoriginal,DPGCASCADINGGRAVITY} and Randall-Sundrum brane models, which will be the main focus of this paper.

The Randall-Sundrum brane theory was first published in 1999 \citep{randall1999large} as an attempt to tackle the hierarchy problem. The hierarchy problem addresses the enormous difference in the scale of gravity (the weakest fundamental force) and the weak scale (the second weakest). By assuming that the universe is embedded in a 5-dimensional manifold with a warp factor with the form of a negative exponential, this theory states that the contributions of this higher-dimensional bulk must be taken into account when considering both the effective 4-dimensional Planck scale and the effective mass of particles. In this first model, two branes are considered, one of them being our universe and the other the ``gravity brane" from which the gravitational force emerges and travels to our brane, being warped by the bulk's metric. Later, a modification to this brane model was proposed \citep{randall1999alternative} where the compactification (the setting of boundaries to the extra dimension) was avoided by just assuming the ``gravity brane" was so far to our universe that it may not be taken into account when studying the physics occurring in our brane. This second Model is what we will be referring to from now on when talking about the Randall-Sundrum brane model (RS).

RS has gained popularity in both the fields of cosmology and particle physics \citep{casagrande2008flavor, reece2008particle} and multiple phenomena ranging from particle symmetries \citep{dando2005clash} to inflation \citep{kim2000inflation} have been studied under this Brane World scenario. Among these works, the case in which a phantom DE energy field is influenced by a RS scenario \citep{acuna2018presence} calls our attention.

In this work they study a Randall-Sundrum Model interacting with a universe with a phantom field as a Dark Energy fluid and perform a dynamic analysis considering two scenarios: one considering a general field with the characteristic equation $w< -1$, and the second one considering the explicit form of the scalar field with a potential with a maximum. The results presented there, denote a strong relation between the magnitude of the brane tension $\lambda$, which is the free parameter of the theory, and the time of the Big Rip caused by the phantom field: the stronger the brane tension, the sooner the Big Rip occurs.

Variable brane tension models (VBT) have been developed and studied in literature such as \citep{gergely2009eotvos}. We especially focus on the works of \citep{garcia2018brane} and \citep{verdugo2024synchronize}. On the first one, the brane tension $\lambda$ of a RS scenario adopts the form of a polynomial function $\lambda(z)= (z+1)^n$ where $n$ is a new free parameter on the model. When constrained using SNIa, H(z), BAO and CMB, $n \approx 6.19$  which arises a cosmological- constant-like term that reproduces the late cosmic acceleration without the need of a DE fluid.

However, on \citep{verdugo2024synchronize} the VBT model is revisited this time constraining $\lambda$ with two lensing datasamples at different scales: the lensing galaxy cluster Abell 1689, and early-type galaxies. Under this procedure, the value estimated for $\lambda \approx 7.8$ which favors a phantom DE field rather than a cosmological constant. Motivated by this finding, we propose a redshift-dependent exponential variable brane tension as a modification to the Randall-Sundrum scenario, from now on Exponential Brane Randall-Sundrum (Exponential Brane RS). This model considers $\lambda$ as a scalar function, in an attempt to test whether the brane tension can be responsible for the late accelerated expansion of the universe accounted by observations \citep{riess1998observational}, without the addition of a dark energy field.

Apart from shedding light into the nature of a higher-dimensional universe and its evolution, Exponential Brane RS also presents an excellent opportunity to test the new points found for both Observational Hubble Data (OHD) \citep{DATAHZ}, \citep{HZNEWDATA32} and Baryonic Acoustic Oscilations (BAO) \citep{DATABAO} datasamples. By performing the parameter constriction using these extended datasamples as well as Type Ia supernovae (Pantheon+SH0ES) \citep{DATASN}, and Strong Lensing Systems (SLS) \citep{DEMODELSSLS}; and then comparing it with $\Lambda$-CDM we can get an idea of both the success of the modification to RS proposed and the performance of this new data for parameter constriction.

The paper is organized as follows: In Section \ref{MF} the whole mathematical formalism involved in a Randall-Sundrum scenario with a variable tension in the brane is discussed and the Friedmann equations are obtained both considering phantom DE and without it. Then in Section \ref{DS} all the formal and technical details regarding the parameter constriction using OHD, SLS, BAO and Pantheon+SH0ES as well as the joint are stated, along with the details of the Bayesian method used. In section  \ref{Results}  we present the method used to test the cosmological constraints with data samples and their respective graphs and results.
In section \ref{comparison} is given the AIC and BIC criteria to test the empirical support of the model compared to $\Lambda$CDM. Finally, Section \ref{conclusions} is devoted to the discussion of the results and the  information of the physics we can conclude out of this scenario. 

%%%%%%%%%%%%%%%%%%%%%%%%%%%%%%%%%%%%
\section{Mathematical Formalism } \label{MF}
%%%%%%%%%%%%%%%%%%%%%%%%%%%%%%%%%%%%

Regarding the notation in this section: Since tensors acting on the bulk (5-dimensional) appear alongside their corresponding versions on the brane (4-dimensional), for clarity, the 5-dimensional versions will have a tilde on top, and their indices will be denoted by Latin letters a, b, c, running from 0 to 4. The 4-dimensional versions are indicated without a tilde and their indices are indicated with greek letters $\mu$, $\nu$, that run from 0 to 3. \\

Assuming that Einstein field equations apply on the 5-dimensional bulk, using the Gauss-Codazzi equations\citep{abdalla2008gauss} and the Israel-Lanczos Junction Conditions \citep{musgrave1996junctions}, it is possible to make a projection of the 5-dimensional Einstein equations into the 4-dimensional brane in a way that only the tangential components survive , leading to

\begin{equation}\label{einsteinsobrebrana}
\begin{aligned}
    G_{\mu\nu} = & -\frac{1}{2} \tilde{\Lambda}g_{\mu\nu} + \frac{2}{3} \tilde{k}^2 F_{\mu\nu} -\frac{1}{12}\tilde{k}^4 \lambda^2g_{\mu\nu} \\
    & + 8\pi G T_{\mu\nu}+\tilde{k}^4 \Pi _{\mu\nu}- E_{\mu\nu}.
\end{aligned}
\end{equation}

In this expression, $\lambda$ is the brane tension, the free parameter of the model which can be interpreted as a brane density, $\tilde{\Lambda}$ is the 5 dimensional anti-de-Sitter cosmological constant of the bulk and $\tilde{k}^4$  is a constant analogous to the one found in the traditional Einstein equation: $ k= 8 \pi G $. Also

\begin{equation}
    E_{\mu\nu}=\tilde{C}^a_{bcd} n_a n^c g^b_\mu g^d_\nu,
\end{equation}

 $\tilde{C}^a_{bcd}$ being the Weyl tensor in the bulk and $n_A$ is the normal unitary vector to the brane;

\begin{equation}
 F_{\mu\nu}=\tilde{T}_{cd} g^c_\mu g^d_\nu + (\tilde{T}_{cd}n^c n^d - \frac{1}{4}\tilde{T})g_{\mu\nu},  
\end{equation}
where $\tilde{T}_{cd} $, $\tilde{T}$ are respectively the energy-momentum tensor and the energy-momentum scalar on the bulk and 
\begin{equation}
\begin{aligned}
\Pi_{\mu\nu} = & -\frac{1}{4} T_{\mu\alpha} T_\nu^\alpha + \frac{1}{12} TT_{\mu\nu} \\
& + \frac{1}{8} g_{\mu\nu} T_{\alpha\beta} T^{\alpha\beta} - \frac{1}{24} g_{\mu\nu} T^2. 
\end{aligned}
\end{equation}

As the universe as a whole is our object of study (concentrated on the brane), Friedmann-Lemaitre-Robertson-Walker (FLRW) metric \citep{melia2022friedmann} is used whose line element is
\begin{equation}\label{friedmannmetadim}
   ds^2= -dt^2 + a^2(t) \left[\frac{dr^2}{1-\kappa r^2} + r^2d\Omega\right], 
\end{equation}
where $a(t)$ is the scale factor and $\kappa$ is the curvature.\\
 From substituting the FLRW metric on the Randall-Sundrum field equations, we get the RS Friedmann equation
\begin{equation}\label{ecfriedmannbrana}
    H^2= \frac{k^2}{3}\left(\rho + \frac{\rho^2}{2\lambda} \right) - \frac{\kappa}{a^2}.
\end{equation}
Where $H$ is the Hubble parameter which indicates the expansion rate of the universe $\rho$ is the fluid density. By differentiating this equation with respect  to coordinate time, we can obtain its analogous acceleration or Raychaudhuri equation as
\begin{equation}\label{raychaudhuribranas}
    \dot{H}= -\frac{k^2}{2} \left(\rho + p\right) \left( 1 + \frac{\rho}{\lambda} \right) + \kappa,
\end{equation}
where $p$ is the pressure of the fluid, and the continuity equation: $\dot{\rho}= -3H(\rho + p)$ has been substituted.\\

From (\ref{raychaudhuribranas}), it is easy to see that a fluid capable of accelerating the universe must have an EoS that satisfies

\begin{equation}\label{wbranas}
    \omega < -\frac{1}{3} \left[\frac{1 + 2\rho / \lambda}{1+ \rho/ \lambda} \right].
\end{equation}

Equations (\ref{ecfriedmannbrana}) and (\ref{raychaudhuribranas}) differ from GR only by the addition of a second term in which the brain tension $lambda$ emerges as a coupling factor of the brane with the bulk. In order to rewrite  (\ref{ecfriedmannbrana}) in terms of the density parameters: $\Omega_\nu= k^2\rho_\nu /3H_0^2$, (where $\nu$ indicates an arbitrary density parameter) the following change of variable was necessary:$\tilde{\lambda}=\lambda_02k^2/3H_0^2$  yielding

\begin{equation}\label{friedmannbranas}
E(z)^2= \sum_{\nu=1}^{n} \Omega_{0\nu}(z+1)^{3(w_{\nu}+1)} + \frac{ \Omega_{0\nu}^2(z+1)^{6(w_{\nu}+1)}}{\tilde{\lambda}}.
\end{equation}
This $\tilde{\lambda}$ is a scalar whose value is to be determined, meaning it is an additional free parameter of the theory. In order to make the brane tension a function of the redshift, the form of $\lambda$ becomes

\begin{equation}
    \lambda= \lambda_0 f(z+1),
\end{equation}
where $\lambda_0$ is the value of the brane tension at present day and $f(z+1)$ is, in principle, an arbitary function of the redshift. With a variable tension then the RS Friedmann equation, in its most general form becomes

\begin{equation}\label{friedmannbranasmasgeneral}
E(z)^2= \sum_{\nu=1}^{n} \Omega_{0\nu}(z+1)^{3(w_{\nu}+1)} + \frac{ \Omega_{0\nu}^2(z+1)^{6(w_{\nu}+1)}}{\tilde{\lambda}f(z+1)}.
\end{equation}

Now the challenge is finding a suitable function $f(z+1)$ in order to reproduce the evolution of the universe as indicated by observations. This function, since it is found in the denominator, cannot be $0$ for any value of $z$. Apart from that, we demand that it is class $C^k$ for convenience and to ensure its evolution for $z>0$ is somehow similar to the well-known deSitter behavior for late times in $\Lambda$-CDM \citep{singh2008cosmological}. A function that fulfills these requirements is $f(z+1)= e^{\alpha(z+1)}$ where the $\alpha$ is a free parameter added to modulate the exponential according to the data.\\

\begin{figure*}
        \centering
\includegraphics[width=0.5\textwidth]{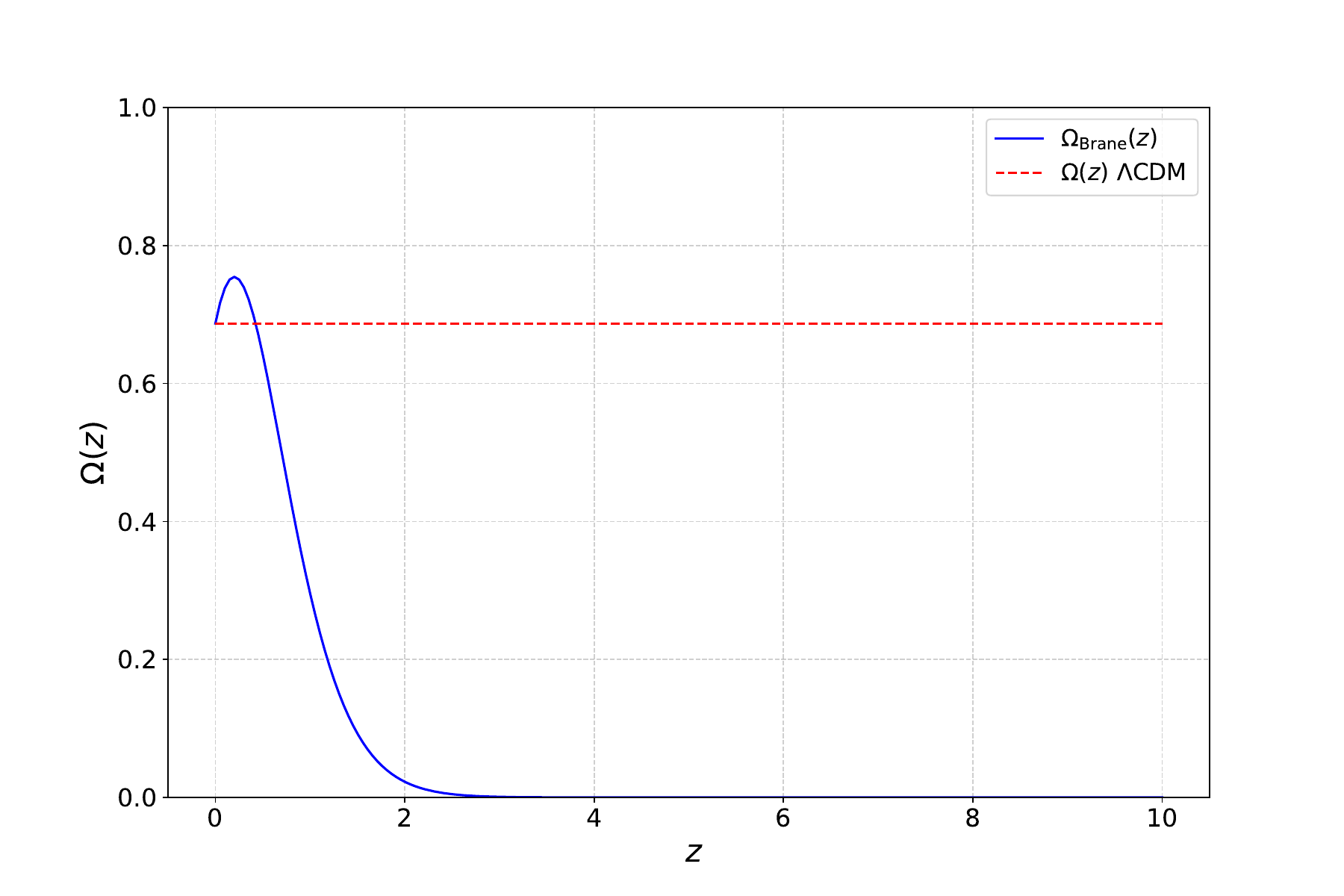}
\caption{$\Omega(z)$ comparison between the Exponential Brane RS model and $\Lambda$CDM model where $\Omega_{Brane}$ (blue line) represents the density parameter of the brane tension and $\Omega_{\Lambda_{\text{CDM}}}$ (red dashed line) shows the dark energy density parameter, which remains constant with redshift. For both models is assumed the density parameter for $h$ and $\Omega_{m}$(dark matter and baryonic matter) of Planck et al. (2018) \citep{Planck2020}}
\label{omegaeffective}
\end{figure*}

Even though other functions might fulfill the requirements previously mentioned, the choice of a negative exponential form for the brane tension, \( \lambda(z) = \lambda_0 e^{\alpha z} \), is motivated by its behavior at different redshifts. As \( z \to 0 \) (late-time Universe), the exponential term \( e^{\alpha z} \) approaches 1, causing the brane tension to behave as a constant \( \lambda_0 \), mimicking the effect of a cosmological constant \( \Lambda \), and thus driving the accelerated expansion of the Universe. Conversely, as \( z \to \infty \) (early Universe), \( \lambda(z) \) tends to 0, effectively ``turning off'' the brane tension. This behavior is physically consistent, as dark energy was subdominant in the early Universe and did not significantly contribute to cosmic acceleration at that time. This behavior is shown in the figure \ref{omegaeffective} where the evolution of the density parameter of the variable brane tension is compared to the one of the cosmological constant of $\Lambda$-CDM. This "emergent" DE behavior (or in this case, variable brane tension) might help provide a better description of the evolution of our universe and even solve some of of its current problems such as the Hubble tension \citep{di2021realm}.
\\Furthermore, the negative exponential form of the brane tension is inspired by the Randall-Sundrum (RS) brane-world scenario itself \citep{randall1999large}, where the warp factor in the extra-dimensional bulk takes the form of a negative exponential so the higher-dimensional gravitational scale is warped down by the exponential factor, leading to an effective 4-dimensional Planck mass that is much smaller than the fundamental scale, as required by observations. This analogy allows us to model the late-time accelerated expansion of the Universe without explicitly invoking dark energy, relying instead on the intrinsic properties of the brane itself to account for the observed cosmic acceleration.\\

For this proposal, (\ref{friedmannbranasmasgeneral}) becomes

\begin{equation}\label{friedmannfinal}
   E(z)^2= \sum_i \Omega_i (z+1)^{3(w_i + 1)}  + \Omega_i^2 (z+1)^{6(w_i + 1)}\frac{\Omega_T}{e^{\alpha z}},
\end{equation}
where,
\begin{equation}
    \Omega_T = \frac{1-(\sum_i \Omega_i )}{ \sum_i \Omega_i^2   }
\end{equation}

and the parameter $\tilde{\lambda}$ has been rewritten in terms of the other free parameters of the model
\begin{equation} \label{lambdadespejada}
    \tilde{\lambda}= \frac{(\Omega_{0m}^2 + \Omega_{0r}^2) e^{-\alpha}}{1-( \Omega_{0r} + \Omega_{0m})}.
\end{equation}
so it does not explicitly appears at (\ref{friedmannfinal}).\\
\begin{equation}
    \Omega_T = \frac{1-(\sum_i \Omega_i )}{ \sum_i \Omega_i^2   }
\end{equation}

and the parameter $\tilde{\lambda}$ has been rewritten in terms of the other free parameters of the model
\begin{equation} \label{lambdadespejada}
    \tilde{\lambda}= \frac{(\Omega_{0m}^2 + \Omega_{0r}^2) e^{-\alpha}}{1-( \Omega_{0r} + \Omega_{0m})}.
\end{equation}
so it does not explicitly appears at (\ref{friedmannfinal}).\\

As usual, the second order cosmographic term is

\begin{equation}
q(z)= \frac{(z+1)}{2E(z)^2}\frac{dE(z)^2}{dz} -1 ,
\end{equation}
which is known as the deceleration parameter.\\

%%%%%%%%%%%%%%%%%%%%%%%%%%%%%%%%%%%%
\section{Datasamples} \label{DS}
%%%%%%%%%%%%%%%%%%%%%%%%%%%%%%%%%%%%%%%%
In this section we present the data to test the Exponential Brane RS model with updated observational data of Observational Hubble Data (OHD), Supernovae Type Ia (SNIa), Baryonic Acoustic Oscillations (BAO) and Strong Lensing System (SLS).
\subsection{Observational Hubble Data}

Cosmic chronometers are massive, passively evolving galaxies with populations of old stellar formations and very low star formation rates, that is, with little contamination of young components. These characteristics allow us to determine its temporal evolution and estimate the expansion rate of the universe with the age difference method \citep{cosmicchronometers}, \citep{borghi2022toward} given by 
\begin{equation}
     H(z)=- \frac{1}{1+z} \frac{dz}{dt}.
\end{equation}
For this, it must be assumed that General Relativity applies within the framework of these galaxies. Apart from that, cosmic clocks are model independent. An OHD homogeneous samples will be used to test the models \citep{DATAHZ}. The chi square function takes the form
\begin{equation}
\chi^2_{\text{OHD}} = \sum\limits_{i=1}^{N} \left( \frac{H_{\text{th}}(z_i) - H_{\text{obs}}(z_i)}{\sigma_{i,\text{obs}}} \right)^2,
\end{equation}
where $H_{\text{th}}(z_i)$ is the theoretical Hubble parameter  $H_{\text{obs}}(z_i) \pm \sigma_{i,\text{obs}}$ is the observational Hubble parameter (from DA method) with its uncertainty at the redshift $z_i$, and $N$ is the number of points used. For this analysis a new point for the OHD data \citep{HZNEWDATA32} will be included.
%%%%%%%%%%%%%%%%%%%%%%%%%%%%%%%%%%%%%%%%%%%%%%%%%%%%%%%%%%%%%%%%%%%%%%%%%%

\subsection{Type Ia Supernovae}
A supernova is a stellar explosion that occurs in the late stages of the life
of a star, when it begins to run out of fuel and then collapses
due to the gravitational force. During this explosion, the luminosity increases by many millions that of the standard star, eclipsing entire galaxies.
Type Ia supernovae occur in binary systems in which one of the stars is a white dwarf. The importance of these supernovae is that their light curves (luminosity vs time) have an easily standarizable peak. In this way, the absolute magnitude \citep{AbsolutemagSNIa} can be measured for each registered event.\\

We can define the luminosity distance $D_L(z)$ as
\begin{equation}
D_L(z)= (z+1)\frac{c}{H_0} \int_{0}^{z} \frac{1}{E(z')} \, dz'.
\end{equation} 
Samples of SNIa \cite{DATASN} provide distance modulus measurements at different redshifts. As the measurements in this kind of samples are correlated, it is appropriate to build the chi-square function as 

\begin{equation}
 \chi^2_{\text{SNIa}} = a + \log\left(\frac{c}{2\pi}\right) - \frac{b^2}{c},
\end{equation}
 where
 
\[
a = (\Delta{\mu})^T \cdot \text{Cov}_P^{-1} \cdot \Delta{\mu},
\]
\[
b = (\Delta{\mu})^T \cdot \text{Cov}_P^{-1} \cdot \Delta1,
\] 
\[
c = \Delta1^T \cdot \text{Cov}_P^{-1} \cdot \Delta1,
\] 
Where the superscript \( T \) is the transpose of the vectors that are in matrix form and $(\Delta{\mu})$ is the vector of residuals between the theoretical distance modulus given by the model and the observed one. \( \Delta1 = (1, 1, \ldots, 1)^T \) is a matrix formed of only by ones, \( \text{Cov}_P \) is the covariance matrix formed by adding the systematic and statistical uncertainties: \( \text{Cov}_P = \text{Cov}_{P,\text{sys}} + \text{Cov}_{P,\text{stat}} \).\\

The theoretical distance modulus is estimated by
\[
\Delta{\mu} = 5 \log_{10} \left[\frac{D_L(z)}{10 \text{pc}}\right],
\]
Where $D_L(z)$ is the luminous distance discussed before.
%%%%%%%%%%%%%%%%%%%%%%%%%%%%%%%%%%%%%%%%%%%%%%%%%%%%%%%%%%%%%%%%%%

\subsection{Baryonic Acoustic Oscillations}

Baryonic acoustic oscillations are the result of two opposing forces:
gravity and pressure. These forces fought at a very early age of the universe in an extremely dense region of primordial plasma. This plasma was formed by electrons and baryons (protons and neutrons). The photons were trapped in this plasma due to Thomson scattering, releasing a large amount of
outward pressure, while at the same time, the density of the plasma attracted
matter gravitationally \citep{BAOFIRSTARTICLE}.
At the epoch of recombination $(z \approx 1100)$, the universe expanded enough
so that the temperature decreased. This allowed the baryonic matter to form
hydrogen, with which, being neutral, photons do not interact with as frequently causing the decoupling of the photons from the plasma, leaving layers of baryonic matter in the form of waves. The distance from the center of the dense region to the first
ripple of baryonic matter (where galaxies later formed) serves
as a standard rule for measuring the expansion of the Universe.
This standard rule is formally defined as the maximum distance traveled
by the sound wave, that is, the sound horizon at the time of baryon decoupling.\\

The theoretical BAO angular scale ($\theta_{\text{th}}$) is estimated as

\begin{equation}
\theta_{\text{th}}(z) = \frac{r_{\text{drag}}}{(1 + z)D_{\text{A}}(z)}, \quad 
\end{equation}

The comoving sound horizon, $r_{s}(z)$, is defined as
\begin{equation}
r_{s}(z) = \frac{c}{H_{0}} \int_{0}^{z} \frac{c_{s}(z_{'})}{E(z_{'})} \, dz_{'}, \quad 
\end{equation}
where the sound speed $c_{s}(z) = \frac{1}{\sqrt{3(1 + R_{\bar{b}}/(1 + z))}}$, with $R_{\bar{b}} = 31500 \Omega_{b}h^{2}(T_{\text{CMB}}/2.7\,\text{K})^{-4}$, and $T_{\text{CMB}}$ is the CMB temperature. The redshift  at the baryon drag epoch is given by \citep{BAOFORMULA} 
\begin{equation}
\begin{aligned}
z_{\text{drag}} = & \ 1291(\Omega_{m0} h^{2})^{0.251} \\
& \hspace{-2em} \cdot \left[ 
    1 + 0.659 (\Omega_{m0} h^{2})^{0.828} 
    \left( 
        1 + b_{1} \left(\frac{\Omega_{b0} h^{2}}{b_{2}}\right)
    \right)
\right],
\end{aligned}
\end{equation}
where
\begin{align*}
b_{1} &= 0.313 (\Omega_{m0} h^{2})^{-0.419} [{1 + 0.607(\Omega_{m0} h^{2})^{0.674}}], \\
b_{2} &= 0.238 (\Omega_{m0} h^{2}).
\end{align*} 

where $\Omega_{\text{m}0}$ and $\Omega_{\text{b}0}$ are the dark matter and baryonic matter components at $z = 0$ respectively.
The most recent compilation of transversal BAO measurements $\theta_{\text{BAO}}(z)$ is presented in \citep{DATABAO}.
Now the $\chi²$ function is built as

\begin{equation}
\chi^2_{\theta} = \sum_{i=1}^{N} \frac{(\theta_{\text{BAO}}^{i} - \theta_{\text{th}}(z_i))^2}{(\sigma_{\theta_{\text{BAO}}^{i}})^2}
\end{equation}
where $\theta_i^{\text{BAO}} \pm \sigma_{\theta_{\text{BAO}}^{i}}$ is the BAO angular scale  and its respective uncertainty at 1$\sigma$ measured at $z$ given by the sample.

As part of the updated measurements, two new points for the ratio \(D_V/r_d\) where DV is the dilation scale and six new points for the ratio \(D_M/r_d\) where $D_{M}$ the comoving distance were added based on the DESI 2024 VI collaboration \citep{desicollaboration2024}. These measurements are specifically used to compute the contributions to the \(\chi^2\) functions for \(D_V/r_d\) and \(D_M/r_d\), respectively, where $r_{d}$ is the sound horizon at the baryon drag epoch $r_{s}(z_{drag})$.

The comoving distance \(D_M(z)\) and the dilation scale \(D_V(z)\) \citep{DILATIONSCALE2005} are defined as follows:

\begin{equation}
D_M(z) = \frac{c}{H_0} \int_0^z \frac{dz'}{E(z')},
\end{equation}

\begin{equation}
D_V(z) = \Big[z\, D_H(z)\, D_M^2(z)\Big]^{1/3},
\end{equation}
\begin{equation}
D_H(z) = \frac{c}{H(z)}.    
\end{equation}

To incorporate the DESI measurements, we define the normalized quantities as:

\begin{equation}
D_{M, rd} = \frac{D_M}{r_d}, \quad D_{V, rd} = \frac{D_V}{r_d}.
\end{equation}

The total \(\chi^2\) function is then constructed as the sum of the BAO angular scale contribution, \(\chi^2_{\theta}\), and the contribution from the DESI measurements:

\begin{equation}
\chi^2_{\text{BAO}} = \chi^2_{\theta} + \chi^2_{D_V, D_M},
\end{equation}

where

\begin{equation}
\chi^2_{D_V, D_M} = \sum_{i} \frac{\Bigl(D_i - D_i^{\text{th}}(z_i)\Bigr)^2}{\sigma_{D_i}^2},
\end{equation}

with the observables defined as

\begin{equation}
D_i = \{ D_{M, rd}, \; D_{V, rd} \}.
\end{equation}

The observed values \(D_i^{\text{obs}}\) and their uncertainties \(\sigma_{D_i}\) are taken from the DESI 2024 VI dataset, which includes six measurements for \(D_{M, rd}\) and two for \(D_{V, rd}\), obtained at different redshifts \(z_i\).

\subsection{Strong Lensing System}
A gravitational lens is the deviation in the path of light due to  the curvature of spacetime caused by a very massive object such as a galaxy or
a black hole. This deviation of light allows us to observe objects that are
located behind the massive object, however, the curved rays make it look as if the source had a different apparent position.
The methodology for using gravitational lensing as a measure of the expansion of the universe has been used through years to constrain cosmological parameters of different models \citep{Cao_2015SLS} \citep{SLSmagaña2018}. This comparison is done through the Einstein radius $\theta_E$, which is the angular distance between the source and the image measured by the observer 
\begin{equation}
\theta_E= 4\pi \frac{\sigma^2_{SIS}}{c^2} \frac{D_{ls}}{Ds}.
\end{equation}

Where $D_{s}$ is the angular diameter distance to the source, and  $D_{ls}$ is the angular diameter distance from the lens to the source.

Solving for $\frac{D_{ls}}{Ds}$, we have:

\begin{equation}
\frac{D_{ls}}{Ds} = \frac{c^2 \theta_E}{4 \pi \sigma_{SIS}^2},
\end{equation}

where $\sigma_{SIS}$ is the stellar dispersion radius in a Singular Isothermal Sphere (SIS) system, which is the assumed model for the lens. This is an experimental parameter. We can define the theoretical distance ratio to calculate $\frac{D_{ls}}{Ds}$ in terms of the model parameters as

\begin{equation}\label{lensdist}
\frac{D_{ls}}{Ds}= \frac{\int_{zl}^{zs} \frac{dz}{H(z, \theta)}}{ \int_{0}^{zs} \frac{dz}{H(z, \theta)} },
\end{equation}
where $\theta$ is the free parameter for any cosmological model and $z_{l}$, $z_{s}$ are the redshifts to the lens and source respectively.\\

The SLS datasample comprises 144 measurements, where for each measurement, the provided data are: $z_l$, $z_s$, $\theta_E$, $d \theta_E$, $\sigma$, and $d \sigma$. Here, $z_l \epsilon (0.0625, 0.958)$ and $z_s \epsilon (0.2172,3.595)$ \citep{DEMODELSSLS}. Using the datasample we can estimate the cosmological parameters with the given $\chi^2$ function
\begin{equation}
\chi^2_{\text{SLS}} = \sum_{i=1}^{144} \left[ \frac{D_{\text{th}}(z_L, z_S) - D_{\text{obs}}(\theta_E, \sigma^2)}{\delta D_{\text{obs}}} \right]^2,    
\end{equation}
where $D_{\text{obs}} = {c^2 \theta_E}/{4\pi \sigma^2}$.
The errors of each $D_{obs}$ measurement which are given by
\begin{equation}
\delta D_{obs} = D_{obs} \sqrt{\left(\frac{\delta \theta_E}{\theta_E}\right)^2 + 4\left(\frac{\delta \sigma}{\sigma}\right)^2}.
\end{equation}
Where $\delta \theta_{E}$ is the error for the Einstein radius and $\delta \sigma$ the velocity dispersion.

%%%%%%%%%%%%%%%%%%%%%%%%%%%%%%%%%%%%%%%

\section{Results}\label{Results}

We performed a Bayesian analysis using the Markov Chain Monte Carlo (MCMC) Ensemble sampler from the emcee Python module \citep{mcmchammer}. 
For this analysis we performed 400 walkers near the maximum probability region and 5000 n-burn steps until convergence is reached using the autocorrelation time proposed by \citep{autocorrelation}. After convergence,  5000 more MCMC steps are done. The priors used are for $\Omega_{dm}$:[0.2,1.0] and $\alpha$:[1.5,10.0].

Tables \ref{parametroswphriess} and \ref{parametroswphplanck} show the best fit values to $1\sigma$ of the parameters of the model with gaussian prior of Planck on $h$ and gaussian prior of Riess on $h$ respectively.\\
We tested the Exponential Brane RS model for two cases: Exponential Brane RS model case Planck with a gaussian prior for the reduced hubble parameter $h=(0.674 \pm 0.005)$ \citep{Planck2020}  and Exponential Brane RS model case Riess with a reduced hubble parameter $h=(0.7330 \pm 0.0104)$ \citep{Riess_2022}. A gaussian prior for baryonic matter $\Omega_{b0} \cdot h^2 = (0.02218 \pm 0.00055)$ \cite{DESIBARYON2024} was used for both cases. We tested $\Lambda$CDM model using a gaussian prior on baryonic matter  
$\Omega_{b0} \cdot h^2 = ({0.02218} \pm {0.00055})$ \cite{DESIBARYON2024} to $1\sigma$ and obtained a value of  
$H(z=0) = 68.017^{+0.7314}_{-0.7196}$ and  
$q(z=0) =-0.5292^{+0.0176}_{-0.0158}$  
with a transition redshift of  
$z_{t} = 0.62262^{+0.0380}_{-0.0421}$.  
For case Planck to $1\sigma$, we obtained  
$H(z=0) = 67.9601^{+0.6605}_{0.5115}$  
and for case Riess $1\sigma$  
$H(z=0) = 70.90807^{+0.954}_{-0.831}$.

\begin{figure*}
    \centering
\includegraphics[width=0.5\textwidth]{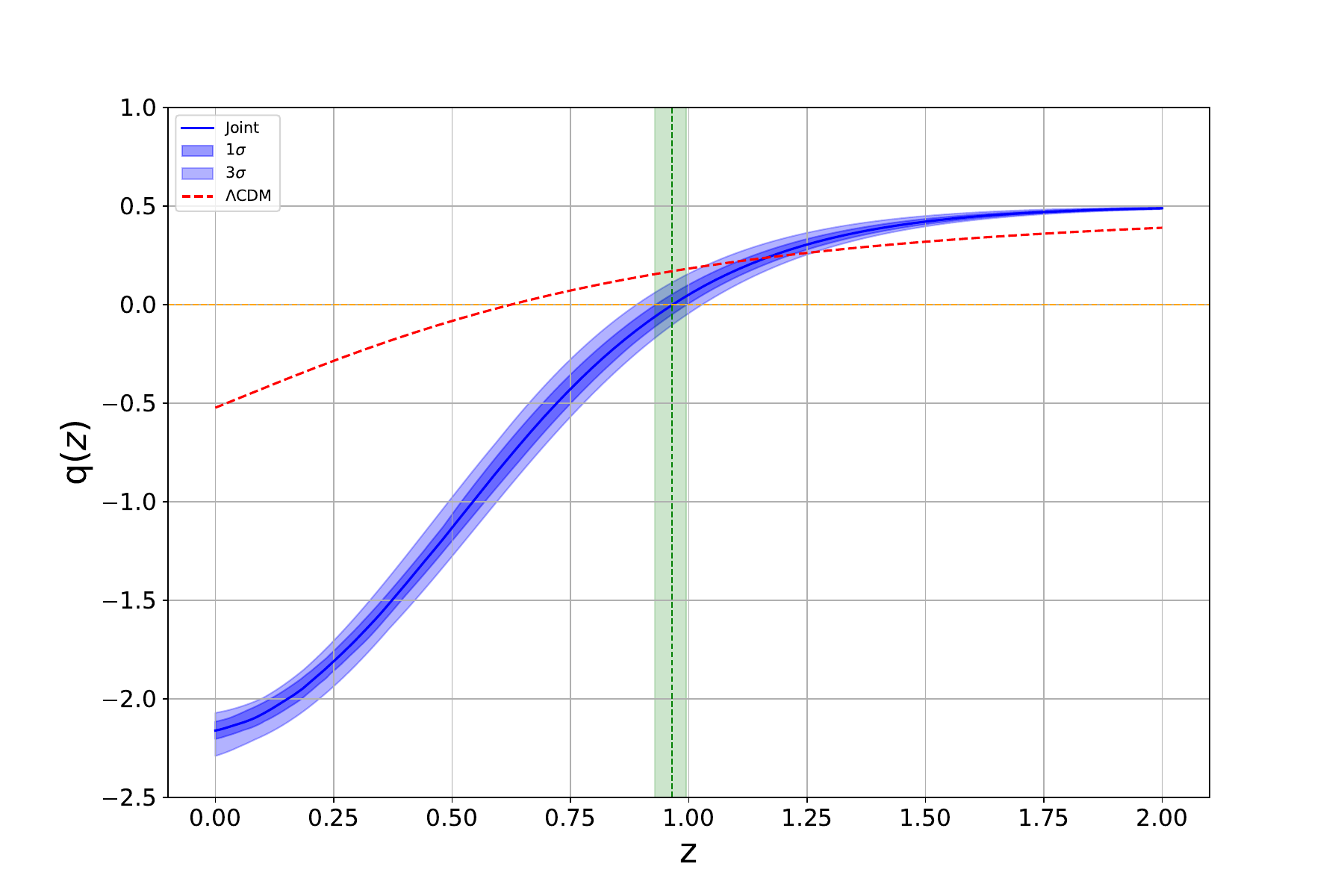}
\caption{ $q(z)$ joint best fit-curve of the Exponential Brane RS model case Planck and its uncertainty to 1$\sigma$ and 3$\sigma$.
}
\label{qzplanckgraph}
\end{figure*}

\begin{figure*}
    \centering
\includegraphics[width=0.5\textwidth]{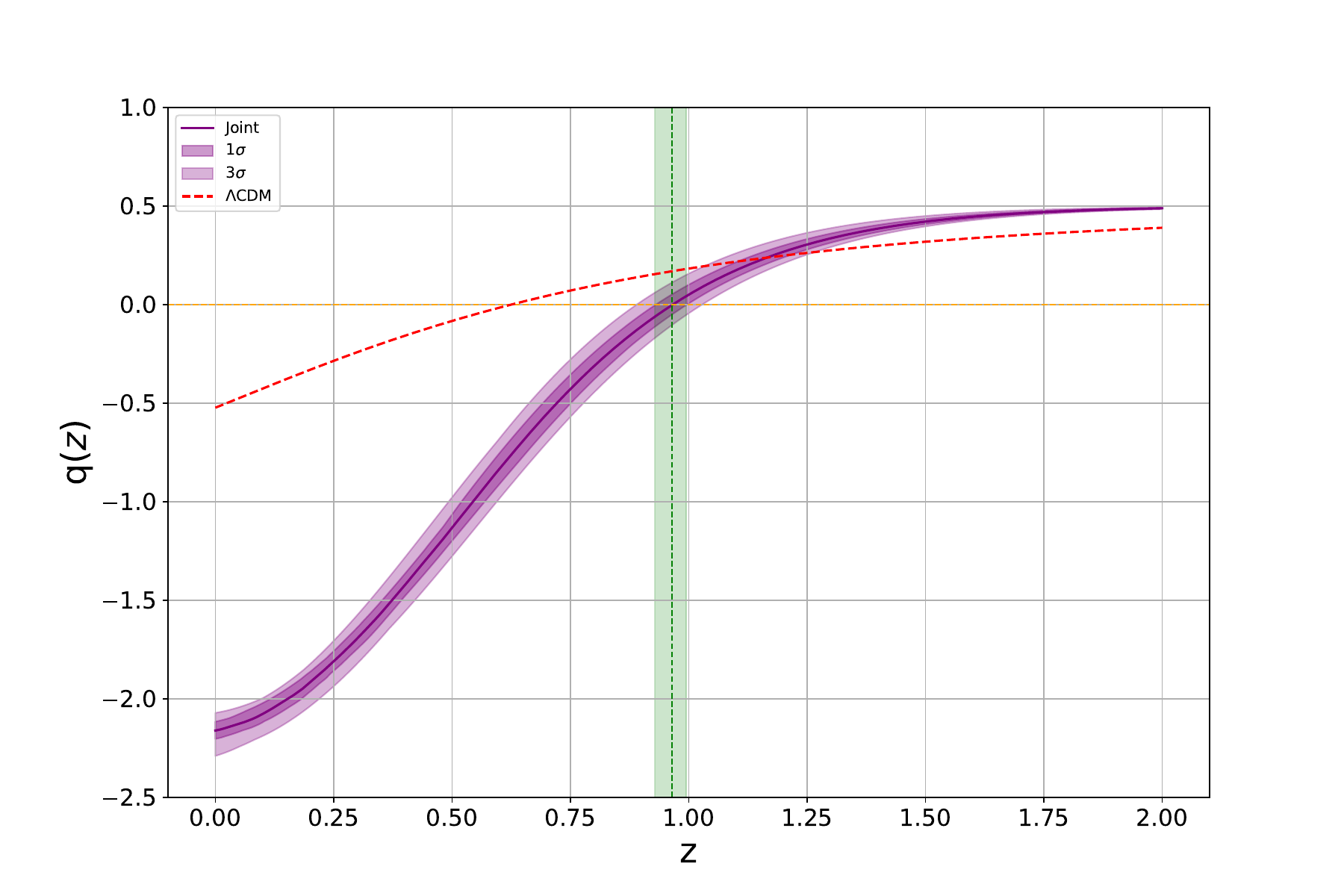}
\caption{ $q(z)$ joint best fit-curve of the Exponential Brane RS model case Riess and its uncertainty to 1$\sigma$ and 3$\sigma$.
}
\label{qzriessgraph}
\end{figure*}

Figures \ref{qzplanckgraph}, \ref{qzriessgraph} show the reconstruction of the deceleration parameter for the Exponential Brane RS model with their respective transition redshift values for both cases, and their transition value for redshift $z_{t}$ to $1\sigma$.

We obtained for the Exponential Brane RS model case Riess to $1\sigma$ $q(z=0) = -2.1606^{+0.0897}_{-0.12914}$  
and a transition value $z_{t} = 0.9649^{+0.020}_{-0.019}$, 
for case Planck to $1\sigma$  
$q(z=0) = -2.21784^{+0.1391}_{-0.1475}$  
and to $3\sigma$  
$z_{t} = 1.143^{+0.124}_{-0.122}$ respectively.

\begin{figure*}
    \centering
\includegraphics[width=0.5\textwidth]{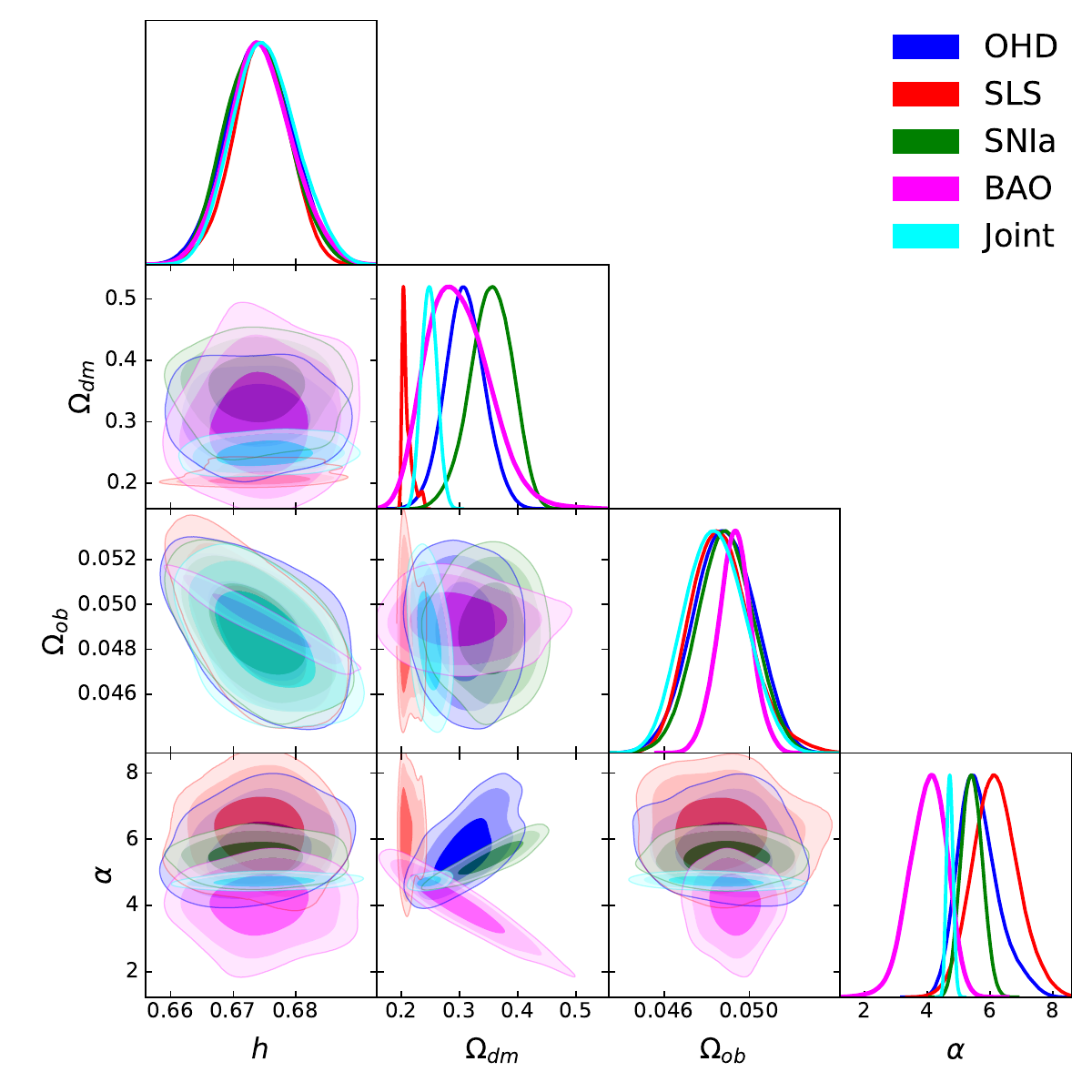}
\caption{Bottom panel: 2D contours of the free model parameters at 1$\sigma$, 2$\sigma$, and 3$\sigma$ (from darker to lighter color bands) confidence levels for OHD, SLS, SNIa, BAO data and Joint of them for the Exponential Brane RS model case Planck.
}
\label{contourswph}
\end{figure*}

\begin{figure*}    
\centering
\includegraphics[width=0.5\textwidth]{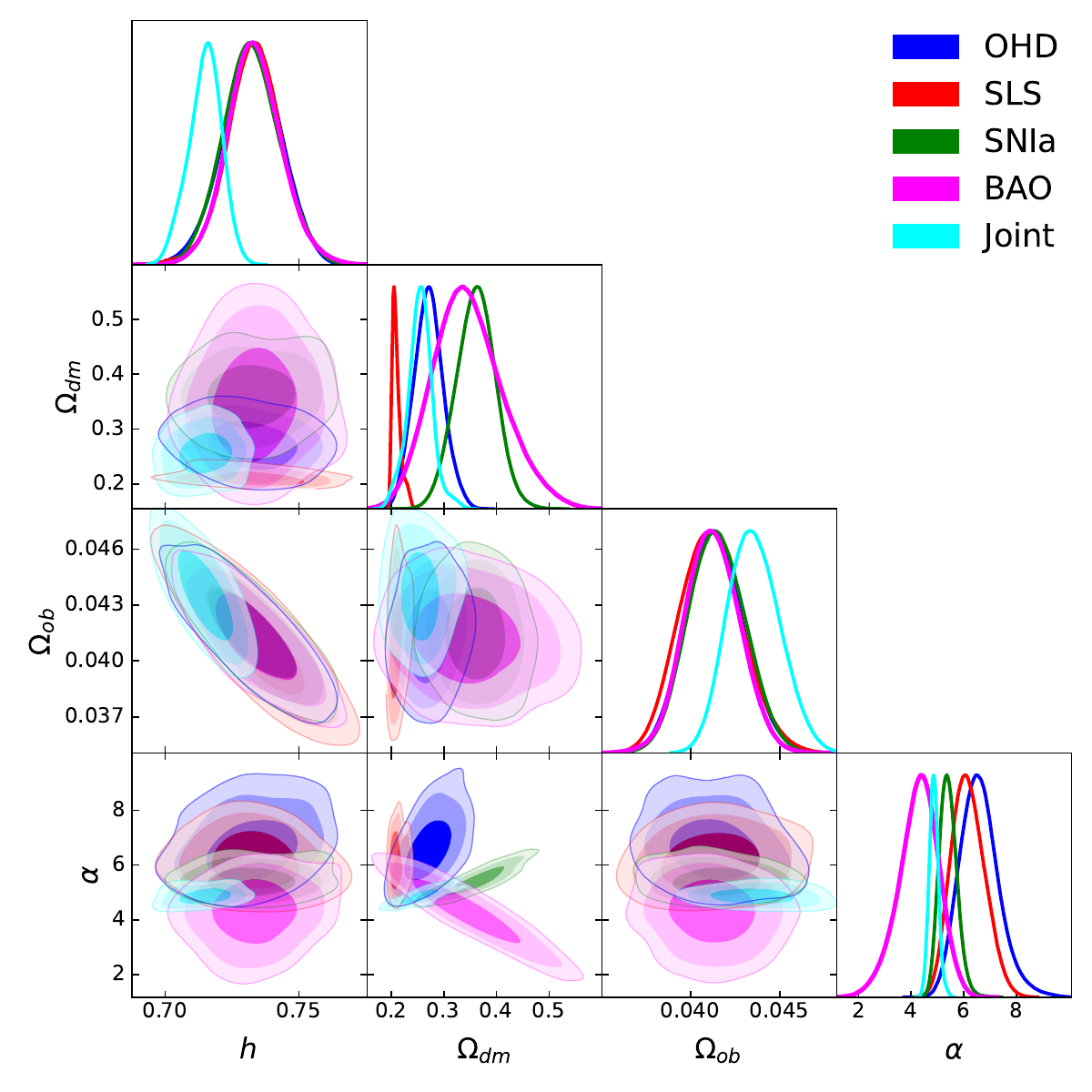}
\caption{Bottom panel: 2D contours of the free model parameters at 1$\sigma$, 2$\sigma$, and 3$\sigma$ (from darker to lighter color bands) confidence levels for OHD, SLS, SNIa, BAO data and Joint of them Exponential Brane RS model case Riess.
}
\label{contoursriess}
\end{figure*}

\begin{figure*}
    \centering
\includegraphics[width=0.5\textwidth]{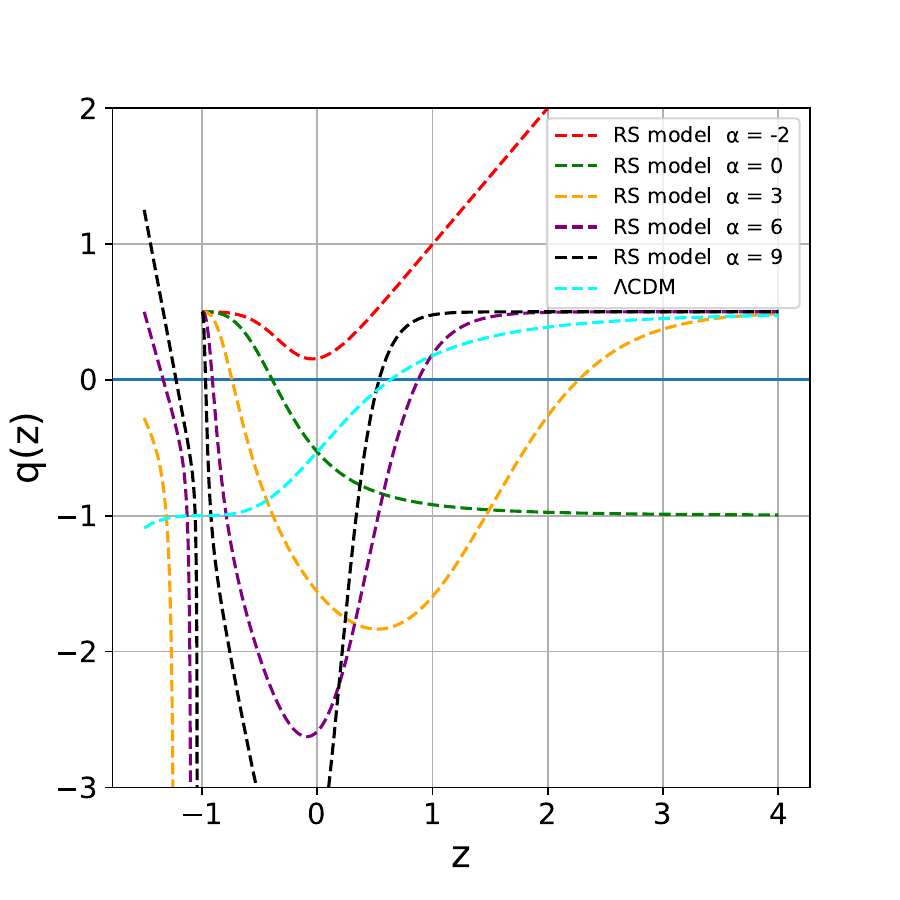} 

\caption{$q(z)$ test of different $\alpha$ in the Exponential Brane RS model assuming gaussian priors of Planck parameters on $h$, $\Omega_{dm}$, $\Omega_{bar}$.}
\label{qzwphalphatests}
\end{figure*}

Figure \ref{contourswph}, \ref{contoursriess} show the parameter probability distribution according to each datasample and the 2D plots confidence contours at $1\sigma$, $2\sigma$ and $3\sigma$ of the Exponential Brane RS model case Planck and case Riess.
Figure \ref{qzwphalphatests} shows the behavior of $q(z)$ assuming different values of the parameter $\alpha$ setting the rest of the parameters to the values given by Planck \citep{Planck2020}. Note that in this graph the $\alpha$ parameter is not a free parameter in the Bayesian method with MCMC chains.

From table \ref{bestmodelcriteria} and using the model criteria selection, it is obtained the values of ${\text{AIC}}_{\text{min}}= 2087.4$ and ${\text{BIC}}_{\text{min}}=2089.6$. With this minimum values, it results for $\Lambda$CDM a value for  
$\Delta\text{AIC}= 12.1$ and $\Delta\text{BIC}=1.01$,  
for Exponential Brane RS model with Planck priors results a value for  
$\Delta\text{AIC}=18.9$ and $\Delta\text{BIC}=18.9$. \\

\begin{figure*}
    \centering
    \begin{tabular}{ccc}
        \includegraphics[width=0.45\textwidth]{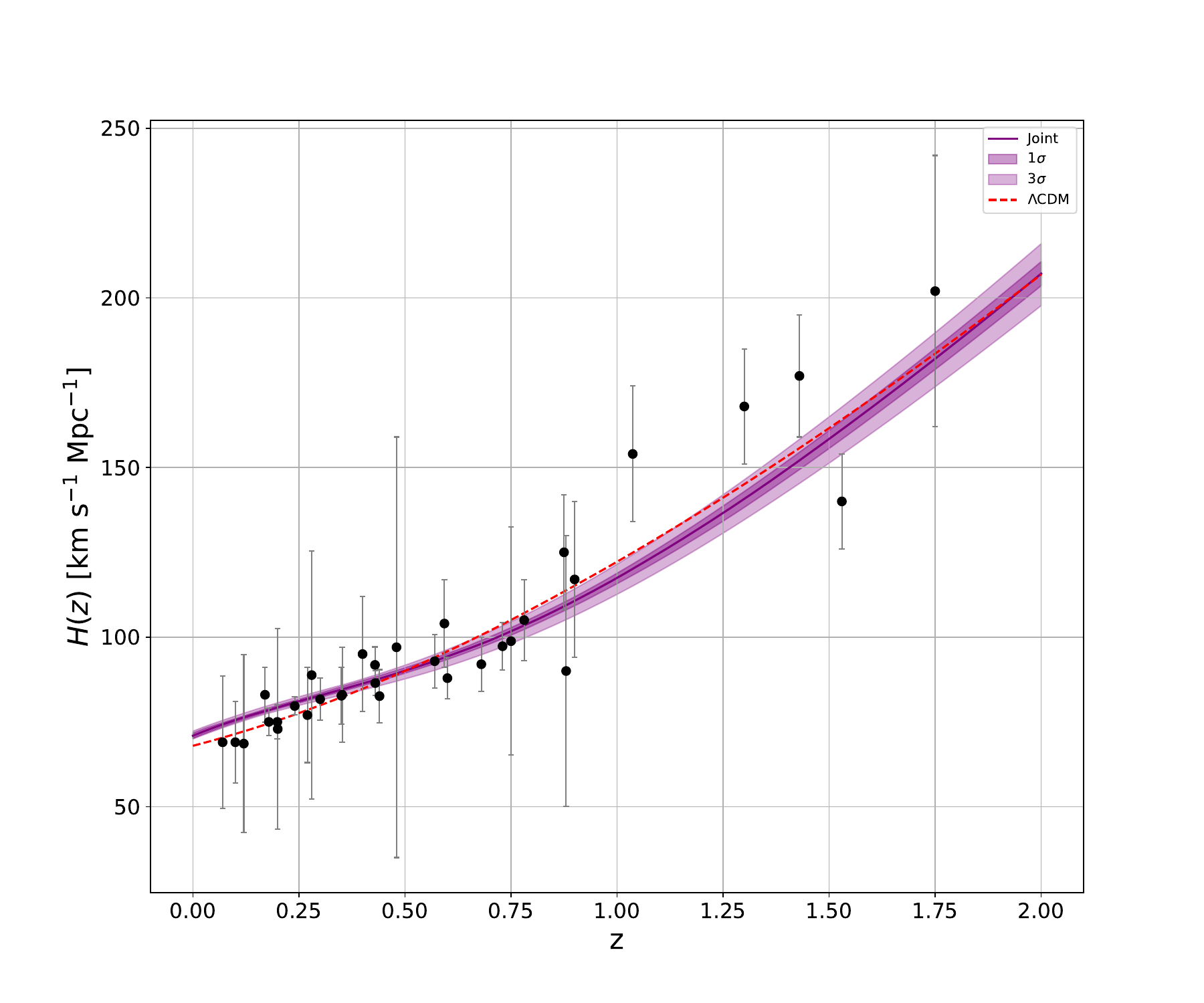} &
        \includegraphics[width=0.45\textwidth]{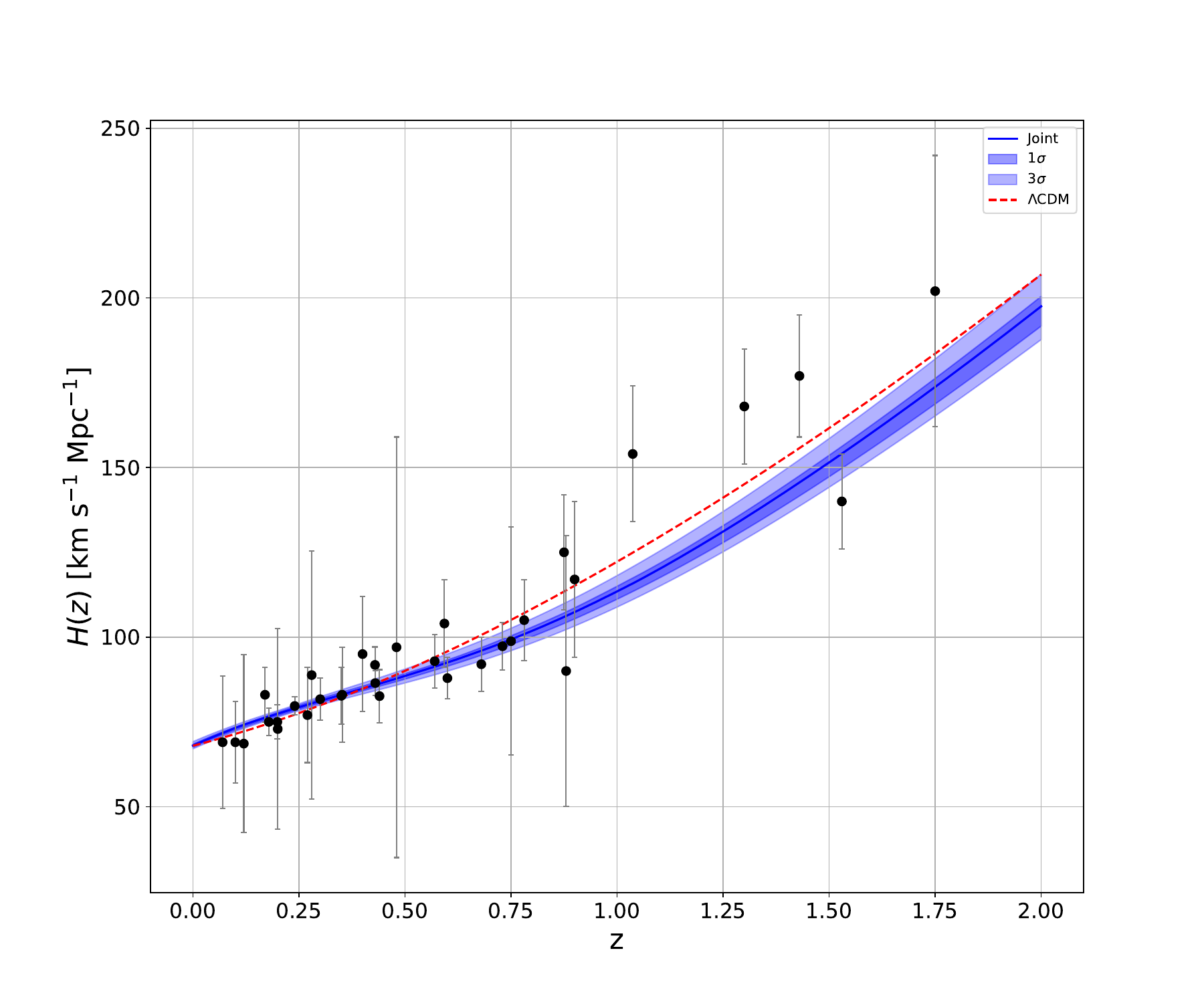} \\
        (a) & (b)
    \end{tabular}
    \caption{(a) $H(z)$ Joint best fit-curve of the Exponential Brane RS model case Riess and its uncertainty to 1$\sigma$ and 3$\sigma$. 
    (b) $H(z)$ Joint best fit-curve of the  Exponential Brane RS model case Planck and its uncertainty to 1$\sigma$ and 3$\sigma$.}
    \label{hzRIESSandPLANCKgraph}
\end{figure*}

\begin{table*}
\centering
\vspace{12pt} % 
\vspace{10pt} % 
{\renewcommand{\arraystretch}{1.6}
\begin{tabular}{|c|c|c|c|c|c|}
\hline
Data set & $h$ & $\Omega_{dm0}$ & $\Omega_{b0}$ & $\alpha$ &  $\chi^2_{min}$ \\
\cline{1-1}\cline{2-2}\cline{3-3}\cline{4-4}\cline{5-5}\cline{6-6}
OHD & $0.7328^{+0.0099}_{-0.0104}$ & $0.2717^{+0.0276}_{-0.0306}$ & $0.0413^{+0.0016}_{-0.0014}$  & $6.4645^{+0.8312}_{-0.5677}$ & $59.76$ \\
    \hline
    Sne Ia  & $0.7232^{+0.0096}_{-0.0099}$  & $0.3605^{+0.0348}_{-0.0416}$  & $0.0416^{+0.0009}_{-0.0010}$  & $5.3548^{+0.3140}_{-0.2759}$  & $1879.89$ \\
    \hline
    SLS & $0.7341^{+0.0095}_{-0.0100}$  & $0.2068^{+0.0129}_{-0.0048}$  & $0.0412^{+0.0017}_{-0.0017}$  & $6.0504^{+0.7524}_{-0.6042}$  & $245.31$ \\
    \hline
    BAO & $0.7341^{+0.0095}_{-0.0100}$  & $0.3150^{+0.0146}_{-0.00183}$  & $0.0412^{+0.0017}_{-0.0017}$  & $5.3503^{+0.624}_{-0.6138}$  & $39.96$ \\
    \hline
    Joint & $0.7090^{+0.0076}_{-0.0079}$  & $0.2638^{+0.0145}_{-0.0103}$  & $0.0433^{+0.0018}_{-0.0013}$  & $5.3270^{+0.1349}_{-0.1283}$  & $2078.935$ \\
    \hline 

\end{tabular}}
\caption{Best fit values to 1$\sigma$ for the Exponential Brane RS parameters using gaussian prior of Riess on $h$ and gaussian prior of DESI on $\Omega_{b0}$ with samples of OHD, SNIa, SLS, BAO and the Joint.}
\label{parametroswphriess}
\end{table*}

\begin{table*}
\centering
\vspace{12pt} % 
\vspace{10pt} %
{\renewcommand{\arraystretch}{1.6}
\begin{tabular}{|c|c|c|c|c|c|}
\hline
Data set & $h$ & $\Omega_{dm0}$ & $\Omega_{b0}$ & $\alpha$ &  $\chi^2_{min}$ \\
\cline{1-1}\cline{2-2}\cline{3-3}\cline{4-4}\cline{5-5}\cline{6-6}
\hline
OHD & $0.6741^{+0.0053}_{-0.0052}$ & $0.3073^{+0.0357}_{-0.0293}$ & $0.0487^{+0.0015}_{-0.0013}$ & $5.5436^{+0.7031}_{-0.5146}$ & $27.670$ \\
\hline
Sne Ia  & $0.6739^{+0.0045}_{-0.0053}$ & $0.3557^{+0.0360}_{-0.0358}$ & $0.0489^{+0.0013}_{-0.0013}$ & $5.4023^{+0.3377}_{-0.3241}$ & $1759.241$ \\
\hline
SLS & $0.6742^{+0.0049}_{-0.0048}$ & $0.2070^{+0.0101}_{-0.0052}$ & $0.0487^{+0.0015}_{-0.0014}$ & $6.0081^{+0.7056}_{-0.5731}$ & $246.421$ \\
\hline
BAO & $0.6755^{+0.0042}_{-0.0043}$ & $0.3902^{+0.0305}_{-0.02756 }$ & $0.0494^{+0.0017}_{-0.0011}$ & $4.5493^{+0.2013}_{-0.1983}$ & \textbf{$41.4732$} \\
\hline
Joint & $0.6755^{+0.0042}_{-0.0043}$ & $0.3056^{+0.0162}_{-0.0132}$ & $0.0484^{+0.0015}_{-0.0015}$ & $4.7188^{+0.059}_{-0.099}$ & $2099.241$ \\
\hline
\end{tabular}
\caption{Best fit values to 1$\sigma$ for the Exponential Brane RS parameters using gaussian prior of Planck on $h$ and gaussian prior of DESI on $\Omega_{b}$ with samples of OHD, SNIa, SLS, BAO and the Joint.}}
\label{parametroswphplanck}
\end{table*}

\begin{figure*}
    \centering
\includegraphics[width=0.5\textwidth]{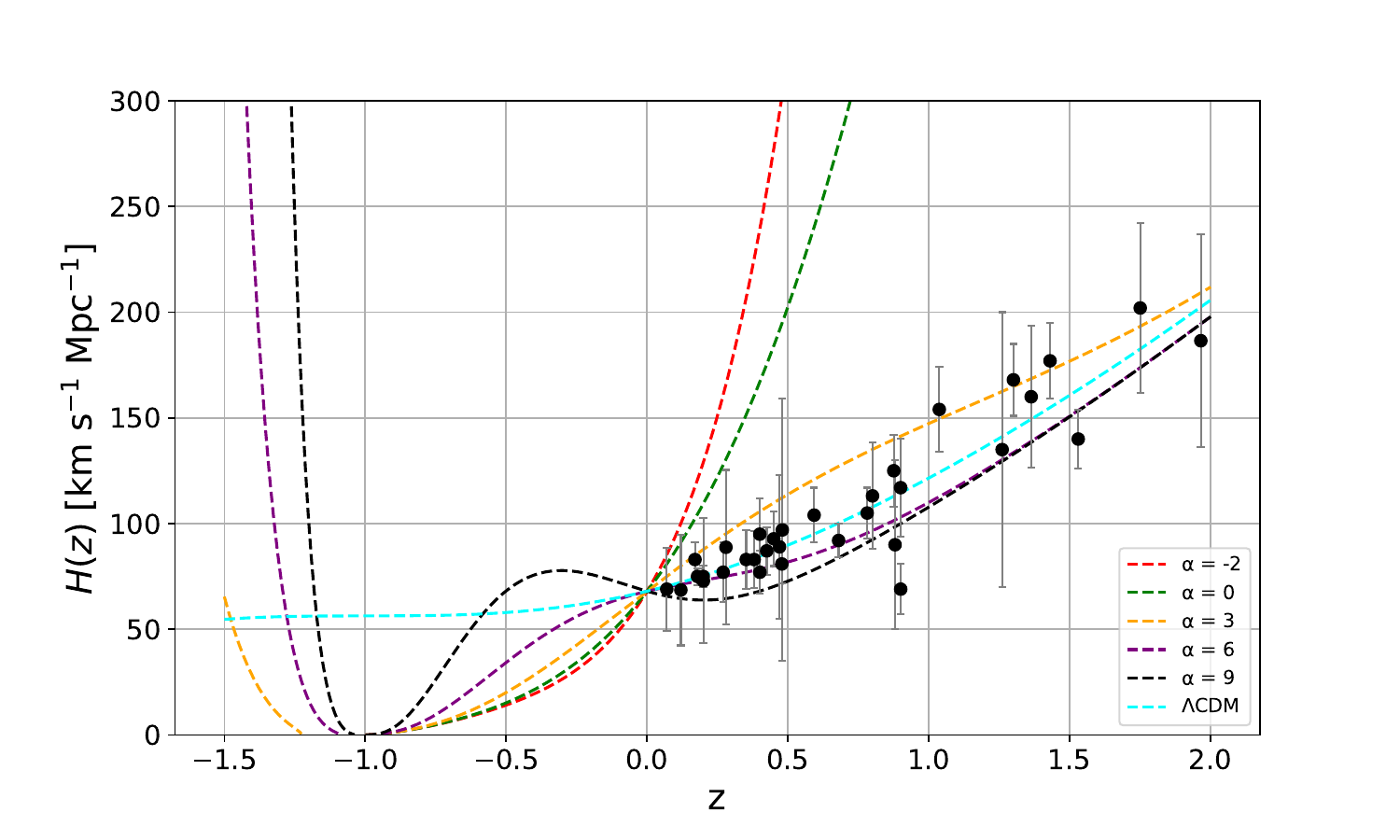} 
\caption{$H(z)$ test of differents $\alpha$ in the RS model assuming Planck gaussian priors for parameters f $h$, $\Omega_{dm}$, $\Omega_{bar}$.}
\label{HZwphalphatests}
\end{figure*}
\begin{table*}
\centering
\begin{tabular}{|c|c|}
\hline
$\Delta$AIC & Empirical support for model $i$ \\
\hline
0 - 2 & Substantial \\
4 - 7 & Considerably less \\
$>$ 10 & Essentially none \\
\hline
$\Delta$BIC & Evidence against model $i$ \\
\hline
0 - 2 & Not worth more than a bare mention \\
2 - 6 & Positive \\
6 - 10 & Strong \\
$>$ 10 & Very strong \\
\hline
\end{tabular}
\caption{$\Delta$AIC and $\Delta$BIC criteria.}
\label{bestmodelcriteria}
\end{table*}

%%%%%%%%%%%%%%%%%%%%%%%%%%%%%%%%%%%%%%%%%%%%%%%%%%%%%%%%%%
\section{Comparison of models}\label{comparison}

To compare among the different DE models, we use the Akaike Information Criterion \cite{1974ITAC...19..716A}, which is a measure of support of the model and Bayesian Information Criterion (BIC)\cite{BIC1974} which is a measure against the model. Both are defined as:

\begin{equation}
\text{AIC} = \chi^2_{\text{min}} + 2\alpha,
\end{equation}
\begin{equation}
\text{BIC} = \chi^2_{\text{min}} + \alpha \ln N,
\end{equation} 
where $\chi^2_{\text{min}}$ is the chi-square obtained from the best fit values of the parameters, $\alpha$ is the number of parameters, and $N$ the number of data points used in the fit. The AIC and BIC criteria  gives supports to the model and evidence against the model respectively. Table \ref{bestmodelcriteria} shows the $\Delta \text{AIC} = {AIC}{i} - \text{AIC}_{\text{min}}$ and $\Delta \text{BIC} = \text{BIC}_i - \text{BIC}_{\text{min}}$ criteria, where $ \text{AIC}_{\text{min}}$ and $\text{BIC}_{\text{min}}$ are minimum among all the models.

%%%%%%%%%%%%%%%%%%%%%%%%%%%%%%%%%%%%%%%%%%%%%%%%%%%%
\section{Discussion and conclusions} \label{conclusions}
The table \ref{parametroswphriess} and table \ref{parametroswphplanck} show that the parameter $\Omega_{dm0}$ is consistent to $ 1\sigma$ with the results published by \cite{Planck2020} and with $\Lambda$CDM. Table \ref{parametroswphriess} shows that the parameter $h$ is consistent to $1\sigma$ with the results published by \cite{Riess2022Hz} and the parameters $\Omega_{dm0}$ and $\Omega_{b0}$ are consistent to $3\sigma$ with \cite{Planck2020} and $\Lambda$CDM.
Fig. \ref{hzRIESSandPLANCKgraph} shows that the Hubble parameter has a similar shape but does not constrains well to $3\sigma$ along the graph compared to the $\Lambda CDM$ until $z\gtrsim 0.5$. The RS model with Riess prior on $h$ is consistent to $1\sigma$ for $\Lambda$CDM in $H(z=0)$ while the RS model with Planck priors is not. 
 
 Fig. \ref{qzplanckgraph} and fig. \ref{qzriessgraph} show that the $q(z)$ for both RS model with Planck priors Riess priors can describe the behavior of the universe at redshifts $z\gtrsim1.5$. They are also consistent with $\Lambda$CDM to 3$\sigma$ at $1.15 \lesssim z\lesssim 1.60$. At $z\lesssim 1.15$ the model does not constraint well to $\Lambda$CDM, where we can also compare with the deceleration parameter for $\Lambda$CDM given by $q(z=0)=-0.51^{+0.024}_{-0.024}$ \cite{Riess_2022}. The Exponential Brane RS case Planck and Riess priors on $h$ does not constraints well at $q(z=0)$ and $z_{t}$ compared to $\Lambda$CDM. Notice that the $q(z)$ for both Planck and Riess priors have a similar shape to the CPL model \citep{CPLmodelqzgraphs}, but are not consistent with it at $q(z=0)$.
 
Figures \ref{qzwphalphatests} and \ref{HZwphalphatests} compare the behavior of both models with different $\alpha$ values and it can be seen that for the RS model assuming parameters values of \cite{Planck2020}, only positive values of $\alpha$ can reproduce the accelerated expansion. It can also be concluded that
 the greater the value of $\alpha$ is the faster the universe will expand after a certain point, for example for $\alpha=9$ at $z\approx 1$ the accelerating expansion of the universe will be bigger. If $\alpha=0$ the universe starts at an accelerated state and then transitions to a decelerated expansion at $z\approx 1$. This model produces at any value of the parameter $\alpha$ produces a divergence at $z=-1$, notice that the value of $H(z)$ does not change independent of the value of $\alpha$ at $z \approx 0 $, with this information we can compare  the result of the free parameter in the case \ref{parametroswphriess} to $1\sigma$ given by the model $\alpha = 5.3270^{+0.1349}_{-0.1283}$ and for the case \ref{parametroswphplanck} to $1\sigma$ given by the model $\alpha = 4.7188^{+0.059}_{-0.099}$. This  results are in between $\alpha=[6,3]$ but closer to $\alpha=5$. This value of $\alpha$ makes the universe starts to accelerate at $z_{t} =  0.62262^{+0.0380}_{-0.0421}$ for the Exponential RS model case Planck and $z_{t} = 0.9649^{+0.020}_{-0.019}$ for the Exponential RS model case Riess, but then decelerates before $z=0$. If we take into account the results to $1\sigma$ the model can describe a universe that starts to accelerate late but the $\alpha$ parameter have a very much stronger influence in the acceleration and starts to decelerate after $z\approx0$.
 
 Table \ref{bestmodelcriteria} shows that the Exponential Brane RS model case Riess represents the  ${\text{AIC}}_{\text{min}}$ and ${\text{BIC}}_{\text{min}}$. We can compare all the  models given that the empirical support for the model $\Lambda$CDM is essentially none and the evidence against it is not worth more than a bare mention. For the Exponential Brane RS model case Planck the empirical support is essentially none and the evidence against it is very strong. We must also take into account that for the $\Lambda$CDM we used a a gaussian prior on baryonic matter $\Omega_{b0} h^2 = (0.02218 \pm 0.00055)$ while for the the Exponential brane RS case Planck and case Riess we used gaussian priors on $h$ and $\Omega_{b0}$, this may be affecting the final $\chi_{min}^2$ and therefore the model comparison criterion. It is important to note that all the samples, OHD, SNIa, SLS and BAO constrained better for the Exponential Brane RS model  case Riess.
 
 We can compare the parameter $n$ which represents the variable brane tension in \citep{GarciaAspeitia_2018} whose value ranges between $n \sim 6.19 \pm 0.12$, and the constraints obtained for the $\alpha$ parameter of this work. In this comparison we notice that the values of these two parameters are consistent to 3$\sigma$ but not to 1$\sigma$. It is argued that $n$ is very close to a cosmological constant $\Lambda$ at late times and for the future it evolves to behave as a phantom dark energy. The results obtained on this work suggest a similar scenario but here the phantom DE behavior occurs at earlier times. This is due to the exponential nature of the brane tension, ($e^{\alpha}$), which produces a \textit{stronger} acceleration of the universe than the power term ($(1+z)^n$) used in \citep{GarciaAspeitia_2018}. Also, fig. \ref{qzwphalphatests} shows that a $\alpha \sim6$ makes the universe accelerate more abruptly than the $n \sim 6$ of the VBT model and in both models the Big Rip at future times is inevitable.

 In conclusion, the Exponential Brane RS brane model is a suitable model to explain the evolution of our universe at late times up to present date showing a high concordance with $\Lambda$CDM, specially for $\Omega_{dm}$. Furthermore, this model can reproduce the accelerated expansion of the universe without the introduction of a DE fluid, accounting it instead to the properties of the Brane World encoded on the free parameter of the model, the brane tension $\Lambda$ to which the parameter $\alpha$ is related.  $\Omega_{dm}$ is consistent with the $\Lambda$CDM model and behaves as if in a phantom DE scenario after the $z_{t}$, which also happens earlier due to the influence of the brane tension . However, this behavior is less pronounced with other functional forms for the brane tension studied in literature, specifically for $\lambda= \lambda_0(1+z)^n$. Also, the exponential brane tension RS model studied in this work shows a higher consistency with the previous work for $\Omega_{dm}$ when using a Riess prior opposed to a Planck prior.
 
 The  future time (negative redshifts) evolution of the universe predicted by this model relies heavily on the constraint for the parameter $\alpha$, for certain values of it lead to divergences in the deceleration parameter. This also happens for the free parameter $n$ of the previous work, which means that certain values could lead either to a Big Crunch or a Big Rip. After comparing this model with the previous works with other proposed functions for the brane tension \citep{garcia2018brane},\citep{verdugo2024synchronize} we can conclude that in all cases the free parameter has a defining impact on the evolution of the universe compared to the other universe components, for it is not only capable of accelerating the expansion but also change the expansion rate at different redshifts, even causing divergences.\\ Knowing that an exponential  brane tension for the Randall Sundrum model can describe the evolution of the universe to an acceptable rate of agreement, for future work, this model can be explored further, especially in the context of tackling the  Hubble tension. To explore this, we will make make use of CMB data and test the evolution of $H_0$ at different universe's times. \\ 
 
\begin{acknowledgements}

The authors acknowledge Dr. Miguel Aspeitia and Dr. Juan Magaña for their crucial support throughout this project, from the development of the mathematical framework and numerical tools to comments for improving this manuscript. 
K.J acknowledges the Physics and Mathematics Department of the Universidad Iberoamericana.
D.A acknowledges the financial support of the Universidad Iberoamericana de Mexico where big part of this work was developed. 
We acknowledge the foreign relations department of Universidad Central de Chile.
\end{acknowledgements}

\bibliographystyle{apsrev4-2}  % O el estilo que estés usando
\bibliography{main}

\begin{thebibliography}{63}%
\makeatletter
\providecommand \@ifxundefined [1]{%
 \@ifx{#1\undefined}
}%
\providecommand \@ifnum [1]{%
 \ifnum #1\expandafter \@firstoftwo
 \else \expandafter \@secondoftwo
 \fi
}%
\providecommand \@ifx [1]{%
 \ifx #1\expandafter \@firstoftwo
 \else \expandafter \@secondoftwo
 \fi
}%
\providecommand \natexlab [1]{#1}%
\providecommand \enquote  [1]{``#1''}%
\providecommand \bibnamefont  [1]{#1}%
\providecommand \bibfnamefont [1]{#1}%
\providecommand \citenamefont [1]{#1}%
\providecommand \href@noop [0]{\@secondoftwo}%
\providecommand \href [0]{\begingroup \@sanitize@url \@href}%
\providecommand \@href[1]{\@@startlink{#1}\@@href}%
\providecommand \@@href[1]{\endgroup#1\@@endlink}%
\providecommand \@sanitize@url [0]{\catcode `\\12\catcode `\$12\catcode
  `\&12\catcode `\#12\catcode `\^12\catcode `\_12\catcode `\%12\relax}%
\providecommand \@@startlink[1]{}%
\providecommand \@@endlink[0]{}%
\providecommand \url  [0]{\begingroup\@sanitize@url \@url }%
\providecommand \@url [1]{\endgroup\@href {#1}{\urlprefix }}%
\providecommand \urlprefix  [0]{URL }%
\providecommand \Eprint [0]{\href }%
\providecommand \doibase [0]{https://doi.org/}%
\providecommand \selectlanguage [0]{\@gobble}%
\providecommand \bibinfo  [0]{\@secondoftwo}%
\providecommand \bibfield  [0]{\@secondoftwo}%
\providecommand \translation [1]{[#1]}%
\providecommand \BibitemOpen [0]{}%
\providecommand \bibitemStop [0]{}%
\providecommand \bibitemNoStop [0]{.\EOS\space}%
\providecommand \EOS [0]{\spacefactor3000\relax}%
\providecommand \BibitemShut  [1]{\csname bibitem#1\endcsname}%
\let\auto@bib@innerbib\@empty
%</preamble>
\bibitem [{\citenamefont {Aghanim}\ \emph {et~al.}(2020)\citenamefont {Aghanim}
  \emph {et~al.}}]{Planck2020}%
  \BibitemOpen
  \bibfield  {author} {\bibinfo {author} {\bibfnamefont {N.}~\bibnamefont
  {Aghanim}} \emph {et~al.} (\bibinfo {collaboration} {Planck}),\ }\href
  {https://doi.org/10.1051/0004-6361/201833910} {\bibfield  {journal} {\bibinfo
   {journal} {Astron. Astrophys.}\ }\textbf {\bibinfo {volume} {641}},\
  \bibinfo {pages} {A6} (\bibinfo {year} {2020})},\ \bibinfo {note} {[Erratum:
  Astron.Astrophys. 652, C4 (2021)]},\ \Eprint
  {https://arxiv.org/abs/1807.06209} {arXiv:1807.06209 [astro-ph.CO]}
  \BibitemShut {NoStop}%
\bibitem [{\citenamefont {Di~Valentino}\ \emph {et~al.}(2021)\citenamefont
  {Di~Valentino}, \citenamefont {Mena}, \citenamefont {Pan}, \citenamefont
  {Visinelli}, \citenamefont {Yang}, \citenamefont {Melchiorri}, \citenamefont
  {Mota}, \citenamefont {Riess},\ and\ \citenamefont {Silk}}]{di2021realm}%
  \BibitemOpen
  \bibfield  {author} {\bibinfo {author} {\bibfnamefont {E.}~\bibnamefont
  {Di~Valentino}}, \bibinfo {author} {\bibfnamefont {O.}~\bibnamefont {Mena}},
  \bibinfo {author} {\bibfnamefont {S.}~\bibnamefont {Pan}}, \bibinfo {author}
  {\bibfnamefont {L.}~\bibnamefont {Visinelli}}, \bibinfo {author}
  {\bibfnamefont {W.}~\bibnamefont {Yang}}, \bibinfo {author} {\bibfnamefont
  {A.}~\bibnamefont {Melchiorri}}, \bibinfo {author} {\bibfnamefont {D.~F.}\
  \bibnamefont {Mota}}, \bibinfo {author} {\bibfnamefont {A.~G.}\ \bibnamefont
  {Riess}},\ and\ \bibinfo {author} {\bibfnamefont {J.}~\bibnamefont {Silk}},\
  }\href@noop {} {\bibfield  {journal} {\bibinfo  {journal} {Classical and
  Quantum Gravity}\ }\textbf {\bibinfo {volume} {38}},\ \bibinfo {pages}
  {153001} (\bibinfo {year} {2021})}\BibitemShut {NoStop}%
\bibitem [{\citenamefont {Friederich}(2017)}]{friederich2017fine}%
  \BibitemOpen
  \bibfield  {author} {\bibinfo {author} {\bibfnamefont {S.}~\bibnamefont
  {Friederich}},\ }\href@noop {} {\bibfield  {journal} {\bibinfo  {journal}
  {The Stanford encyclopedia of philosophy}\ } (\bibinfo {year}
  {2017})}\BibitemShut {NoStop}%
\bibitem [{\citenamefont {Lima}(2004)}]{lima2004alternative}%
  \BibitemOpen
  \bibfield  {author} {\bibinfo {author} {\bibfnamefont {J.~A.~S.}\
  \bibnamefont {Lima}},\ }\href@noop {} {\bibfield  {journal} {\bibinfo
  {journal} {Brazilian Journal of Physics}\ }\textbf {\bibinfo {volume} {34}},\
  \bibinfo {pages} {194} (\bibinfo {year} {2004})}\BibitemShut {NoStop}%
\bibitem [{\citenamefont {Perivolaropoulos}\ and\ \citenamefont
  {Skara}(2022)}]{LCDMPROBLEMS}%
  \BibitemOpen
  \bibfield  {author} {\bibinfo {author} {\bibfnamefont {L.}~\bibnamefont
  {Perivolaropoulos}}\ and\ \bibinfo {author} {\bibfnamefont {F.}~\bibnamefont
  {Skara}},\ }\href {https://doi.org/10.1016/j.newar.2022.101659} {\bibfield
  {journal} {\bibinfo  {journal} {New Astronomy Reviews}\ }\textbf {\bibinfo
  {volume} {95}},\ \bibinfo {pages} {101659} (\bibinfo {year}
  {2022})}\BibitemShut {NoStop}%
\bibitem [{\citenamefont {Motta}\ \emph {et~al.}(2021)\citenamefont {Motta},
  \citenamefont {Garc{\'\i}a-Aspeitia}, \citenamefont {Hern{\'a}ndez-Almada},
  \citenamefont {Magana},\ and\ \citenamefont {Verdugo}}]{motta2021taxonomy}%
  \BibitemOpen
  \bibfield  {author} {\bibinfo {author} {\bibfnamefont {V.}~\bibnamefont
  {Motta}}, \bibinfo {author} {\bibfnamefont {M.~A.}\ \bibnamefont
  {Garc{\'\i}a-Aspeitia}}, \bibinfo {author} {\bibfnamefont {A.}~\bibnamefont
  {Hern{\'a}ndez-Almada}}, \bibinfo {author} {\bibfnamefont {J.}~\bibnamefont
  {Magana}},\ and\ \bibinfo {author} {\bibfnamefont {T.}~\bibnamefont
  {Verdugo}},\ }\href@noop {} {\bibfield  {journal} {\bibinfo  {journal}
  {Universe}\ }\textbf {\bibinfo {volume} {7}},\ \bibinfo {pages} {163}
  (\bibinfo {year} {2021})}\BibitemShut {NoStop}%
\bibitem [{\citenamefont {Caldwell}(2002)}]{caldwell2002phantom}%
  \BibitemOpen
  \bibfield  {author} {\bibinfo {author} {\bibfnamefont {R.~R.}\ \bibnamefont
  {Caldwell}},\ }\href@noop {} {\bibfield  {journal} {\bibinfo  {journal}
  {Physics Letters B}\ }\textbf {\bibinfo {volume} {545}},\ \bibinfo {pages}
  {23} (\bibinfo {year} {2002})}\BibitemShut {NoStop}%
\bibitem [{\citenamefont {Goswami}\ \emph {et~al.}(2019)\citenamefont
  {Goswami}, \citenamefont {Pradhan},\ and\ \citenamefont
  {Beesham}}]{quintessencemodelpaper}%
  \BibitemOpen
  \bibfield  {author} {\bibinfo {author} {\bibfnamefont {G.~K.}\ \bibnamefont
  {Goswami}}, \bibinfo {author} {\bibfnamefont {A.}~\bibnamefont {Pradhan}},\
  and\ \bibinfo {author} {\bibfnamefont {A.}~\bibnamefont {Beesham}},\ }\href
  {https://doi.org/10.1142/s0217732320500029} {\bibfield  {journal} {\bibinfo
  {journal} {Modern Physics Letters A}\ }\textbf {\bibinfo {volume} {35}},\
  \bibinfo {pages} {2050002} (\bibinfo {year} {2019})}\BibitemShut {NoStop}%
\bibitem [{\citenamefont {Roy}\ and\ \citenamefont {Bhadra}(2018)}]{phantomDE}%
  \BibitemOpen
  \bibfield  {author} {\bibinfo {author} {\bibfnamefont {N.}~\bibnamefont
  {Roy}}\ and\ \bibinfo {author} {\bibfnamefont {N.}~\bibnamefont {Bhadra}},\
  }\href {https://doi.org/10.1088/1475-7516/2018/06/002} {\bibfield  {journal}
  {\bibinfo  {journal} {Journal of Cosmology and Astroparticle Physics}\
  }\textbf {\bibinfo {volume} {2018}}\bibinfo  {number} { (06)},\ \bibinfo
  {pages} {002}}\BibitemShut {NoStop}%
\bibitem [{\citenamefont {Pawar}\ \emph {et~al.}(2018)\citenamefont {Pawar},
  \citenamefont {Bhuttampalle},\ and\ \citenamefont
  {Agrawal}}]{Kaluza-Kleinmodel}%
  \BibitemOpen
\bibfield  {number} {  }\bibfield  {author} {\bibinfo {author} {\bibfnamefont
  {D.}~\bibnamefont {Pawar}}, \bibinfo {author} {\bibfnamefont
  {G.}~\bibnamefont {Bhuttampalle}},\ and\ \bibinfo {author} {\bibfnamefont
  {P.}~\bibnamefont {Agrawal}},\ }\href
  {https://doi.org/10.1016/j.newast.2018.05.002} {\bibfield  {journal}
  {\bibinfo  {journal} {New Astronomy}\ }\textbf {\bibinfo {volume} {65}},\
  \bibinfo {pages} {1–6} (\bibinfo {year} {2018})}\BibitemShut {NoStop}%
\bibitem [{\citenamefont {Bona}\ and\ \citenamefont
  {Bezares}(2019)}]{Kaluza-Kleinmodel2}%
  \BibitemOpen
  \bibfield  {author} {\bibinfo {author} {\bibfnamefont {C.}~\bibnamefont
  {Bona}}\ and\ \bibinfo {author} {\bibfnamefont {M.}~\bibnamefont {Bezares}},\
  }\href {https://doi.org/10.1103/PhysRevD.100.043509} {\bibfield  {journal}
  {\bibinfo  {journal} {Phys. Rev. D}\ }\textbf {\bibinfo {volume} {100}},\
  \bibinfo {pages} {043509} (\bibinfo {year} {2019})}\BibitemShut {NoStop}%
\bibitem [{\citenamefont {Crisostomi}\ \emph {et~al.}(2016)\citenamefont
  {Crisostomi}, \citenamefont {Koyama},\ and\ \citenamefont
  {Tasinato}}]{scalar-thensortheories}%
  \BibitemOpen
  \bibfield  {author} {\bibinfo {author} {\bibfnamefont {M.}~\bibnamefont
  {Crisostomi}}, \bibinfo {author} {\bibfnamefont {K.}~\bibnamefont {Koyama}},\
  and\ \bibinfo {author} {\bibfnamefont {G.}~\bibnamefont {Tasinato}},\ }\href
  {https://doi.org/10.1088/1475-7516/2016/04/044} {\bibfield  {journal}
  {\bibinfo  {journal} {Journal of Cosmology and Astroparticle Physics}\
  }\textbf {\bibinfo {volume} {2016}}\bibinfo  {number} { (04)},\ \bibinfo
  {pages} {044–044}}\BibitemShut {NoStop}%
\bibitem [{\citenamefont {Adam}\ \emph {et~al.}(2022)\citenamefont {Adam},
  \citenamefont {Figueras}, \citenamefont {Jacobson},\ and\ \citenamefont
  {Wiseman}}]{Einstein-aethertheory}%
  \BibitemOpen
\bibfield  {number} {  }\bibfield  {author} {\bibinfo {author} {\bibfnamefont
  {A.}~\bibnamefont {Adam}}, \bibinfo {author} {\bibfnamefont {P.}~\bibnamefont
  {Figueras}}, \bibinfo {author} {\bibfnamefont {T.}~\bibnamefont {Jacobson}},\
  and\ \bibinfo {author} {\bibfnamefont {T.}~\bibnamefont {Wiseman}},\ }\href
  {https://doi.org/10.1088/1361-6382/ac5053} {\bibfield  {journal} {\bibinfo
  {journal} {Classical and Quantum Gravity}\ }\textbf {\bibinfo {volume}
  {39}},\ \bibinfo {pages} {125001} (\bibinfo {year} {2022})}\BibitemShut
  {NoStop}%
\bibitem [{\citenamefont {Amendola}\ \emph {et~al.}(2007)\citenamefont
  {Amendola}, \citenamefont {Gannouji}, \citenamefont {Polarski},\ and\
  \citenamefont {Tsujikawa}}]{f(r)gravitypaper1}%
  \BibitemOpen
  \bibfield  {author} {\bibinfo {author} {\bibfnamefont {L.}~\bibnamefont
  {Amendola}}, \bibinfo {author} {\bibfnamefont {R.}~\bibnamefont {Gannouji}},
  \bibinfo {author} {\bibfnamefont {D.}~\bibnamefont {Polarski}},\ and\
  \bibinfo {author} {\bibfnamefont {S.}~\bibnamefont {Tsujikawa}},\ }\bibfield
  {journal} {\bibinfo  {journal} {Physical Review D}\ }\textbf {\bibinfo
  {volume} {75}},\ \href {https://doi.org/10.1103/physrevd.75.083504}
  {10.1103/physrevd.75.083504} (\bibinfo {year} {2007})\BibitemShut {NoStop}%
\bibitem [{\citenamefont {Oikonomou}\ and\ \citenamefont
  {Giannakoudi}(2022)}]{f(r)gravitypaper2}%
  \BibitemOpen
  \bibfield  {author} {\bibinfo {author} {\bibfnamefont {V.~K.}\ \bibnamefont
  {Oikonomou}}\ and\ \bibinfo {author} {\bibfnamefont {I.}~\bibnamefont
  {Giannakoudi}},\ }\bibfield  {journal} {\bibinfo  {journal} {International
  Journal of Modern Physics D}\ }\textbf {\bibinfo {volume} {31}},\ \href
  {https://doi.org/10.1142/s0218271822500754} {10.1142/s0218271822500754}
  (\bibinfo {year} {2022})\BibitemShut {NoStop}%
\bibitem [{\citenamefont {Hořava}(2009)}]{Horavapaper1}%
  \BibitemOpen
  \bibfield  {author} {\bibinfo {author} {\bibfnamefont {P.}~\bibnamefont
  {Hořava}},\ }\bibfield  {journal} {\bibinfo  {journal} {Physical Review D}\
  }\textbf {\bibinfo {volume} {79}},\ \href
  {https://doi.org/10.1103/physrevd.79.084008} {10.1103/physrevd.79.084008}
  (\bibinfo {year} {2009})\BibitemShut {NoStop}%
\bibitem [{\citenamefont {Frusciante}\ and\ \citenamefont
  {Benetti}(2021)}]{Horavapaper2}%
  \BibitemOpen
  \bibfield  {author} {\bibinfo {author} {\bibfnamefont {N.}~\bibnamefont
  {Frusciante}}\ and\ \bibinfo {author} {\bibfnamefont {M.}~\bibnamefont
  {Benetti}},\ }\bibfield  {journal} {\bibinfo  {journal} {Physical Review D}\
  }\textbf {\bibinfo {volume} {103}},\ \href
  {https://doi.org/10.1103/physrevd.103.104060} {10.1103/physrevd.103.104060}
  (\bibinfo {year} {2021})\BibitemShut {NoStop}%
\bibitem [{\citenamefont {Baker}\ \emph {et~al.}(2013)\citenamefont {Baker},
  \citenamefont {Ferreira},\ and\ \citenamefont
  {Skordis}}]{post-friedmannpaper}%
  \BibitemOpen
  \bibfield  {author} {\bibinfo {author} {\bibfnamefont {T.}~\bibnamefont
  {Baker}}, \bibinfo {author} {\bibfnamefont {P.~G.}\ \bibnamefont
  {Ferreira}},\ and\ \bibinfo {author} {\bibfnamefont {C.}~\bibnamefont
  {Skordis}},\ }\bibfield  {journal} {\bibinfo  {journal} {Physical Review D}\
  }\textbf {\bibinfo {volume} {87}},\ \href
  {https://doi.org/10.1103/physrevd.87.024015} {10.1103/physrevd.87.024015}
  (\bibinfo {year} {2013})\BibitemShut {NoStop}%
\bibitem [{\citenamefont {Yekta}\ \emph {et~al.}(2021)\citenamefont {Yekta},
  \citenamefont {Alavi},\ and\ \citenamefont
  {Karimabadi}}]{higherdimensionspaper1}%
  \BibitemOpen
  \bibfield  {author} {\bibinfo {author} {\bibfnamefont {D.~M.}\ \bibnamefont
  {Yekta}}, \bibinfo {author} {\bibfnamefont {S.~A.}\ \bibnamefont {Alavi}},\
  and\ \bibinfo {author} {\bibfnamefont {M.}~\bibnamefont {Karimabadi}},\
  }\bibfield  {journal} {\bibinfo  {journal} {Galaxies}\ }\textbf {\bibinfo
  {volume} {9}},\ \href {https://doi.org/10.3390/galaxies9010004}
  {10.3390/galaxies9010004} (\bibinfo {year} {2021})\BibitemShut {NoStop}%
\bibitem [{\citenamefont {Al-Haysah}\ and\ \citenamefont
  {Hasmani}(2021)}]{higherdimensionalpaper2}%
  \BibitemOpen
  \bibfield  {author} {\bibinfo {author} {\bibfnamefont {A.~M.}\ \bibnamefont
  {Al-Haysah}}\ and\ \bibinfo {author} {\bibfnamefont {A.}~\bibnamefont
  {Hasmani}},\ }\href
  {https://doi.org/https://doi.org/10.1016/j.heliyon.2021.e08063} {\bibfield
  {journal} {\bibinfo  {journal} {Heliyon}\ }\textbf {\bibinfo {volume} {7}},\
  \bibinfo {pages} {e08063} (\bibinfo {year} {2021})}\BibitemShut {NoStop}%
\bibitem [{\citenamefont {Kaluza}(1921)}]{Kaluza:1921}%
  \BibitemOpen
  \bibfield  {author} {\bibinfo {author} {\bibfnamefont {T.}~\bibnamefont
  {Kaluza}},\ }\href {https://doi.org/10.1142/S0218271818700017} {\bibfield
  {journal} {\bibinfo  {journal} {Sitzungsber. Preuss. Akad. Wiss. Berlin
  (Math. Phys. )}\ }\textbf {\bibinfo {volume} {1921}},\ \bibinfo {pages} {966}
  (\bibinfo {year} {1921})},\ \Eprint {https://arxiv.org/abs/1803.08616}
  {arXiv:1803.08616 [physics.hist-ph]} \BibitemShut {NoStop}%
\bibitem [{\citenamefont {Bailin}\ and\ \citenamefont
  {Love}(1987)}]{Kaluzakleintheories}%
  \BibitemOpen
  \bibfield  {author} {\bibinfo {author} {\bibfnamefont {D.}~\bibnamefont
  {Bailin}}\ and\ \bibinfo {author} {\bibfnamefont {A.}~\bibnamefont {Love}},\
  }\href {https://doi.org/10.1088/0034-4885/50/9/001} {\bibfield  {journal}
  {\bibinfo  {journal} {Rept. Prog. Phys.}\ }\textbf {\bibinfo {volume} {50}},\
  \bibinfo {pages} {1087} (\bibinfo {year} {1987})}\BibitemShut {NoStop}%
\bibitem [{\citenamefont {Oli}\ and\ \citenamefont
  {Joshi}(2024)}]{Kaluza-kleinpaper2}%
  \BibitemOpen
  \bibfield  {author} {\bibinfo {author} {\bibfnamefont {S.}~\bibnamefont
  {Oli}}\ and\ \bibinfo {author} {\bibfnamefont {B.~P.}\ \bibnamefont
  {Joshi}},\ }\href {https://doi.org/10.1140/epjp/s13360-024-04912-x}
  {\bibfield  {journal} {\bibinfo  {journal} {Eur. Phys. J. Plus}\ }\textbf
  {\bibinfo {volume} {139}},\ \bibinfo {pages} {138} (\bibinfo {year}
  {2024})}\BibitemShut {NoStop}%
\bibitem [{\citenamefont {Dvali}\ \emph {et~al.}(2000)\citenamefont {Dvali},
  \citenamefont {Gabadadze},\ and\ \citenamefont
  {Porrati}}]{DGPgravityoriginal}%
  \BibitemOpen
  \bibfield  {author} {\bibinfo {author} {\bibfnamefont {G.}~\bibnamefont
  {Dvali}}, \bibinfo {author} {\bibfnamefont {G.}~\bibnamefont {Gabadadze}},\
  and\ \bibinfo {author} {\bibfnamefont {M.}~\bibnamefont {Porrati}},\ }\href
  {https://doi.org/10.1016/s0370-2693(00)00669-9} {\bibfield  {journal}
  {\bibinfo  {journal} {Physics Letters B}\ }\textbf {\bibinfo {volume}
  {485}},\ \bibinfo {pages} {208–214} (\bibinfo {year} {2000})}\BibitemShut
  {NoStop}%
\bibitem [{\citenamefont {de~Rham}\ \emph {et~al.}(2008)\citenamefont
  {de~Rham}, \citenamefont {Dvali}, \citenamefont {Hofmann}, \citenamefont
  {Khoury}, \citenamefont {Pujol\`as}, \citenamefont {Redi},\ and\
  \citenamefont {Tolley}}]{DPGCASCADINGGRAVITY}%
  \BibitemOpen
  \bibfield  {author} {\bibinfo {author} {\bibfnamefont {C.}~\bibnamefont
  {de~Rham}}, \bibinfo {author} {\bibfnamefont {G.}~\bibnamefont {Dvali}},
  \bibinfo {author} {\bibfnamefont {S.}~\bibnamefont {Hofmann}}, \bibinfo
  {author} {\bibfnamefont {J.}~\bibnamefont {Khoury}}, \bibinfo {author}
  {\bibfnamefont {O.}~\bibnamefont {Pujol\`as}}, \bibinfo {author}
  {\bibfnamefont {M.}~\bibnamefont {Redi}},\ and\ \bibinfo {author}
  {\bibfnamefont {A.~J.}\ \bibnamefont {Tolley}},\ }\href
  {https://doi.org/10.1103/PhysRevLett.100.251603} {\bibfield  {journal}
  {\bibinfo  {journal} {Phys. Rev. Lett.}\ }\textbf {\bibinfo {volume} {100}},\
  \bibinfo {pages} {251603} (\bibinfo {year} {2008})}\BibitemShut {NoStop}%
\bibitem [{\citenamefont {Randall}\ and\ \citenamefont
  {Sundrum}(1999{\natexlab{a}})}]{randall1999large}%
  \BibitemOpen
  \bibfield  {author} {\bibinfo {author} {\bibfnamefont {L.}~\bibnamefont
  {Randall}}\ and\ \bibinfo {author} {\bibfnamefont {R.}~\bibnamefont
  {Sundrum}},\ }\href@noop {} {\bibfield  {journal} {\bibinfo  {journal}
  {Physical review letters}\ }\textbf {\bibinfo {volume} {83}},\ \bibinfo
  {pages} {3370} (\bibinfo {year} {1999}{\natexlab{a}})}\BibitemShut {NoStop}%
\bibitem [{\citenamefont {Randall}\ and\ \citenamefont
  {Sundrum}(1999{\natexlab{b}})}]{randall1999alternative}%
  \BibitemOpen
  \bibfield  {author} {\bibinfo {author} {\bibfnamefont {L.}~\bibnamefont
  {Randall}}\ and\ \bibinfo {author} {\bibfnamefont {R.}~\bibnamefont
  {Sundrum}},\ }\href@noop {} {\bibfield  {journal} {\bibinfo  {journal}
  {Physical Review Letters}\ }\textbf {\bibinfo {volume} {83}},\ \bibinfo
  {pages} {4690} (\bibinfo {year} {1999}{\natexlab{b}})}\BibitemShut {NoStop}%
\bibitem [{\citenamefont {Casagrande}\ \emph {et~al.}(2008)\citenamefont
  {Casagrande}, \citenamefont {Goertz}, \citenamefont {Haisch}, \citenamefont
  {Neubert},\ and\ \citenamefont {Pfoh}}]{casagrande2008flavor}%
  \BibitemOpen
  \bibfield  {author} {\bibinfo {author} {\bibfnamefont {S.}~\bibnamefont
  {Casagrande}}, \bibinfo {author} {\bibfnamefont {F.}~\bibnamefont {Goertz}},
  \bibinfo {author} {\bibfnamefont {U.}~\bibnamefont {Haisch}}, \bibinfo
  {author} {\bibfnamefont {M.}~\bibnamefont {Neubert}},\ and\ \bibinfo {author}
  {\bibfnamefont {T.}~\bibnamefont {Pfoh}},\ }\href@noop {} {\bibfield
  {journal} {\bibinfo  {journal} {Journal of High Energy Physics}\ }\textbf
  {\bibinfo {volume} {2008}},\ \bibinfo {pages} {094} (\bibinfo {year}
  {2008})}\BibitemShut {NoStop}%
\bibitem [{\citenamefont {Reece}(2008)}]{reece2008particle}%
  \BibitemOpen
  \bibfield  {author} {\bibinfo {author} {\bibfnamefont {M.}~\bibnamefont
  {Reece}},\ }\href@noop {} {\  (\bibinfo {year} {2008})}\BibitemShut {NoStop}%
\bibitem [{\citenamefont {Dando}\ \emph {et~al.}(2005)\citenamefont {Dando},
  \citenamefont {Davidson}, \citenamefont {George}, \citenamefont {Volkas},\
  and\ \citenamefont {Wali}}]{dando2005clash}%
  \BibitemOpen
  \bibfield  {author} {\bibinfo {author} {\bibfnamefont {G.}~\bibnamefont
  {Dando}}, \bibinfo {author} {\bibfnamefont {A.}~\bibnamefont {Davidson}},
  \bibinfo {author} {\bibfnamefont {D.~P.}\ \bibnamefont {George}}, \bibinfo
  {author} {\bibfnamefont {R.~R.}\ \bibnamefont {Volkas}},\ and\ \bibinfo
  {author} {\bibfnamefont {K.~C.}\ \bibnamefont {Wali}},\ }\href@noop {}
  {\bibfield  {journal} {\bibinfo  {journal} {Physical Review D}\ }\textbf
  {\bibinfo {volume} {72}},\ \bibinfo {pages} {045016} (\bibinfo {year}
  {2005})}\BibitemShut {NoStop}%
\bibitem [{\citenamefont {Kim}\ and\ \citenamefont
  {Do~Kim}(2000)}]{kim2000inflation}%
  \BibitemOpen
  \bibfield  {author} {\bibinfo {author} {\bibfnamefont {H.~B.}\ \bibnamefont
  {Kim}}\ and\ \bibinfo {author} {\bibfnamefont {H.}~\bibnamefont {Do~Kim}},\
  }\href@noop {} {\bibfield  {journal} {\bibinfo  {journal} {Physical Review
  D}\ }\textbf {\bibinfo {volume} {61}},\ \bibinfo {pages} {064003} (\bibinfo
  {year} {2000})}\BibitemShut {NoStop}%
\bibitem [{\citenamefont {Acu{\~n}a-C{\'a}rdenas}\ \emph
  {et~al.}(2018)\citenamefont {Acu{\~n}a-C{\'a}rdenas}, \citenamefont
  {Astorga-Moreno}, \citenamefont {Garc{\'\i}a-Aspeitia},\ and\ \citenamefont
  {L{\'o}pez-Dom{\'\i}nguez}}]{acuna2018presence}%
  \BibitemOpen
  \bibfield  {author} {\bibinfo {author} {\bibfnamefont {R.~O.}\ \bibnamefont
  {Acu{\~n}a-C{\'a}rdenas}}, \bibinfo {author} {\bibfnamefont {J.}~\bibnamefont
  {Astorga-Moreno}}, \bibinfo {author} {\bibfnamefont {M.~A.}\ \bibnamefont
  {Garc{\'\i}a-Aspeitia}},\ and\ \bibinfo {author} {\bibfnamefont
  {J.}~\bibnamefont {L{\'o}pez-Dom{\'\i}nguez}},\ }\href@noop {} {\bibfield
  {journal} {\bibinfo  {journal} {Classical and Quantum Gravity}\ }\textbf
  {\bibinfo {volume} {35}},\ \bibinfo {pages} {035008} (\bibinfo {year}
  {2018})}\BibitemShut {NoStop}%
\bibitem [{\citenamefont {Gergely}(2009)}]{gergely2009eotvos}%
  \BibitemOpen
  \bibfield  {author} {\bibinfo {author} {\bibfnamefont {L.~A.}\ \bibnamefont
  {Gergely}},\ }\href@noop {} {\bibfield  {journal} {\bibinfo  {journal}
  {Physical Review D—Particles, Fields, Gravitation, and Cosmology}\ }\textbf
  {\bibinfo {volume} {79}},\ \bibinfo {pages} {086007} (\bibinfo {year}
  {2009})}\BibitemShut {NoStop}%
\bibitem [{\citenamefont {Garcia-Aspeitia}\ \emph {et~al.}(2018)\citenamefont
  {Garcia-Aspeitia}, \citenamefont {Hernandez-Almada}, \citenamefont {Magana},
  \citenamefont {Amante}, \citenamefont {Motta},\ and\ \citenamefont
  {Mart{\'\i}nez-Robles}}]{garcia2018brane}%
  \BibitemOpen
  \bibfield  {author} {\bibinfo {author} {\bibfnamefont {M.~A.}\ \bibnamefont
  {Garcia-Aspeitia}}, \bibinfo {author} {\bibfnamefont {A.}~\bibnamefont
  {Hernandez-Almada}}, \bibinfo {author} {\bibfnamefont {J.}~\bibnamefont
  {Magana}}, \bibinfo {author} {\bibfnamefont {M.~H.}\ \bibnamefont {Amante}},
  \bibinfo {author} {\bibfnamefont {V.}~\bibnamefont {Motta}},\ and\ \bibinfo
  {author} {\bibfnamefont {C.}~\bibnamefont {Mart{\'\i}nez-Robles}},\
  }\href@noop {} {\bibfield  {journal} {\bibinfo  {journal} {Physical Review
  D}\ }\textbf {\bibinfo {volume} {97}},\ \bibinfo {pages} {101301} (\bibinfo
  {year} {2018})}\BibitemShut {NoStop}%
\bibitem [{\citenamefont {Verdugo}\ \emph {et~al.}(2024)\citenamefont
  {Verdugo}, \citenamefont {Amante}, \citenamefont {Maga\~na}, \citenamefont
  {Garc\'\i{}a-Aspeitia}, \citenamefont {Hern\'andez-Almada},\ and\
  \citenamefont {Motta}}]{verdugo2024synchronize}%
  \BibitemOpen
  \bibfield  {author} {\bibinfo {author} {\bibfnamefont {T.}~\bibnamefont
  {Verdugo}}, \bibinfo {author} {\bibfnamefont {M.~H.}\ \bibnamefont {Amante}},
  \bibinfo {author} {\bibfnamefont {J.}~\bibnamefont {Maga\~na}}, \bibinfo
  {author} {\bibfnamefont {M.~A.}\ \bibnamefont {Garc\'\i{}a-Aspeitia}},
  \bibinfo {author} {\bibfnamefont {A.}~\bibnamefont {Hern\'andez-Almada}},\
  and\ \bibinfo {author} {\bibfnamefont {V.}~\bibnamefont {Motta}},\ }\href
  {https://doi.org/10.1140/epjc/s10052-024-12434-0} {\bibfield  {journal}
  {\bibinfo  {journal} {Eur. Phys. J. C}\ }\textbf {\bibinfo {volume} {84}},\
  \bibinfo {pages} {93} (\bibinfo {year} {2024})},\ \Eprint
  {https://arxiv.org/abs/2401.06376} {arXiv:2401.06376 [astro-ph.CO]}
  \BibitemShut {NoStop}%
\bibitem [{\citenamefont {Riess}\ \emph {et~al.}(1998)\citenamefont {Riess},
  \citenamefont {Filippenko}, \citenamefont {Challis}, \citenamefont
  {Clocchiatti}, \citenamefont {Diercks}, \citenamefont {Garnavich},
  \citenamefont {Gilliland}, \citenamefont {Hogan}, \citenamefont {Jha},
  \citenamefont {Kirshner} \emph {et~al.}}]{riess1998observational}%
  \BibitemOpen
  \bibfield  {author} {\bibinfo {author} {\bibfnamefont {A.~G.}\ \bibnamefont
  {Riess}}, \bibinfo {author} {\bibfnamefont {A.~V.}\ \bibnamefont
  {Filippenko}}, \bibinfo {author} {\bibfnamefont {P.}~\bibnamefont {Challis}},
  \bibinfo {author} {\bibfnamefont {A.}~\bibnamefont {Clocchiatti}}, \bibinfo
  {author} {\bibfnamefont {A.}~\bibnamefont {Diercks}}, \bibinfo {author}
  {\bibfnamefont {P.~M.}\ \bibnamefont {Garnavich}}, \bibinfo {author}
  {\bibfnamefont {R.~L.}\ \bibnamefont {Gilliland}}, \bibinfo {author}
  {\bibfnamefont {C.~J.}\ \bibnamefont {Hogan}}, \bibinfo {author}
  {\bibfnamefont {S.}~\bibnamefont {Jha}}, \bibinfo {author} {\bibfnamefont
  {R.~P.}\ \bibnamefont {Kirshner}}, \emph {et~al.},\ }\href@noop {} {\bibfield
   {journal} {\bibinfo  {journal} {The astronomical journal}\ }\textbf
  {\bibinfo {volume} {116}},\ \bibinfo {pages} {1009} (\bibinfo {year}
  {1998})}\BibitemShut {NoStop}%
\bibitem [{\citenamefont {Moresco}(2012)}]{DATAHZ}%
  \BibitemOpen
  \bibfield  {author} {\bibinfo {author} {\bibfnamefont {M.~e.~a.}\
  \bibnamefont {Moresco}},\ }\href@noop {} {\bibfield  {journal} {\bibinfo
  {journal} {Journal of Cosmology and Astroparticle Physics}\ }\textbf
  {\bibinfo {volume} {2012}}}\BibitemShut {NoStop}%
\bibitem [{\citenamefont {Borghi}\ \emph
  {et~al.}(2022{\natexlab{a}})\citenamefont {Borghi}, \citenamefont {Moresco},
  ,\ and\ \citenamefont {Cimatti}}]{HZNEWDATA32}%
  \BibitemOpen
  \bibfield  {author} {\bibinfo {author} {\bibfnamefont {N.}~\bibnamefont
  {Borghi}}, \bibinfo {author} {\bibfnamefont {M.}~\bibnamefont {Moresco}}, ,\
  and\ \bibinfo {author} {\bibfnamefont {A.}~\bibnamefont {Cimatti}},\
  }\href@noop {} {\bibfield  {journal} {\bibinfo  {journal} {The Astrophysical
  Journal Letters, Volume 928, Number 1}\ }\textbf {\bibinfo {volume} {928}},\
  \bibinfo {pages} {L4} (\bibinfo {year} {2022}{\natexlab{a}})}\BibitemShut
  {NoStop}%
\bibitem [{\citenamefont {C.~Nunes}\ \emph {et~al.}(2020)\citenamefont
  {C.~Nunes}, \citenamefont {K.~Yada}, \citenamefont {Jesus},\ and\
  \citenamefont {Bernui}}]{DATABAO}%
  \BibitemOpen
  \bibfield  {author} {\bibinfo {author} {\bibfnamefont {R.}~\bibnamefont
  {C.~Nunes}}, \bibinfo {author} {\bibfnamefont {S.}~\bibnamefont {K.~Yada}},
  \bibinfo {author} {\bibfnamefont {J.}~\bibnamefont {Jesus}},\ and\ \bibinfo
  {author} {\bibfnamefont {A.}~\bibnamefont {Bernui}},\ }\href@noop {}
  {\bibfield  {journal} {\bibinfo  {journal} {Monthly Notices of the Royal
  Astronomical Society}\ }\textbf {\bibinfo {volume} {497}},\ \bibinfo {pages}
  {2133–2141} (\bibinfo {year} {2020})}\BibitemShut {NoStop}%
\bibitem [{\citenamefont {Scolnic}(2022)}]{DATASN}%
  \BibitemOpen
  \bibfield  {author} {\bibinfo {author} {\bibfnamefont {D.~e.~a.}\
  \bibnamefont {Scolnic}},\ }\href@noop {} {\bibfield  {journal} {\bibinfo
  {journal} {The Astrophysical Journal}\ }\textbf {\bibinfo {volume} {938}},\
  \bibinfo {pages} {L2} (\bibinfo {year} {2022})}\BibitemShut {NoStop}%
\bibitem [{\citenamefont {Amante}\ \emph {et~al.}(2020)\citenamefont {Amante},
  \citenamefont {Magaña}, \citenamefont {Motta}, \citenamefont
  {García-Aspeitia},\ and\ \citenamefont {Verdugo}}]{DEMODELSSLS}%
  \BibitemOpen
  \bibfield  {author} {\bibinfo {author} {\bibfnamefont {M.~H.}\ \bibnamefont
  {Amante}}, \bibinfo {author} {\bibfnamefont {J.}~\bibnamefont {Magaña}},
  \bibinfo {author} {\bibfnamefont {V.}~\bibnamefont {Motta}}, \bibinfo
  {author} {\bibfnamefont {M.~A.}\ \bibnamefont {García-Aspeitia}},\ and\
  \bibinfo {author} {\bibfnamefont {T.}~\bibnamefont {Verdugo}},\ }\href
  {https://doi.org/10.1093/mnras/staa2760} {\bibfield  {journal} {\bibinfo
  {journal} {Monthly Notices of the Royal Astronomical Society}\ }\textbf
  {\bibinfo {volume} {498}},\ \bibinfo {pages} {6013} (\bibinfo {year}
  {2020})},\ \Eprint
  {https://arxiv.org/abs/https://academic.oup.com/mnras/article-pdf/498/4/6013/33838751/staa2760.pdf}
  {https://academic.oup.com/mnras/article-pdf/498/4/6013/33838751/staa2760.pdf}
  \BibitemShut {NoStop}%
\bibitem [{\citenamefont {Abdalla}\ \emph {et~al.}(2008)\citenamefont
  {Abdalla}, \citenamefont {Guimaraes},\ and\ \citenamefont {Hoff~da
  Silva}}]{abdalla2008gauss}%
  \BibitemOpen
  \bibfield  {author} {\bibinfo {author} {\bibfnamefont {M.}~\bibnamefont
  {Abdalla}}, \bibinfo {author} {\bibfnamefont {M.}~\bibnamefont {Guimaraes}},\
  and\ \bibinfo {author} {\bibfnamefont {J.}~\bibnamefont {Hoff~da Silva}},\
  }\href@noop {} {\bibfield  {journal} {\bibinfo  {journal} {The European
  Physical Journal C}\ }\textbf {\bibinfo {volume} {55}},\ \bibinfo {pages}
  {337} (\bibinfo {year} {2008})}\BibitemShut {NoStop}%
\bibitem [{\citenamefont {Musgrave}\ and\ \citenamefont
  {Lake}(1996)}]{musgrave1996junctions}%
  \BibitemOpen
  \bibfield  {author} {\bibinfo {author} {\bibfnamefont {P.}~\bibnamefont
  {Musgrave}}\ and\ \bibinfo {author} {\bibfnamefont {K.}~\bibnamefont
  {Lake}},\ }\href@noop {} {\bibfield  {journal} {\bibinfo  {journal}
  {Classical and Quantum Gravity}\ }\textbf {\bibinfo {volume} {13}},\ \bibinfo
  {pages} {1885} (\bibinfo {year} {1996})}\BibitemShut {NoStop}%
\bibitem [{\citenamefont {Melia}(2022)}]{melia2022friedmann}%
  \BibitemOpen
  \bibfield  {author} {\bibinfo {author} {\bibfnamefont {F.}~\bibnamefont
  {Melia}},\ }\href@noop {} {\bibfield  {journal} {\bibinfo  {journal} {Modern
  Physics Letters A}\ }\textbf {\bibinfo {volume} {37}},\ \bibinfo {pages}
  {2250016} (\bibinfo {year} {2022})}\BibitemShut {NoStop}%
\bibitem [{\citenamefont {Singh}(2008)}]{singh2008cosmological}%
  \BibitemOpen
  \bibfield  {author} {\bibinfo {author} {\bibfnamefont {J.}~\bibnamefont
  {Singh}},\ }\href@noop {} {\bibfield  {journal} {\bibinfo  {journal}
  {Astrophysics and space science}\ }\textbf {\bibinfo {volume} {318}},\
  \bibinfo {pages} {103} (\bibinfo {year} {2008})}\BibitemShut {NoStop}%
\bibitem [{\citenamefont {Favale}\ \emph {et~al.}(2023)\citenamefont {Favale},
  \citenamefont {Gómez-Valent},\ and\ \citenamefont
  {Migliaccio}}]{cosmicchronometers}%
  \BibitemOpen
  \bibfield  {author} {\bibinfo {author} {\bibfnamefont {A.}~\bibnamefont
  {Favale}}, \bibinfo {author} {\bibfnamefont {A.}~\bibnamefont
  {Gómez-Valent}},\ and\ \bibinfo {author} {\bibfnamefont {M.}~\bibnamefont
  {Migliaccio}},\ }\href {https://doi.org/10.1093/mnras/stad1621} {\bibfield
  {journal} {\bibinfo  {journal} {Monthly Notices of the Royal Astronomical
  Society}\ }\textbf {\bibinfo {volume} {523}},\ \bibinfo {pages} {3406–3422}
  (\bibinfo {year} {2023})}\BibitemShut {NoStop}%
\bibitem [{\citenamefont {Borghi}\ \emph
  {et~al.}(2022{\natexlab{b}})\citenamefont {Borghi}, \citenamefont {Moresco},\
  and\ \citenamefont {Cimatti}}]{borghi2022toward}%
  \BibitemOpen
  \bibfield  {author} {\bibinfo {author} {\bibfnamefont {N.}~\bibnamefont
  {Borghi}}, \bibinfo {author} {\bibfnamefont {M.}~\bibnamefont {Moresco}},\
  and\ \bibinfo {author} {\bibfnamefont {A.}~\bibnamefont {Cimatti}},\
  }\href@noop {} {\bibfield  {journal} {\bibinfo  {journal} {The Astrophysical
  Journal Letters}\ }\textbf {\bibinfo {volume} {928}},\ \bibinfo {pages} {L4}
  (\bibinfo {year} {2022}{\natexlab{b}})}\BibitemShut {NoStop}%
\bibitem [{\citenamefont {{Mahtessian}}\ \emph {et~al.}(2023)\citenamefont
  {{Mahtessian}}, \citenamefont {{Karapetian}}, \citenamefont
  {{Hovhannisyan}},\ and\ \citenamefont {{Mahtessian}}}]{AbsolutemagSNIa}%
  \BibitemOpen
  \bibfield  {author} {\bibinfo {author} {\bibfnamefont {A.~P.}\ \bibnamefont
  {{Mahtessian}}}, \bibinfo {author} {\bibfnamefont {G.~S.}\ \bibnamefont
  {{Karapetian}}}, \bibinfo {author} {\bibfnamefont {M.~A.}\ \bibnamefont
  {{Hovhannisyan}}},\ and\ \bibinfo {author} {\bibfnamefont {L.~A.}\
  \bibnamefont {{Mahtessian}}},\ }\href
  {https://doi.org/10.4236/ijaa.2023.132003} {\bibfield  {journal} {\bibinfo
  {journal} {International Journal of Astronomy and Astrophysics}\ }\textbf
  {\bibinfo {volume} {13}},\ \bibinfo {pages} {39} (\bibinfo {year} {2023})},\
  \Eprint {https://arxiv.org/abs/2209.00549} {arXiv:2209.00549 [astro-ph.CO]}
  \BibitemShut {NoStop}%
\bibitem [{\citenamefont {Peebles}\ and\ \citenamefont
  {Yu}(1970)}]{BAOFIRSTARTICLE}%
  \BibitemOpen
  \bibfield  {author} {\bibinfo {author} {\bibfnamefont {P.~J.~E.}\
  \bibnamefont {Peebles}}\ and\ \bibinfo {author} {\bibfnamefont {J.~T.}\
  \bibnamefont {Yu}},\ }\href {https://doi.org/10.1086/150713} {\bibfield
  {journal} {\bibinfo  {journal} {Astrophysical Journal}\ }\textbf {\bibinfo
  {volume} {162}},\ \bibinfo {pages} {815} (\bibinfo {year}
  {1970})}\BibitemShut {NoStop}%
\bibitem [{\citenamefont {Eisenstein}\ and\ \citenamefont
  {Hu}(1998)}]{BAOFORMULA}%
  \BibitemOpen
  \bibfield  {author} {\bibinfo {author} {\bibfnamefont {D.~J.}\ \bibnamefont
  {Eisenstein}}\ and\ \bibinfo {author} {\bibfnamefont {W.}~\bibnamefont
  {Hu}},\ }\href@noop {} {\bibfield  {journal} {\bibinfo  {journal} {The
  Astrophysical Journal}\ }\textbf {\bibinfo {volume} {496}},\ \bibinfo {pages}
  {L2} (\bibinfo {year} {1998})}\BibitemShut {NoStop}%
\bibitem [{\citenamefont {Collaboration}\ \emph
  {et~al.}(2024{\natexlab{a}})\citenamefont {Collaboration}, \citenamefont
  {Adame}, \citenamefont {Aguilar}, \citenamefont {Ahlen}, \citenamefont
  {Alam}, \citenamefont {Alexander}, \citenamefont {Alvarez}, \citenamefont
  {Alves}, \citenamefont {Anand}, \citenamefont {Andrade}, \citenamefont
  {Armengaud}, \citenamefont {Avila}, \citenamefont {Aviles}, \citenamefont
  {Awan}, \citenamefont {Bahr-Kalus}, \citenamefont {Bailey}, \citenamefont
  {Baltay}, \citenamefont {Bault}, \citenamefont {Behera}, \citenamefont
  {BenZvi}, \citenamefont {Bera}, \citenamefont {Beutler}, \citenamefont
  {Bianchi}, \citenamefont {Blake}, \citenamefont {Blum}, \citenamefont
  {Brieden}, \citenamefont {Brodzeller}, \citenamefont {Brooks}, \citenamefont
  {Buckley-Geer}, \citenamefont {Burtin}, \citenamefont {Calderon},
  \citenamefont {Canning}, \citenamefont {Rosell}, \citenamefont {Cereskaite},
  \citenamefont {Cervantes-Cota}, \citenamefont {Chabanier}, \citenamefont
  {Chaussidon}, \citenamefont {Chaves-Montero}, \citenamefont {Chen},
  \citenamefont {Chen}, \citenamefont {Claybaugh}, \citenamefont {Cole},
  \citenamefont {Cuceu}, \citenamefont {Davis}, \citenamefont {Dawson},
  \citenamefont {de~la Macorra}, \citenamefont {de~Mattia}, \citenamefont
  {Deiosso}, \citenamefont {Dey}, \citenamefont {Dey}, \citenamefont {Ding},
  \citenamefont {Doel}, \citenamefont {Edelstein}, \citenamefont
  {Eftekharzadeh}, \citenamefont {Eisenstein}, \citenamefont {Elliott},
  \citenamefont {Fagrelius}, \citenamefont {Fanning}, \citenamefont {Ferraro},
  \citenamefont {Ereza}, \citenamefont {Findlay}, \citenamefont {Flaugher},
  \citenamefont {Font-Ribera}, \citenamefont {Forero-Sánchez}, \citenamefont
  {Forero-Romero}, \citenamefont {Frenk}, \citenamefont {Garcia-Quintero},
  \citenamefont {Gaztañaga}, \citenamefont {Gil-Marín}, \citenamefont
  {Gontcho}, \citenamefont {Gonzalez-Morales}, \citenamefont {Gonzalez-Perez},
  \citenamefont {Gordon}, \citenamefont {Green}, \citenamefont {Gruen},
  \citenamefont {Gsponer}, \citenamefont {Gutierrez}, \citenamefont {Guy},
  \citenamefont {Hadzhiyska}, \citenamefont {Hahn}, \citenamefont {Hanif},
  \citenamefont {Herrera-Alcantar}, \citenamefont {Honscheid}, \citenamefont
  {Howlett}, \citenamefont {Huterer}, \citenamefont {Iršič}, \citenamefont
  {Ishak}, \citenamefont {Juneau}, \citenamefont {Karaçaylı}, \citenamefont
  {Kehoe}, \citenamefont {Kent}, \citenamefont {Kirkby}, \citenamefont
  {Kremin}, \citenamefont {Krolewski}, \citenamefont {Lai}, \citenamefont
  {Lan}, \citenamefont {Landriau}, \citenamefont {Lang}, \citenamefont
  {Lasker}, \citenamefont {Goff}, \citenamefont {Guillou}, \citenamefont
  {Leauthaud}, \citenamefont {Levi}, \citenamefont {Li}, \citenamefont
  {Linder}, \citenamefont {Lodha}, \citenamefont {Magneville}, \citenamefont
  {Manera}, \citenamefont {Margala}, \citenamefont {Martini}, \citenamefont
  {Maus}, \citenamefont {McDonald}, \citenamefont {Medina-Varela},
  \citenamefont {Meisner}, \citenamefont {Mena-Fernández}, \citenamefont
  {Miquel}, \citenamefont {Moon}, \citenamefont {Moore}, \citenamefont
  {Moustakas}, \citenamefont {Mudur}, \citenamefont {Mueller}, \citenamefont
  {Muñoz-Gutiérrez}, \citenamefont {Myers}, \citenamefont {Nadathur},
  \citenamefont {Napolitano}, \citenamefont {Neveux}, \citenamefont {Newman},
  \citenamefont {Nguyen}, \citenamefont {Nie}, \citenamefont {Niz},
  \citenamefont {Noriega}, \citenamefont {Padmanabhan}, \citenamefont
  {Paillas}, \citenamefont {Palanque-Delabrouille}, \citenamefont {Pan},
  \citenamefont {Penmetsa}, \citenamefont {Percival}, \citenamefont {Pieri},
  \citenamefont {Pinon}, \citenamefont {Poppett}, \citenamefont {Porredon},
  \citenamefont {Prada}, \citenamefont {Pérez-Fernández}, \citenamefont
  {Pérez-Ràfols}, \citenamefont {Rabinowitz}, \citenamefont {Raichoor},
  \citenamefont {Ramírez-Pérez}, \citenamefont {Ramirez-Solano},
  \citenamefont {Ravoux}, \citenamefont {Rashkovetskyi}, \citenamefont
  {Rezaie}, \citenamefont {Rich}, \citenamefont {Rocher}, \citenamefont
  {Rockosi}, \citenamefont {Roe}, \citenamefont {Rosado-Marin}, \citenamefont
  {Ross}, \citenamefont {Rossi}, \citenamefont {Ruggeri}, \citenamefont
  {Ruhlmann-Kleider}, \citenamefont {Samushia}, \citenamefont {Sanchez},
  \citenamefont {Saulder}, \citenamefont {Schlafly}, \citenamefont {Schlegel},
  \citenamefont {Schubnell}, \citenamefont {Seo}, \citenamefont {Shafieloo},
  \citenamefont {Sharples}, \citenamefont {Silber}, \citenamefont {Slosar},
  \citenamefont {Smith}, \citenamefont {Sprayberry}, \citenamefont {Tan},
  \citenamefont {Tarlé}, \citenamefont {Taylor}, \citenamefont {Trusov},
  \citenamefont {Ureña-López}, \citenamefont {Vaisakh}, \citenamefont
  {Valcin}, \citenamefont {Valdes}, \citenamefont {Vargas-Magaña},
  \citenamefont {Verde}, \citenamefont {Walther}, \citenamefont {Wang},
  \citenamefont {Wang}, \citenamefont {Weaver}, \citenamefont {Weaverdyck},
  \citenamefont {Wechsler}, \citenamefont {Weinberg}, \citenamefont {White},
  \citenamefont {Yu}, \citenamefont {Yu}, \citenamefont {Yuan}, \citenamefont
  {Yèche}, \citenamefont {Zaborowski}, \citenamefont {Zarrouk}, \citenamefont
  {Zhang}, \citenamefont {Zhao}, \citenamefont {Zhao}, \citenamefont {Zhou},
  \citenamefont {Zhuang},\ and\ \citenamefont {Zou}}]{desicollaboration2024}%
  \BibitemOpen
  \bibfield  {author} {\bibinfo {author} {\bibfnamefont {D.}~\bibnamefont
  {Collaboration}}, \bibinfo {author} {\bibfnamefont {A.~G.}\ \bibnamefont
  {Adame}}, \bibinfo {author} {\bibfnamefont {J.}~\bibnamefont {Aguilar}},
  \bibinfo {author} {\bibfnamefont {S.}~\bibnamefont {Ahlen}}, \bibinfo
  {author} {\bibfnamefont {S.}~\bibnamefont {Alam}}, \bibinfo {author}
  {\bibfnamefont {D.~M.}\ \bibnamefont {Alexander}}, \bibinfo {author}
  {\bibfnamefont {M.}~\bibnamefont {Alvarez}}, \bibinfo {author} {\bibfnamefont
  {O.}~\bibnamefont {Alves}}, \bibinfo {author} {\bibfnamefont
  {A.}~\bibnamefont {Anand}}, \bibinfo {author} {\bibfnamefont
  {U.}~\bibnamefont {Andrade}}, \bibinfo {author} {\bibfnamefont
  {E.}~\bibnamefont {Armengaud}}, \bibinfo {author} {\bibfnamefont
  {S.}~\bibnamefont {Avila}}, \bibinfo {author} {\bibfnamefont
  {A.}~\bibnamefont {Aviles}}, \bibinfo {author} {\bibfnamefont
  {H.}~\bibnamefont {Awan}}, \bibinfo {author} {\bibfnamefont {B.}~\bibnamefont
  {Bahr-Kalus}}, \bibinfo {author} {\bibfnamefont {S.}~\bibnamefont {Bailey}},
  \bibinfo {author} {\bibfnamefont {C.}~\bibnamefont {Baltay}}, \bibinfo
  {author} {\bibfnamefont {A.}~\bibnamefont {Bault}}, \bibinfo {author}
  {\bibfnamefont {J.}~\bibnamefont {Behera}}, \bibinfo {author} {\bibfnamefont
  {S.}~\bibnamefont {BenZvi}}, \bibinfo {author} {\bibfnamefont
  {A.}~\bibnamefont {Bera}}, \bibinfo {author} {\bibfnamefont {F.}~\bibnamefont
  {Beutler}}, \bibinfo {author} {\bibfnamefont {D.}~\bibnamefont {Bianchi}},
  \bibinfo {author} {\bibfnamefont {C.}~\bibnamefont {Blake}}, \bibinfo
  {author} {\bibfnamefont {R.}~\bibnamefont {Blum}}, \bibinfo {author}
  {\bibfnamefont {S.}~\bibnamefont {Brieden}}, \bibinfo {author} {\bibfnamefont
  {A.}~\bibnamefont {Brodzeller}}, \bibinfo {author} {\bibfnamefont
  {D.}~\bibnamefont {Brooks}}, \bibinfo {author} {\bibfnamefont
  {E.}~\bibnamefont {Buckley-Geer}}, \bibinfo {author} {\bibfnamefont
  {E.}~\bibnamefont {Burtin}}, \bibinfo {author} {\bibfnamefont
  {R.}~\bibnamefont {Calderon}}, \bibinfo {author} {\bibfnamefont
  {R.}~\bibnamefont {Canning}}, \bibinfo {author} {\bibfnamefont {A.~C.}\
  \bibnamefont {Rosell}}, \bibinfo {author} {\bibfnamefont {R.}~\bibnamefont
  {Cereskaite}}, \bibinfo {author} {\bibfnamefont {J.~L.}\ \bibnamefont
  {Cervantes-Cota}}, \bibinfo {author} {\bibfnamefont {S.}~\bibnamefont
  {Chabanier}}, \bibinfo {author} {\bibfnamefont {E.}~\bibnamefont
  {Chaussidon}}, \bibinfo {author} {\bibfnamefont {J.}~\bibnamefont
  {Chaves-Montero}}, \bibinfo {author} {\bibfnamefont {S.}~\bibnamefont
  {Chen}}, \bibinfo {author} {\bibfnamefont {X.}~\bibnamefont {Chen}}, \bibinfo
  {author} {\bibfnamefont {T.}~\bibnamefont {Claybaugh}}, \bibinfo {author}
  {\bibfnamefont {S.}~\bibnamefont {Cole}}, \bibinfo {author} {\bibfnamefont
  {A.}~\bibnamefont {Cuceu}}, \bibinfo {author} {\bibfnamefont {T.~M.}\
  \bibnamefont {Davis}}, \bibinfo {author} {\bibfnamefont {K.}~\bibnamefont
  {Dawson}}, \bibinfo {author} {\bibfnamefont {A.}~\bibnamefont {de~la
  Macorra}}, \bibinfo {author} {\bibfnamefont {A.}~\bibnamefont {de~Mattia}},
  \bibinfo {author} {\bibfnamefont {N.}~\bibnamefont {Deiosso}}, \bibinfo
  {author} {\bibfnamefont {A.}~\bibnamefont {Dey}}, \bibinfo {author}
  {\bibfnamefont {B.}~\bibnamefont {Dey}}, \bibinfo {author} {\bibfnamefont
  {Z.}~\bibnamefont {Ding}}, \bibinfo {author} {\bibfnamefont {P.}~\bibnamefont
  {Doel}}, \bibinfo {author} {\bibfnamefont {J.}~\bibnamefont {Edelstein}},
  \bibinfo {author} {\bibfnamefont {S.}~\bibnamefont {Eftekharzadeh}}, \bibinfo
  {author} {\bibfnamefont {D.~J.}\ \bibnamefont {Eisenstein}}, \bibinfo
  {author} {\bibfnamefont {A.}~\bibnamefont {Elliott}}, \bibinfo {author}
  {\bibfnamefont {P.}~\bibnamefont {Fagrelius}}, \bibinfo {author}
  {\bibfnamefont {K.}~\bibnamefont {Fanning}}, \bibinfo {author} {\bibfnamefont
  {S.}~\bibnamefont {Ferraro}}, \bibinfo {author} {\bibfnamefont
  {J.}~\bibnamefont {Ereza}}, \bibinfo {author} {\bibfnamefont
  {N.}~\bibnamefont {Findlay}}, \bibinfo {author} {\bibfnamefont
  {B.}~\bibnamefont {Flaugher}}, \bibinfo {author} {\bibfnamefont
  {A.}~\bibnamefont {Font-Ribera}}, \bibinfo {author} {\bibfnamefont
  {D.}~\bibnamefont {Forero-Sánchez}}, \bibinfo {author} {\bibfnamefont
  {J.~E.}\ \bibnamefont {Forero-Romero}}, \bibinfo {author} {\bibfnamefont
  {C.~S.}\ \bibnamefont {Frenk}}, \bibinfo {author} {\bibfnamefont
  {C.}~\bibnamefont {Garcia-Quintero}}, \bibinfo {author} {\bibfnamefont
  {E.}~\bibnamefont {Gaztañaga}}, \bibinfo {author} {\bibfnamefont
  {H.}~\bibnamefont {Gil-Marín}}, \bibinfo {author} {\bibfnamefont {S.~G.~A.}\
  \bibnamefont {Gontcho}}, \bibinfo {author} {\bibfnamefont {A.~X.}\
  \bibnamefont {Gonzalez-Morales}}, \bibinfo {author} {\bibfnamefont
  {V.}~\bibnamefont {Gonzalez-Perez}}, \bibinfo {author} {\bibfnamefont
  {C.}~\bibnamefont {Gordon}}, \bibinfo {author} {\bibfnamefont
  {D.}~\bibnamefont {Green}}, \bibinfo {author} {\bibfnamefont
  {D.}~\bibnamefont {Gruen}}, \bibinfo {author} {\bibfnamefont
  {R.}~\bibnamefont {Gsponer}}, \bibinfo {author} {\bibfnamefont
  {G.}~\bibnamefont {Gutierrez}}, \bibinfo {author} {\bibfnamefont
  {J.}~\bibnamefont {Guy}}, \bibinfo {author} {\bibfnamefont {B.}~\bibnamefont
  {Hadzhiyska}}, \bibinfo {author} {\bibfnamefont {C.}~\bibnamefont {Hahn}},
  \bibinfo {author} {\bibfnamefont {M.~M.~S.}\ \bibnamefont {Hanif}}, \bibinfo
  {author} {\bibfnamefont {H.~K.}\ \bibnamefont {Herrera-Alcantar}}, \bibinfo
  {author} {\bibfnamefont {K.}~\bibnamefont {Honscheid}}, \bibinfo {author}
  {\bibfnamefont {C.}~\bibnamefont {Howlett}}, \bibinfo {author} {\bibfnamefont
  {D.}~\bibnamefont {Huterer}}, \bibinfo {author} {\bibfnamefont
  {V.}~\bibnamefont {Iršič}}, \bibinfo {author} {\bibfnamefont
  {M.}~\bibnamefont {Ishak}}, \bibinfo {author} {\bibfnamefont
  {S.}~\bibnamefont {Juneau}}, \bibinfo {author} {\bibfnamefont {N.~G.}\
  \bibnamefont {Karaçaylı}}, \bibinfo {author} {\bibfnamefont
  {R.}~\bibnamefont {Kehoe}}, \bibinfo {author} {\bibfnamefont
  {S.}~\bibnamefont {Kent}}, \bibinfo {author} {\bibfnamefont {D.}~\bibnamefont
  {Kirkby}}, \bibinfo {author} {\bibfnamefont {A.}~\bibnamefont {Kremin}},
  \bibinfo {author} {\bibfnamefont {A.}~\bibnamefont {Krolewski}}, \bibinfo
  {author} {\bibfnamefont {Y.}~\bibnamefont {Lai}}, \bibinfo {author}
  {\bibfnamefont {T.~W.}\ \bibnamefont {Lan}}, \bibinfo {author} {\bibfnamefont
  {M.}~\bibnamefont {Landriau}}, \bibinfo {author} {\bibfnamefont
  {D.}~\bibnamefont {Lang}}, \bibinfo {author} {\bibfnamefont {J.}~\bibnamefont
  {Lasker}}, \bibinfo {author} {\bibfnamefont {J.~M.~L.}\ \bibnamefont {Goff}},
  \bibinfo {author} {\bibfnamefont {L.~L.}\ \bibnamefont {Guillou}}, \bibinfo
  {author} {\bibfnamefont {A.}~\bibnamefont {Leauthaud}}, \bibinfo {author}
  {\bibfnamefont {M.~E.}\ \bibnamefont {Levi}}, \bibinfo {author}
  {\bibfnamefont {T.~S.}\ \bibnamefont {Li}}, \bibinfo {author} {\bibfnamefont
  {E.}~\bibnamefont {Linder}}, \bibinfo {author} {\bibfnamefont
  {K.}~\bibnamefont {Lodha}}, \bibinfo {author} {\bibfnamefont
  {C.}~\bibnamefont {Magneville}}, \bibinfo {author} {\bibfnamefont
  {M.}~\bibnamefont {Manera}}, \bibinfo {author} {\bibfnamefont
  {D.}~\bibnamefont {Margala}}, \bibinfo {author} {\bibfnamefont
  {P.}~\bibnamefont {Martini}}, \bibinfo {author} {\bibfnamefont
  {M.}~\bibnamefont {Maus}}, \bibinfo {author} {\bibfnamefont {P.}~\bibnamefont
  {McDonald}}, \bibinfo {author} {\bibfnamefont {L.}~\bibnamefont
  {Medina-Varela}}, \bibinfo {author} {\bibfnamefont {A.}~\bibnamefont
  {Meisner}}, \bibinfo {author} {\bibfnamefont {J.}~\bibnamefont
  {Mena-Fernández}}, \bibinfo {author} {\bibfnamefont {R.}~\bibnamefont
  {Miquel}}, \bibinfo {author} {\bibfnamefont {J.}~\bibnamefont {Moon}},
  \bibinfo {author} {\bibfnamefont {S.}~\bibnamefont {Moore}}, \bibinfo
  {author} {\bibfnamefont {J.}~\bibnamefont {Moustakas}}, \bibinfo {author}
  {\bibfnamefont {N.}~\bibnamefont {Mudur}}, \bibinfo {author} {\bibfnamefont
  {E.}~\bibnamefont {Mueller}}, \bibinfo {author} {\bibfnamefont
  {A.}~\bibnamefont {Muñoz-Gutiérrez}}, \bibinfo {author} {\bibfnamefont
  {A.~D.}\ \bibnamefont {Myers}}, \bibinfo {author} {\bibfnamefont
  {S.}~\bibnamefont {Nadathur}}, \bibinfo {author} {\bibfnamefont
  {L.}~\bibnamefont {Napolitano}}, \bibinfo {author} {\bibfnamefont
  {R.}~\bibnamefont {Neveux}}, \bibinfo {author} {\bibfnamefont {J.~A.}\
  \bibnamefont {Newman}}, \bibinfo {author} {\bibfnamefont {N.~M.}\
  \bibnamefont {Nguyen}}, \bibinfo {author} {\bibfnamefont {J.}~\bibnamefont
  {Nie}}, \bibinfo {author} {\bibfnamefont {G.}~\bibnamefont {Niz}}, \bibinfo
  {author} {\bibfnamefont {H.~E.}\ \bibnamefont {Noriega}}, \bibinfo {author}
  {\bibfnamefont {N.}~\bibnamefont {Padmanabhan}}, \bibinfo {author}
  {\bibfnamefont {E.}~\bibnamefont {Paillas}}, \bibinfo {author} {\bibfnamefont
  {N.}~\bibnamefont {Palanque-Delabrouille}}, \bibinfo {author} {\bibfnamefont
  {J.}~\bibnamefont {Pan}}, \bibinfo {author} {\bibfnamefont {S.}~\bibnamefont
  {Penmetsa}}, \bibinfo {author} {\bibfnamefont {W.~J.}\ \bibnamefont
  {Percival}}, \bibinfo {author} {\bibfnamefont {M.~M.}\ \bibnamefont {Pieri}},
  \bibinfo {author} {\bibfnamefont {M.}~\bibnamefont {Pinon}}, \bibinfo
  {author} {\bibfnamefont {C.}~\bibnamefont {Poppett}}, \bibinfo {author}
  {\bibfnamefont {A.}~\bibnamefont {Porredon}}, \bibinfo {author}
  {\bibfnamefont {F.}~\bibnamefont {Prada}}, \bibinfo {author} {\bibfnamefont
  {A.}~\bibnamefont {Pérez-Fernández}}, \bibinfo {author} {\bibfnamefont
  {I.}~\bibnamefont {Pérez-Ràfols}}, \bibinfo {author} {\bibfnamefont
  {D.}~\bibnamefont {Rabinowitz}}, \bibinfo {author} {\bibfnamefont
  {A.}~\bibnamefont {Raichoor}}, \bibinfo {author} {\bibfnamefont
  {C.}~\bibnamefont {Ramírez-Pérez}}, \bibinfo {author} {\bibfnamefont
  {S.}~\bibnamefont {Ramirez-Solano}}, \bibinfo {author} {\bibfnamefont
  {C.}~\bibnamefont {Ravoux}}, \bibinfo {author} {\bibfnamefont
  {M.}~\bibnamefont {Rashkovetskyi}}, \bibinfo {author} {\bibfnamefont
  {M.}~\bibnamefont {Rezaie}}, \bibinfo {author} {\bibfnamefont
  {J.}~\bibnamefont {Rich}}, \bibinfo {author} {\bibfnamefont {A.}~\bibnamefont
  {Rocher}}, \bibinfo {author} {\bibfnamefont {C.}~\bibnamefont {Rockosi}},
  \bibinfo {author} {\bibfnamefont {N.~A.}\ \bibnamefont {Roe}}, \bibinfo
  {author} {\bibfnamefont {A.}~\bibnamefont {Rosado-Marin}}, \bibinfo {author}
  {\bibfnamefont {A.~J.}\ \bibnamefont {Ross}}, \bibinfo {author}
  {\bibfnamefont {G.}~\bibnamefont {Rossi}}, \bibinfo {author} {\bibfnamefont
  {R.}~\bibnamefont {Ruggeri}}, \bibinfo {author} {\bibfnamefont
  {V.}~\bibnamefont {Ruhlmann-Kleider}}, \bibinfo {author} {\bibfnamefont
  {L.}~\bibnamefont {Samushia}}, \bibinfo {author} {\bibfnamefont
  {E.}~\bibnamefont {Sanchez}}, \bibinfo {author} {\bibfnamefont
  {C.}~\bibnamefont {Saulder}}, \bibinfo {author} {\bibfnamefont {E.~F.}\
  \bibnamefont {Schlafly}}, \bibinfo {author} {\bibfnamefont {D.}~\bibnamefont
  {Schlegel}}, \bibinfo {author} {\bibfnamefont {M.}~\bibnamefont {Schubnell}},
  \bibinfo {author} {\bibfnamefont {H.}~\bibnamefont {Seo}}, \bibinfo {author}
  {\bibfnamefont {A.}~\bibnamefont {Shafieloo}}, \bibinfo {author}
  {\bibfnamefont {R.}~\bibnamefont {Sharples}}, \bibinfo {author}
  {\bibfnamefont {J.}~\bibnamefont {Silber}}, \bibinfo {author} {\bibfnamefont
  {A.}~\bibnamefont {Slosar}}, \bibinfo {author} {\bibfnamefont
  {A.}~\bibnamefont {Smith}}, \bibinfo {author} {\bibfnamefont
  {D.}~\bibnamefont {Sprayberry}}, \bibinfo {author} {\bibfnamefont
  {T.}~\bibnamefont {Tan}}, \bibinfo {author} {\bibfnamefont {G.}~\bibnamefont
  {Tarlé}}, \bibinfo {author} {\bibfnamefont {P.}~\bibnamefont {Taylor}},
  \bibinfo {author} {\bibfnamefont {S.}~\bibnamefont {Trusov}}, \bibinfo
  {author} {\bibfnamefont {L.~A.}\ \bibnamefont {Ureña-López}}, \bibinfo
  {author} {\bibfnamefont {R.}~\bibnamefont {Vaisakh}}, \bibinfo {author}
  {\bibfnamefont {D.}~\bibnamefont {Valcin}}, \bibinfo {author} {\bibfnamefont
  {F.}~\bibnamefont {Valdes}}, \bibinfo {author} {\bibfnamefont
  {M.}~\bibnamefont {Vargas-Magaña}}, \bibinfo {author} {\bibfnamefont
  {L.}~\bibnamefont {Verde}}, \bibinfo {author} {\bibfnamefont
  {M.}~\bibnamefont {Walther}}, \bibinfo {author} {\bibfnamefont
  {B.}~\bibnamefont {Wang}}, \bibinfo {author} {\bibfnamefont {M.~S.}\
  \bibnamefont {Wang}}, \bibinfo {author} {\bibfnamefont {B.~A.}\ \bibnamefont
  {Weaver}}, \bibinfo {author} {\bibfnamefont {N.}~\bibnamefont {Weaverdyck}},
  \bibinfo {author} {\bibfnamefont {R.~H.}\ \bibnamefont {Wechsler}}, \bibinfo
  {author} {\bibfnamefont {D.~H.}\ \bibnamefont {Weinberg}}, \bibinfo {author}
  {\bibfnamefont {M.}~\bibnamefont {White}}, \bibinfo {author} {\bibfnamefont
  {J.}~\bibnamefont {Yu}}, \bibinfo {author} {\bibfnamefont {Y.}~\bibnamefont
  {Yu}}, \bibinfo {author} {\bibfnamefont {S.}~\bibnamefont {Yuan}}, \bibinfo
  {author} {\bibfnamefont {C.}~\bibnamefont {Yèche}}, \bibinfo {author}
  {\bibfnamefont {E.~A.}\ \bibnamefont {Zaborowski}}, \bibinfo {author}
  {\bibfnamefont {P.}~\bibnamefont {Zarrouk}}, \bibinfo {author} {\bibfnamefont
  {H.}~\bibnamefont {Zhang}}, \bibinfo {author} {\bibfnamefont
  {C.}~\bibnamefont {Zhao}}, \bibinfo {author} {\bibfnamefont {R.}~\bibnamefont
  {Zhao}}, \bibinfo {author} {\bibfnamefont {R.}~\bibnamefont {Zhou}}, \bibinfo
  {author} {\bibfnamefont {T.}~\bibnamefont {Zhuang}},\ and\ \bibinfo {author}
  {\bibfnamefont {H.}~\bibnamefont {Zou}},\ }\href
  {https://arxiv.org/abs/2404.03002} {\bibinfo {title} {Desi 2024 vi:
  Cosmological constraints from the measurements of baryon acoustic
  oscillations}} (\bibinfo {year} {2024}{\natexlab{a}}),\ \Eprint
  {https://arxiv.org/abs/2404.03002} {arXiv:2404.03002 [astro-ph.CO]}
  \BibitemShut {NoStop}%
\bibitem [{\citenamefont {Eisenstein}\ \emph {et~al.}(2005)\citenamefont
  {Eisenstein}, \citenamefont {Zehavi}, \citenamefont {Hogg}, \citenamefont
  {Scoccimarro}, \citenamefont {Blanton}, \citenamefont {Nichol}, \citenamefont
  {Scranton}, \citenamefont {Seo}, \citenamefont {Tegmark}, \citenamefont
  {Zheng}, \citenamefont {Anderson}, \citenamefont {Annis}, \citenamefont
  {Bahcall}, \citenamefont {Brinkmann}, \citenamefont {Burles}, \citenamefont
  {Castander}, \citenamefont {Connolly}, \citenamefont {Csabai}, \citenamefont
  {Doi}, \citenamefont {Fukugita}, \citenamefont {Frieman}, \citenamefont
  {Glazebrook}, \citenamefont {Gunn}, \citenamefont {Hendry}, \citenamefont
  {Hennessy}, \citenamefont {Ivezić}, \citenamefont {Kent}, \citenamefont
  {Knapp}, \citenamefont {Lin}, \citenamefont {Loh}, \citenamefont {Lupton},
  \citenamefont {Margon}, \citenamefont {McKay}, \citenamefont {Meiksin},
  \citenamefont {Munn}, \citenamefont {Pope}, \citenamefont {Richmond},
  \citenamefont {Schlegel}, \citenamefont {Schneider}, \citenamefont
  {Shimasaku}, \citenamefont {Stoughton}, \citenamefont {Strauss},
  \citenamefont {SubbaRao}, \citenamefont {Szalay}, \citenamefont {Szapudi},
  \citenamefont {Tucker}, \citenamefont {Yanny},\ and\ \citenamefont
  {York}}]{DILATIONSCALE2005}%
  \BibitemOpen
  \bibfield  {author} {\bibinfo {author} {\bibfnamefont {D.~J.}\ \bibnamefont
  {Eisenstein}}, \bibinfo {author} {\bibfnamefont {I.}~\bibnamefont {Zehavi}},
  \bibinfo {author} {\bibfnamefont {D.~W.}\ \bibnamefont {Hogg}}, \bibinfo
  {author} {\bibfnamefont {R.}~\bibnamefont {Scoccimarro}}, \bibinfo {author}
  {\bibfnamefont {M.~R.}\ \bibnamefont {Blanton}}, \bibinfo {author}
  {\bibfnamefont {R.~C.}\ \bibnamefont {Nichol}}, \bibinfo {author}
  {\bibfnamefont {R.}~\bibnamefont {Scranton}}, \bibinfo {author}
  {\bibfnamefont {H.}~\bibnamefont {Seo}}, \bibinfo {author} {\bibfnamefont
  {M.}~\bibnamefont {Tegmark}}, \bibinfo {author} {\bibfnamefont
  {Z.}~\bibnamefont {Zheng}}, \bibinfo {author} {\bibfnamefont {S.~F.}\
  \bibnamefont {Anderson}}, \bibinfo {author} {\bibfnamefont {J.}~\bibnamefont
  {Annis}}, \bibinfo {author} {\bibfnamefont {N.}~\bibnamefont {Bahcall}},
  \bibinfo {author} {\bibfnamefont {J.}~\bibnamefont {Brinkmann}}, \bibinfo
  {author} {\bibfnamefont {S.}~\bibnamefont {Burles}}, \bibinfo {author}
  {\bibfnamefont {F.~J.}\ \bibnamefont {Castander}}, \bibinfo {author}
  {\bibfnamefont {A.}~\bibnamefont {Connolly}}, \bibinfo {author}
  {\bibfnamefont {I.}~\bibnamefont {Csabai}}, \bibinfo {author} {\bibfnamefont
  {M.}~\bibnamefont {Doi}}, \bibinfo {author} {\bibfnamefont {M.}~\bibnamefont
  {Fukugita}}, \bibinfo {author} {\bibfnamefont {J.~A.}\ \bibnamefont
  {Frieman}}, \bibinfo {author} {\bibfnamefont {K.}~\bibnamefont {Glazebrook}},
  \bibinfo {author} {\bibfnamefont {J.~E.}\ \bibnamefont {Gunn}}, \bibinfo
  {author} {\bibfnamefont {J.~S.}\ \bibnamefont {Hendry}}, \bibinfo {author}
  {\bibfnamefont {G.}~\bibnamefont {Hennessy}}, \bibinfo {author}
  {\bibfnamefont {Z.}~\bibnamefont {Ivezić}}, \bibinfo {author} {\bibfnamefont
  {S.}~\bibnamefont {Kent}}, \bibinfo {author} {\bibfnamefont {G.~R.}\
  \bibnamefont {Knapp}}, \bibinfo {author} {\bibfnamefont {H.}~\bibnamefont
  {Lin}}, \bibinfo {author} {\bibfnamefont {Y.}~\bibnamefont {Loh}}, \bibinfo
  {author} {\bibfnamefont {R.~H.}\ \bibnamefont {Lupton}}, \bibinfo {author}
  {\bibfnamefont {B.}~\bibnamefont {Margon}}, \bibinfo {author} {\bibfnamefont
  {T.~A.}\ \bibnamefont {McKay}}, \bibinfo {author} {\bibfnamefont
  {A.}~\bibnamefont {Meiksin}}, \bibinfo {author} {\bibfnamefont {J.~A.}\
  \bibnamefont {Munn}}, \bibinfo {author} {\bibfnamefont {A.}~\bibnamefont
  {Pope}}, \bibinfo {author} {\bibfnamefont {M.~W.}\ \bibnamefont {Richmond}},
  \bibinfo {author} {\bibfnamefont {D.}~\bibnamefont {Schlegel}}, \bibinfo
  {author} {\bibfnamefont {D.~P.}\ \bibnamefont {Schneider}}, \bibinfo {author}
  {\bibfnamefont {K.}~\bibnamefont {Shimasaku}}, \bibinfo {author}
  {\bibfnamefont {C.}~\bibnamefont {Stoughton}}, \bibinfo {author}
  {\bibfnamefont {M.~A.}\ \bibnamefont {Strauss}}, \bibinfo {author}
  {\bibfnamefont {M.}~\bibnamefont {SubbaRao}}, \bibinfo {author}
  {\bibfnamefont {A.~S.}\ \bibnamefont {Szalay}}, \bibinfo {author}
  {\bibfnamefont {I.}~\bibnamefont {Szapudi}}, \bibinfo {author} {\bibfnamefont
  {D.~L.}\ \bibnamefont {Tucker}}, \bibinfo {author} {\bibfnamefont
  {B.}~\bibnamefont {Yanny}},\ and\ \bibinfo {author} {\bibfnamefont {D.~G.}\
  \bibnamefont {York}},\ }\href {https://doi.org/10.1086/466512} {\bibfield
  {journal} {\bibinfo  {journal} {The Astrophysical Journal}\ }\textbf
  {\bibinfo {volume} {633}},\ \bibinfo {pages} {560–574} (\bibinfo {year}
  {2005})}\BibitemShut {NoStop}%
\bibitem [{\citenamefont {Cao}\ \emph {et~al.}(2015)\citenamefont {Cao},
  \citenamefont {Biesiada}, \citenamefont {Gavazzi}, \citenamefont
  {Piórkowska},\ and\ \citenamefont {Zhu}}]{Cao_2015SLS}%
  \BibitemOpen
  \bibfield  {author} {\bibinfo {author} {\bibfnamefont {S.}~\bibnamefont
  {Cao}}, \bibinfo {author} {\bibfnamefont {M.}~\bibnamefont {Biesiada}},
  \bibinfo {author} {\bibfnamefont {R.}~\bibnamefont {Gavazzi}}, \bibinfo
  {author} {\bibfnamefont {A.}~\bibnamefont {Piórkowska}},\ and\ \bibinfo
  {author} {\bibfnamefont {Z.-H.}\ \bibnamefont {Zhu}},\ }\href
  {https://doi.org/10.1088/0004-637X/806/2/185} {\bibfield  {journal} {\bibinfo
   {journal} {The Astrophysical Journal}\ }\textbf {\bibinfo {volume} {806}},\
  \bibinfo {pages} {185} (\bibinfo {year} {2015})}\BibitemShut {NoStop}%
\bibitem [{\citenamefont {Magaña}\ \emph {et~al.}(2018)\citenamefont
  {Magaña}, \citenamefont {Acebrón}, \citenamefont {Motta}, \citenamefont
  {Verdugo}, \citenamefont {Jullo},\ and\ \citenamefont
  {Limousin}}]{SLSmagaña2018}%
  \BibitemOpen
  \bibfield  {author} {\bibinfo {author} {\bibfnamefont {J.}~\bibnamefont
  {Magaña}}, \bibinfo {author} {\bibfnamefont {A.}~\bibnamefont {Acebrón}},
  \bibinfo {author} {\bibfnamefont {V.}~\bibnamefont {Motta}}, \bibinfo
  {author} {\bibfnamefont {T.}~\bibnamefont {Verdugo}}, \bibinfo {author}
  {\bibfnamefont {E.}~\bibnamefont {Jullo}},\ and\ \bibinfo {author}
  {\bibfnamefont {M.}~\bibnamefont {Limousin}},\ }\href
  {https://doi.org/10.3847/1538-4357/aada7d} {\bibfield  {journal} {\bibinfo
  {journal} {The Astrophysical Journal}\ }\textbf {\bibinfo {volume} {865}},\
  \bibinfo {pages} {122} (\bibinfo {year} {2018})}\BibitemShut {NoStop}%
\bibitem [{\citenamefont {Foreman-Mackey}\ \emph {et~al.}(2013)\citenamefont
  {Foreman-Mackey}, \citenamefont {Hogg}, \citenamefont {Lang},\ and\
  \citenamefont {Goodman}}]{mcmchammer}%
  \BibitemOpen
  \bibfield  {author} {\bibinfo {author} {\bibfnamefont {D.}~\bibnamefont
  {Foreman-Mackey}}, \bibinfo {author} {\bibfnamefont {D.~W.}\ \bibnamefont
  {Hogg}}, \bibinfo {author} {\bibfnamefont {D.}~\bibnamefont {Lang}},\ and\
  \bibinfo {author} {\bibfnamefont {J.}~\bibnamefont {Goodman}},\ }\href
  {https://doi.org/10.1086/670067} {\bibfield  {journal} {\bibinfo  {journal}
  {Publ. Astron. Soc. Pac.}\ }\textbf {\bibinfo {volume} {125}},\ \bibinfo
  {pages} {306} (\bibinfo {year} {2013})},\ \Eprint
  {https://arxiv.org/abs/1202.3665} {arXiv:1202.3665 [astro-ph.IM]}
  \BibitemShut {NoStop}%
\bibitem [{\citenamefont {{Goodman}}\ and\ \citenamefont
  {{Weare}}(2010)}]{autocorrelation}%
  \BibitemOpen
  \bibfield  {author} {\bibinfo {author} {\bibfnamefont {J.}~\bibnamefont
  {{Goodman}}}\ and\ \bibinfo {author} {\bibfnamefont {J.}~\bibnamefont
  {{Weare}}},\ }\href {https://doi.org/10.2140/camcos.2010.5.65} {\bibfield
  {journal} {\bibinfo  {journal} {Communications in Applied Mathematics and
  Computational Science}\ }\textbf {\bibinfo {volume} {5}},\ \bibinfo {pages}
  {65} (\bibinfo {year} {2010})}\BibitemShut {NoStop}%
\bibitem [{\citenamefont {Riess}\ \emph {et~al.}(2022)\citenamefont {Riess},
  \citenamefont {Yuan}, \citenamefont {Macri}, \citenamefont {Scolnic},
  \citenamefont {Brout}, \citenamefont {Casertano}, \citenamefont {Jones},
  \citenamefont {Murakami}, \citenamefont {Anand}, \citenamefont {Breuval},
  \citenamefont {Brink}, \citenamefont {Filippenko}, \citenamefont {Hoffmann},
  \citenamefont {Jha}, \citenamefont {D’arcy~Kenworthy}, \citenamefont
  {Mackenty}, \citenamefont {Stahl},\ and\ \citenamefont {Zheng}}]{Riess_2022}%
  \BibitemOpen
  \bibfield  {author} {\bibinfo {author} {\bibfnamefont {A.~G.}\ \bibnamefont
  {Riess}}, \bibinfo {author} {\bibfnamefont {W.}~\bibnamefont {Yuan}},
  \bibinfo {author} {\bibfnamefont {L.~M.}\ \bibnamefont {Macri}}, \bibinfo
  {author} {\bibfnamefont {D.}~\bibnamefont {Scolnic}}, \bibinfo {author}
  {\bibfnamefont {D.}~\bibnamefont {Brout}}, \bibinfo {author} {\bibfnamefont
  {S.}~\bibnamefont {Casertano}}, \bibinfo {author} {\bibfnamefont {D.~O.}\
  \bibnamefont {Jones}}, \bibinfo {author} {\bibfnamefont {Y.}~\bibnamefont
  {Murakami}}, \bibinfo {author} {\bibfnamefont {G.~S.}\ \bibnamefont {Anand}},
  \bibinfo {author} {\bibfnamefont {L.}~\bibnamefont {Breuval}}, \bibinfo
  {author} {\bibfnamefont {T.~G.}\ \bibnamefont {Brink}}, \bibinfo {author}
  {\bibfnamefont {A.~V.}\ \bibnamefont {Filippenko}}, \bibinfo {author}
  {\bibfnamefont {S.}~\bibnamefont {Hoffmann}}, \bibinfo {author}
  {\bibfnamefont {S.~W.}\ \bibnamefont {Jha}}, \bibinfo {author} {\bibfnamefont
  {W.}~\bibnamefont {D’arcy~Kenworthy}}, \bibinfo {author} {\bibfnamefont
  {J.}~\bibnamefont {Mackenty}}, \bibinfo {author} {\bibfnamefont {B.~E.}\
  \bibnamefont {Stahl}},\ and\ \bibinfo {author} {\bibfnamefont
  {W.}~\bibnamefont {Zheng}},\ }\href
  {https://doi.org/10.3847/2041-8213/ac5c5b} {\bibfield  {journal} {\bibinfo
  {journal} {The Astrophysical Journal Letters}\ }\textbf {\bibinfo {volume}
  {934}},\ \bibinfo {pages} {L7} (\bibinfo {year} {2022})}\BibitemShut
  {NoStop}%
\bibitem [{\citenamefont {Collaboration}\ \emph
  {et~al.}(2024{\natexlab{b}})\citenamefont {Collaboration}, \citenamefont
  {Adame}, \citenamefont {Aguilar}, \citenamefont {Ahlen}, \citenamefont
  {Alam}, \citenamefont {Alexander}, \citenamefont {Alvarez}, \citenamefont
  {Alves}, \citenamefont {Anand}, \citenamefont {Andrade}, \citenamefont
  {Armengaud}, \citenamefont {Avila}, \citenamefont {Aviles}, \citenamefont
  {Awan}, \citenamefont {Bahr-Kalus}, \citenamefont {Bailey}, \citenamefont
  {Baltay}, \citenamefont {Bault}, \citenamefont {Behera}, \citenamefont
  {BenZvi}, \citenamefont {Bera}, \citenamefont {Beutler}, \citenamefont
  {Bianchi}, \citenamefont {Blake}, \citenamefont {Blum}, \citenamefont
  {Brieden}, \citenamefont {Brodzeller}, \citenamefont {Brooks}, \citenamefont
  {Buckley-Geer}, \citenamefont {Burtin}, \citenamefont {Calderon},
  \citenamefont {Canning}, \citenamefont {Rosell}, \citenamefont {Cereskaite},
  \citenamefont {Cervantes-Cota}, \citenamefont {Chabanier}, \citenamefont
  {Chaussidon}, \citenamefont {Chaves-Montero}, \citenamefont {Chen},
  \citenamefont {Chen}, \citenamefont {Claybaugh}, \citenamefont {Cole},
  \citenamefont {Cuceu}, \citenamefont {Davis}, \citenamefont {Dawson},
  \citenamefont {de~la Macorra}, \citenamefont {de~Mattia}, \citenamefont
  {Deiosso}, \citenamefont {Dey}, \citenamefont {Dey}, \citenamefont {Ding},
  \citenamefont {Doel}, \citenamefont {Edelstein}, \citenamefont
  {Eftekharzadeh}, \citenamefont {Eisenstein}, \citenamefont {Elliott},
  \citenamefont {Fagrelius}, \citenamefont {Fanning}, \citenamefont {Ferraro},
  \citenamefont {Ereza}, \citenamefont {Findlay}, \citenamefont {Flaugher},
  \citenamefont {Font-Ribera}, \citenamefont {Forero-Sánchez}, \citenamefont
  {Forero-Romero}, \citenamefont {Frenk}, \citenamefont {Garcia-Quintero},
  \citenamefont {Gaztañaga}, \citenamefont {Gil-Marín}, \citenamefont
  {Gontcho}, \citenamefont {Gonzalez-Morales}, \citenamefont {Gonzalez-Perez},
  \citenamefont {Gordon}, \citenamefont {Green}, \citenamefont {Gruen},
  \citenamefont {Gsponer}, \citenamefont {Gutierrez}, \citenamefont {Guy},
  \citenamefont {Hadzhiyska}, \citenamefont {Hahn}, \citenamefont {Hanif},
  \citenamefont {Herrera-Alcantar}, \citenamefont {Honscheid}, \citenamefont
  {Howlett}, \citenamefont {Huterer}, \citenamefont {Iršič}, \citenamefont
  {Ishak}, \citenamefont {Juneau}, \citenamefont {Karaçaylı}, \citenamefont
  {Kehoe}, \citenamefont {Kent}, \citenamefont {Kirkby}, \citenamefont
  {Kremin}, \citenamefont {Krolewski}, \citenamefont {Lai}, \citenamefont
  {Lan}, \citenamefont {Landriau}, \citenamefont {Lang}, \citenamefont
  {Lasker}, \citenamefont {Goff}, \citenamefont {Guillou}, \citenamefont
  {Leauthaud}, \citenamefont {Levi}, \citenamefont {Li}, \citenamefont
  {Linder}, \citenamefont {Lodha}, \citenamefont {Magneville}, \citenamefont
  {Manera}, \citenamefont {Margala}, \citenamefont {Martini}, \citenamefont
  {Maus}, \citenamefont {McDonald}, \citenamefont {Medina-Varela},
  \citenamefont {Meisner}, \citenamefont {Mena-Fernández}, \citenamefont
  {Miquel}, \citenamefont {Moon}, \citenamefont {Moore}, \citenamefont
  {Moustakas}, \citenamefont {Mudur}, \citenamefont {Mueller}, \citenamefont
  {Muñoz-Gutiérrez}, \citenamefont {Myers}, \citenamefont {Nadathur},
  \citenamefont {Napolitano}, \citenamefont {Neveux}, \citenamefont {Newman},
  \citenamefont {Nguyen}, \citenamefont {Nie}, \citenamefont {Niz},
  \citenamefont {Noriega}, \citenamefont {Padmanabhan}, \citenamefont
  {Paillas}, \citenamefont {Palanque-Delabrouille}, \citenamefont {Pan},
  \citenamefont {Penmetsa}, \citenamefont {Percival}, \citenamefont {Pieri},
  \citenamefont {Pinon}, \citenamefont {Poppett}, \citenamefont {Porredon},
  \citenamefont {Prada}, \citenamefont {Pérez-Fernández}, \citenamefont
  {Pérez-Ràfols}, \citenamefont {Rabinowitz}, \citenamefont {Raichoor},
  \citenamefont {Ramírez-Pérez}, \citenamefont {Ramirez-Solano},
  \citenamefont {Ravoux}, \citenamefont {Rashkovetskyi}, \citenamefont
  {Rezaie}, \citenamefont {Rich}, \citenamefont {Rocher}, \citenamefont
  {Rockosi}, \citenamefont {Roe}, \citenamefont {Rosado-Marin}, \citenamefont
  {Ross}, \citenamefont {Rossi}, \citenamefont {Ruggeri}, \citenamefont
  {Ruhlmann-Kleider}, \citenamefont {Samushia}, \citenamefont {Sanchez},
  \citenamefont {Saulder}, \citenamefont {Schlafly}, \citenamefont {Schlegel},
  \citenamefont {Schubnell}, \citenamefont {Seo}, \citenamefont {Shafieloo},
  \citenamefont {Sharples}, \citenamefont {Silber}, \citenamefont {Slosar},
  \citenamefont {Smith}, \citenamefont {Sprayberry}, \citenamefont {Tan},
  \citenamefont {Tarlé}, \citenamefont {Taylor}, \citenamefont {Trusov},
  \citenamefont {Ureña-López}, \citenamefont {Vaisakh}, \citenamefont
  {Valcin}, \citenamefont {Valdes}, \citenamefont {Vargas-Magaña},
  \citenamefont {Verde}, \citenamefont {Walther}, \citenamefont {Wang},
  \citenamefont {Wang}, \citenamefont {Weaver}, \citenamefont {Weaverdyck},
  \citenamefont {Wechsler}, \citenamefont {Weinberg}, \citenamefont {White},
  \citenamefont {Yu}, \citenamefont {Yu}, \citenamefont {Yuan}, \citenamefont
  {Yèche}, \citenamefont {Zaborowski}, \citenamefont {Zarrouk}, \citenamefont
  {Zhang}, \citenamefont {Zhao}, \citenamefont {Zhao}, \citenamefont {Zhou},
  \citenamefont {Zhuang},\ and\ \citenamefont {Zou}}]{DESIBARYON2024}%
  \BibitemOpen
  \bibfield  {author} {\bibinfo {author} {\bibfnamefont {D.}~\bibnamefont
  {Collaboration}}, \bibinfo {author} {\bibfnamefont {A.~G.}\ \bibnamefont
  {Adame}}, \bibinfo {author} {\bibfnamefont {J.}~\bibnamefont {Aguilar}},
  \bibinfo {author} {\bibfnamefont {S.}~\bibnamefont {Ahlen}}, \bibinfo
  {author} {\bibfnamefont {S.}~\bibnamefont {Alam}}, \bibinfo {author}
  {\bibfnamefont {D.~M.}\ \bibnamefont {Alexander}}, \bibinfo {author}
  {\bibfnamefont {M.}~\bibnamefont {Alvarez}}, \bibinfo {author} {\bibfnamefont
  {O.}~\bibnamefont {Alves}}, \bibinfo {author} {\bibfnamefont
  {A.}~\bibnamefont {Anand}}, \bibinfo {author} {\bibfnamefont
  {U.}~\bibnamefont {Andrade}}, \bibinfo {author} {\bibfnamefont
  {E.}~\bibnamefont {Armengaud}}, \bibinfo {author} {\bibfnamefont
  {S.}~\bibnamefont {Avila}}, \bibinfo {author} {\bibfnamefont
  {A.}~\bibnamefont {Aviles}}, \bibinfo {author} {\bibfnamefont
  {H.}~\bibnamefont {Awan}}, \bibinfo {author} {\bibfnamefont {B.}~\bibnamefont
  {Bahr-Kalus}}, \bibinfo {author} {\bibfnamefont {S.}~\bibnamefont {Bailey}},
  \bibinfo {author} {\bibfnamefont {C.}~\bibnamefont {Baltay}}, \bibinfo
  {author} {\bibfnamefont {A.}~\bibnamefont {Bault}}, \bibinfo {author}
  {\bibfnamefont {J.}~\bibnamefont {Behera}}, \bibinfo {author} {\bibfnamefont
  {S.}~\bibnamefont {BenZvi}}, \bibinfo {author} {\bibfnamefont
  {A.}~\bibnamefont {Bera}}, \bibinfo {author} {\bibfnamefont {F.}~\bibnamefont
  {Beutler}}, \bibinfo {author} {\bibfnamefont {D.}~\bibnamefont {Bianchi}},
  \bibinfo {author} {\bibfnamefont {C.}~\bibnamefont {Blake}}, \bibinfo
  {author} {\bibfnamefont {R.}~\bibnamefont {Blum}}, \bibinfo {author}
  {\bibfnamefont {S.}~\bibnamefont {Brieden}}, \bibinfo {author} {\bibfnamefont
  {A.}~\bibnamefont {Brodzeller}}, \bibinfo {author} {\bibfnamefont
  {D.}~\bibnamefont {Brooks}}, \bibinfo {author} {\bibfnamefont
  {E.}~\bibnamefont {Buckley-Geer}}, \bibinfo {author} {\bibfnamefont
  {E.}~\bibnamefont {Burtin}}, \bibinfo {author} {\bibfnamefont
  {R.}~\bibnamefont {Calderon}}, \bibinfo {author} {\bibfnamefont
  {R.}~\bibnamefont {Canning}}, \bibinfo {author} {\bibfnamefont {A.~C.}\
  \bibnamefont {Rosell}}, \bibinfo {author} {\bibfnamefont {R.}~\bibnamefont
  {Cereskaite}}, \bibinfo {author} {\bibfnamefont {J.~L.}\ \bibnamefont
  {Cervantes-Cota}}, \bibinfo {author} {\bibfnamefont {S.}~\bibnamefont
  {Chabanier}}, \bibinfo {author} {\bibfnamefont {E.}~\bibnamefont
  {Chaussidon}}, \bibinfo {author} {\bibfnamefont {J.}~\bibnamefont
  {Chaves-Montero}}, \bibinfo {author} {\bibfnamefont {S.}~\bibnamefont
  {Chen}}, \bibinfo {author} {\bibfnamefont {X.}~\bibnamefont {Chen}}, \bibinfo
  {author} {\bibfnamefont {T.}~\bibnamefont {Claybaugh}}, \bibinfo {author}
  {\bibfnamefont {S.}~\bibnamefont {Cole}}, \bibinfo {author} {\bibfnamefont
  {A.}~\bibnamefont {Cuceu}}, \bibinfo {author} {\bibfnamefont {T.~M.}\
  \bibnamefont {Davis}}, \bibinfo {author} {\bibfnamefont {K.}~\bibnamefont
  {Dawson}}, \bibinfo {author} {\bibfnamefont {A.}~\bibnamefont {de~la
  Macorra}}, \bibinfo {author} {\bibfnamefont {A.}~\bibnamefont {de~Mattia}},
  \bibinfo {author} {\bibfnamefont {N.}~\bibnamefont {Deiosso}}, \bibinfo
  {author} {\bibfnamefont {A.}~\bibnamefont {Dey}}, \bibinfo {author}
  {\bibfnamefont {B.}~\bibnamefont {Dey}}, \bibinfo {author} {\bibfnamefont
  {Z.}~\bibnamefont {Ding}}, \bibinfo {author} {\bibfnamefont {P.}~\bibnamefont
  {Doel}}, \bibinfo {author} {\bibfnamefont {J.}~\bibnamefont {Edelstein}},
  \bibinfo {author} {\bibfnamefont {S.}~\bibnamefont {Eftekharzadeh}}, \bibinfo
  {author} {\bibfnamefont {D.~J.}\ \bibnamefont {Eisenstein}}, \bibinfo
  {author} {\bibfnamefont {A.}~\bibnamefont {Elliott}}, \bibinfo {author}
  {\bibfnamefont {P.}~\bibnamefont {Fagrelius}}, \bibinfo {author}
  {\bibfnamefont {K.}~\bibnamefont {Fanning}}, \bibinfo {author} {\bibfnamefont
  {S.}~\bibnamefont {Ferraro}}, \bibinfo {author} {\bibfnamefont
  {J.}~\bibnamefont {Ereza}}, \bibinfo {author} {\bibfnamefont
  {N.}~\bibnamefont {Findlay}}, \bibinfo {author} {\bibfnamefont
  {B.}~\bibnamefont {Flaugher}}, \bibinfo {author} {\bibfnamefont
  {A.}~\bibnamefont {Font-Ribera}}, \bibinfo {author} {\bibfnamefont
  {D.}~\bibnamefont {Forero-Sánchez}}, \bibinfo {author} {\bibfnamefont
  {J.~E.}\ \bibnamefont {Forero-Romero}}, \bibinfo {author} {\bibfnamefont
  {C.~S.}\ \bibnamefont {Frenk}}, \bibinfo {author} {\bibfnamefont
  {C.}~\bibnamefont {Garcia-Quintero}}, \bibinfo {author} {\bibfnamefont
  {E.}~\bibnamefont {Gaztañaga}}, \bibinfo {author} {\bibfnamefont
  {H.}~\bibnamefont {Gil-Marín}}, \bibinfo {author} {\bibfnamefont {S.~G.~A.}\
  \bibnamefont {Gontcho}}, \bibinfo {author} {\bibfnamefont {A.~X.}\
  \bibnamefont {Gonzalez-Morales}}, \bibinfo {author} {\bibfnamefont
  {V.}~\bibnamefont {Gonzalez-Perez}}, \bibinfo {author} {\bibfnamefont
  {C.}~\bibnamefont {Gordon}}, \bibinfo {author} {\bibfnamefont
  {D.}~\bibnamefont {Green}}, \bibinfo {author} {\bibfnamefont
  {D.}~\bibnamefont {Gruen}}, \bibinfo {author} {\bibfnamefont
  {R.}~\bibnamefont {Gsponer}}, \bibinfo {author} {\bibfnamefont
  {G.}~\bibnamefont {Gutierrez}}, \bibinfo {author} {\bibfnamefont
  {J.}~\bibnamefont {Guy}}, \bibinfo {author} {\bibfnamefont {B.}~\bibnamefont
  {Hadzhiyska}}, \bibinfo {author} {\bibfnamefont {C.}~\bibnamefont {Hahn}},
  \bibinfo {author} {\bibfnamefont {M.~M.~S.}\ \bibnamefont {Hanif}}, \bibinfo
  {author} {\bibfnamefont {H.~K.}\ \bibnamefont {Herrera-Alcantar}}, \bibinfo
  {author} {\bibfnamefont {K.}~\bibnamefont {Honscheid}}, \bibinfo {author}
  {\bibfnamefont {C.}~\bibnamefont {Howlett}}, \bibinfo {author} {\bibfnamefont
  {D.}~\bibnamefont {Huterer}}, \bibinfo {author} {\bibfnamefont
  {V.}~\bibnamefont {Iršič}}, \bibinfo {author} {\bibfnamefont
  {M.}~\bibnamefont {Ishak}}, \bibinfo {author} {\bibfnamefont
  {S.}~\bibnamefont {Juneau}}, \bibinfo {author} {\bibfnamefont {N.~G.}\
  \bibnamefont {Karaçaylı}}, \bibinfo {author} {\bibfnamefont
  {R.}~\bibnamefont {Kehoe}}, \bibinfo {author} {\bibfnamefont
  {S.}~\bibnamefont {Kent}}, \bibinfo {author} {\bibfnamefont {D.}~\bibnamefont
  {Kirkby}}, \bibinfo {author} {\bibfnamefont {A.}~\bibnamefont {Kremin}},
  \bibinfo {author} {\bibfnamefont {A.}~\bibnamefont {Krolewski}}, \bibinfo
  {author} {\bibfnamefont {Y.}~\bibnamefont {Lai}}, \bibinfo {author}
  {\bibfnamefont {T.~W.}\ \bibnamefont {Lan}}, \bibinfo {author} {\bibfnamefont
  {M.}~\bibnamefont {Landriau}}, \bibinfo {author} {\bibfnamefont
  {D.}~\bibnamefont {Lang}}, \bibinfo {author} {\bibfnamefont {J.}~\bibnamefont
  {Lasker}}, \bibinfo {author} {\bibfnamefont {J.~M.~L.}\ \bibnamefont {Goff}},
  \bibinfo {author} {\bibfnamefont {L.~L.}\ \bibnamefont {Guillou}}, \bibinfo
  {author} {\bibfnamefont {A.}~\bibnamefont {Leauthaud}}, \bibinfo {author}
  {\bibfnamefont {M.~E.}\ \bibnamefont {Levi}}, \bibinfo {author}
  {\bibfnamefont {T.~S.}\ \bibnamefont {Li}}, \bibinfo {author} {\bibfnamefont
  {E.}~\bibnamefont {Linder}}, \bibinfo {author} {\bibfnamefont
  {K.}~\bibnamefont {Lodha}}, \bibinfo {author} {\bibfnamefont
  {C.}~\bibnamefont {Magneville}}, \bibinfo {author} {\bibfnamefont
  {M.}~\bibnamefont {Manera}}, \bibinfo {author} {\bibfnamefont
  {D.}~\bibnamefont {Margala}}, \bibinfo {author} {\bibfnamefont
  {P.}~\bibnamefont {Martini}}, \bibinfo {author} {\bibfnamefont
  {M.}~\bibnamefont {Maus}}, \bibinfo {author} {\bibfnamefont {P.}~\bibnamefont
  {McDonald}}, \bibinfo {author} {\bibfnamefont {L.}~\bibnamefont
  {Medina-Varela}}, \bibinfo {author} {\bibfnamefont {A.}~\bibnamefont
  {Meisner}}, \bibinfo {author} {\bibfnamefont {J.}~\bibnamefont
  {Mena-Fernández}}, \bibinfo {author} {\bibfnamefont {R.}~\bibnamefont
  {Miquel}}, \bibinfo {author} {\bibfnamefont {J.}~\bibnamefont {Moon}},
  \bibinfo {author} {\bibfnamefont {S.}~\bibnamefont {Moore}}, \bibinfo
  {author} {\bibfnamefont {J.}~\bibnamefont {Moustakas}}, \bibinfo {author}
  {\bibfnamefont {N.}~\bibnamefont {Mudur}}, \bibinfo {author} {\bibfnamefont
  {E.}~\bibnamefont {Mueller}}, \bibinfo {author} {\bibfnamefont
  {A.}~\bibnamefont {Muñoz-Gutiérrez}}, \bibinfo {author} {\bibfnamefont
  {A.~D.}\ \bibnamefont {Myers}}, \bibinfo {author} {\bibfnamefont
  {S.}~\bibnamefont {Nadathur}}, \bibinfo {author} {\bibfnamefont
  {L.}~\bibnamefont {Napolitano}}, \bibinfo {author} {\bibfnamefont
  {R.}~\bibnamefont {Neveux}}, \bibinfo {author} {\bibfnamefont {J.~A.}\
  \bibnamefont {Newman}}, \bibinfo {author} {\bibfnamefont {N.~M.}\
  \bibnamefont {Nguyen}}, \bibinfo {author} {\bibfnamefont {J.}~\bibnamefont
  {Nie}}, \bibinfo {author} {\bibfnamefont {G.}~\bibnamefont {Niz}}, \bibinfo
  {author} {\bibfnamefont {H.~E.}\ \bibnamefont {Noriega}}, \bibinfo {author}
  {\bibfnamefont {N.}~\bibnamefont {Padmanabhan}}, \bibinfo {author}
  {\bibfnamefont {E.}~\bibnamefont {Paillas}}, \bibinfo {author} {\bibfnamefont
  {N.}~\bibnamefont {Palanque-Delabrouille}}, \bibinfo {author} {\bibfnamefont
  {J.}~\bibnamefont {Pan}}, \bibinfo {author} {\bibfnamefont {S.}~\bibnamefont
  {Penmetsa}}, \bibinfo {author} {\bibfnamefont {W.~J.}\ \bibnamefont
  {Percival}}, \bibinfo {author} {\bibfnamefont {M.~M.}\ \bibnamefont {Pieri}},
  \bibinfo {author} {\bibfnamefont {M.}~\bibnamefont {Pinon}}, \bibinfo
  {author} {\bibfnamefont {C.}~\bibnamefont {Poppett}}, \bibinfo {author}
  {\bibfnamefont {A.}~\bibnamefont {Porredon}}, \bibinfo {author}
  {\bibfnamefont {F.}~\bibnamefont {Prada}}, \bibinfo {author} {\bibfnamefont
  {A.}~\bibnamefont {Pérez-Fernández}}, \bibinfo {author} {\bibfnamefont
  {I.}~\bibnamefont {Pérez-Ràfols}}, \bibinfo {author} {\bibfnamefont
  {D.}~\bibnamefont {Rabinowitz}}, \bibinfo {author} {\bibfnamefont
  {A.}~\bibnamefont {Raichoor}}, \bibinfo {author} {\bibfnamefont
  {C.}~\bibnamefont {Ramírez-Pérez}}, \bibinfo {author} {\bibfnamefont
  {S.}~\bibnamefont {Ramirez-Solano}}, \bibinfo {author} {\bibfnamefont
  {C.}~\bibnamefont {Ravoux}}, \bibinfo {author} {\bibfnamefont
  {M.}~\bibnamefont {Rashkovetskyi}}, \bibinfo {author} {\bibfnamefont
  {M.}~\bibnamefont {Rezaie}}, \bibinfo {author} {\bibfnamefont
  {J.}~\bibnamefont {Rich}}, \bibinfo {author} {\bibfnamefont {A.}~\bibnamefont
  {Rocher}}, \bibinfo {author} {\bibfnamefont {C.}~\bibnamefont {Rockosi}},
  \bibinfo {author} {\bibfnamefont {N.~A.}\ \bibnamefont {Roe}}, \bibinfo
  {author} {\bibfnamefont {A.}~\bibnamefont {Rosado-Marin}}, \bibinfo {author}
  {\bibfnamefont {A.~J.}\ \bibnamefont {Ross}}, \bibinfo {author}
  {\bibfnamefont {G.}~\bibnamefont {Rossi}}, \bibinfo {author} {\bibfnamefont
  {R.}~\bibnamefont {Ruggeri}}, \bibinfo {author} {\bibfnamefont
  {V.}~\bibnamefont {Ruhlmann-Kleider}}, \bibinfo {author} {\bibfnamefont
  {L.}~\bibnamefont {Samushia}}, \bibinfo {author} {\bibfnamefont
  {E.}~\bibnamefont {Sanchez}}, \bibinfo {author} {\bibfnamefont
  {C.}~\bibnamefont {Saulder}}, \bibinfo {author} {\bibfnamefont {E.~F.}\
  \bibnamefont {Schlafly}}, \bibinfo {author} {\bibfnamefont {D.}~\bibnamefont
  {Schlegel}}, \bibinfo {author} {\bibfnamefont {M.}~\bibnamefont {Schubnell}},
  \bibinfo {author} {\bibfnamefont {H.}~\bibnamefont {Seo}}, \bibinfo {author}
  {\bibfnamefont {A.}~\bibnamefont {Shafieloo}}, \bibinfo {author}
  {\bibfnamefont {R.}~\bibnamefont {Sharples}}, \bibinfo {author}
  {\bibfnamefont {J.}~\bibnamefont {Silber}}, \bibinfo {author} {\bibfnamefont
  {A.}~\bibnamefont {Slosar}}, \bibinfo {author} {\bibfnamefont
  {A.}~\bibnamefont {Smith}}, \bibinfo {author} {\bibfnamefont
  {D.}~\bibnamefont {Sprayberry}}, \bibinfo {author} {\bibfnamefont
  {T.}~\bibnamefont {Tan}}, \bibinfo {author} {\bibfnamefont {G.}~\bibnamefont
  {Tarlé}}, \bibinfo {author} {\bibfnamefont {P.}~\bibnamefont {Taylor}},
  \bibinfo {author} {\bibfnamefont {S.}~\bibnamefont {Trusov}}, \bibinfo
  {author} {\bibfnamefont {L.~A.}\ \bibnamefont {Ureña-López}}, \bibinfo
  {author} {\bibfnamefont {R.}~\bibnamefont {Vaisakh}}, \bibinfo {author}
  {\bibfnamefont {D.}~\bibnamefont {Valcin}}, \bibinfo {author} {\bibfnamefont
  {F.}~\bibnamefont {Valdes}}, \bibinfo {author} {\bibfnamefont
  {M.}~\bibnamefont {Vargas-Magaña}}, \bibinfo {author} {\bibfnamefont
  {L.}~\bibnamefont {Verde}}, \bibinfo {author} {\bibfnamefont
  {M.}~\bibnamefont {Walther}}, \bibinfo {author} {\bibfnamefont
  {B.}~\bibnamefont {Wang}}, \bibinfo {author} {\bibfnamefont {M.~S.}\
  \bibnamefont {Wang}}, \bibinfo {author} {\bibfnamefont {B.~A.}\ \bibnamefont
  {Weaver}}, \bibinfo {author} {\bibfnamefont {N.}~\bibnamefont {Weaverdyck}},
  \bibinfo {author} {\bibfnamefont {R.~H.}\ \bibnamefont {Wechsler}}, \bibinfo
  {author} {\bibfnamefont {D.~H.}\ \bibnamefont {Weinberg}}, \bibinfo {author}
  {\bibfnamefont {M.}~\bibnamefont {White}}, \bibinfo {author} {\bibfnamefont
  {J.}~\bibnamefont {Yu}}, \bibinfo {author} {\bibfnamefont {Y.}~\bibnamefont
  {Yu}}, \bibinfo {author} {\bibfnamefont {S.}~\bibnamefont {Yuan}}, \bibinfo
  {author} {\bibfnamefont {C.}~\bibnamefont {Yèche}}, \bibinfo {author}
  {\bibfnamefont {E.~A.}\ \bibnamefont {Zaborowski}}, \bibinfo {author}
  {\bibfnamefont {P.}~\bibnamefont {Zarrouk}}, \bibinfo {author} {\bibfnamefont
  {H.}~\bibnamefont {Zhang}}, \bibinfo {author} {\bibfnamefont
  {C.}~\bibnamefont {Zhao}}, \bibinfo {author} {\bibfnamefont {R.}~\bibnamefont
  {Zhao}}, \bibinfo {author} {\bibfnamefont {R.}~\bibnamefont {Zhou}}, \bibinfo
  {author} {\bibfnamefont {T.}~\bibnamefont {Zhuang}},\ and\ \bibinfo {author}
  {\bibfnamefont {H.}~\bibnamefont {Zou}},\ }\href
  {https://arxiv.org/abs/2404.03002} {\bibinfo {title} {Desi 2024 vi:
  Cosmological constraints from the measurements of baryon acoustic
  oscillations}} (\bibinfo {year} {2024}{\natexlab{b}}),\ \Eprint
  {https://arxiv.org/abs/2404.03002} {arXiv:2404.03002 [astro-ph.CO]}
  \BibitemShut {NoStop}%
\bibitem [{\citenamefont {{Akaike}}(1974)}]{1974ITAC...19..716A}%
  \BibitemOpen
  \bibfield  {author} {\bibinfo {author} {\bibfnamefont {H.}~\bibnamefont
  {{Akaike}}},\ }\href@noop {} {\bibfield  {journal} {\bibinfo  {journal} {IEEE
  Transactions on Automatic Control}\ }\textbf {\bibinfo {volume} {19}},\
  \bibinfo {pages} {716} (\bibinfo {year} {1974})}\BibitemShut {NoStop}%
\bibitem [{\citenamefont {Schwarz}(1978)}]{BIC1974}%
  \BibitemOpen
  \bibfield  {author} {\bibinfo {author} {\bibfnamefont {G.}~\bibnamefont
  {Schwarz}},\ }\href {https://doi.org/10.1214/aos/1176344136} {\bibfield
  {journal} {\bibinfo  {journal} {The Annals of Statistics}\ }\textbf {\bibinfo
  {volume} {6}},\ \bibinfo {pages} {461 } (\bibinfo {year} {1978})}\BibitemShut
  {NoStop}%
\bibitem [{\citenamefont {{Riess}}\ \emph {et~al.}(2022)\citenamefont
  {{Riess}}, \citenamefont {{Yuan}}, \citenamefont {{Macri}}, \citenamefont
  {{Scolnic}}, \citenamefont {{Brout}}, \citenamefont {{Casertano}},
  \citenamefont {{Jones}}, \citenamefont {{Murakami}}, \citenamefont {{Anand}},
  \citenamefont {{Breuval}}, \citenamefont {{Brink}}, \citenamefont
  {{Filippenko}}, \citenamefont {{Hoffmann}}, \citenamefont {{Jha}},
  \citenamefont {{D'arcy Kenworthy}}, \citenamefont {{Mackenty}}, \citenamefont
  {{Stahl}},\ and\ \citenamefont {{Zheng}}}]{Riess2022Hz}%
  \BibitemOpen
  \bibfield  {author} {\bibinfo {author} {\bibfnamefont {A.~G.}\ \bibnamefont
  {{Riess}}}, \bibinfo {author} {\bibfnamefont {W.}~\bibnamefont {{Yuan}}},
  \bibinfo {author} {\bibfnamefont {L.~M.}\ \bibnamefont {{Macri}}}, \bibinfo
  {author} {\bibfnamefont {D.}~\bibnamefont {{Scolnic}}}, \bibinfo {author}
  {\bibfnamefont {D.}~\bibnamefont {{Brout}}}, \bibinfo {author} {\bibfnamefont
  {S.}~\bibnamefont {{Casertano}}}, \bibinfo {author} {\bibfnamefont {D.~O.}\
  \bibnamefont {{Jones}}}, \bibinfo {author} {\bibfnamefont {Y.}~\bibnamefont
  {{Murakami}}}, \bibinfo {author} {\bibfnamefont {G.~S.}\ \bibnamefont
  {{Anand}}}, \bibinfo {author} {\bibfnamefont {L.}~\bibnamefont {{Breuval}}},
  \bibinfo {author} {\bibfnamefont {T.~G.}\ \bibnamefont {{Brink}}}, \bibinfo
  {author} {\bibfnamefont {A.~V.}\ \bibnamefont {{Filippenko}}}, \bibinfo
  {author} {\bibfnamefont {S.}~\bibnamefont {{Hoffmann}}}, \bibinfo {author}
  {\bibfnamefont {S.~W.}\ \bibnamefont {{Jha}}}, \bibinfo {author}
  {\bibfnamefont {W.}~\bibnamefont {{D'arcy Kenworthy}}}, \bibinfo {author}
  {\bibfnamefont {J.}~\bibnamefont {{Mackenty}}}, \bibinfo {author}
  {\bibfnamefont {B.~E.}\ \bibnamefont {{Stahl}}},\ and\ \bibinfo {author}
  {\bibfnamefont {W.}~\bibnamefont {{Zheng}}},\ }\href
  {https://doi.org/10.3847/2041-8213/ac5c5b} {\bibfield  {journal} {\bibinfo
  {journal} {The Astrophysical Journal Letters}\ }\textbf {\bibinfo {volume}
  {934}},\ \bibinfo {pages} {L7} (\bibinfo {year} {2022})},\ \Eprint
  {https://arxiv.org/abs/2112.04510} {arXiv:2112.04510 [astro-ph.CO]}
  \BibitemShut {NoStop}%
\bibitem [{\citenamefont {Rudra}\ \emph {et~al.}(2015)\citenamefont {Rudra},
  \citenamefont {Ranjit},\ and\ \citenamefont {Kundu}}]{CPLmodelqzgraphs}%
  \BibitemOpen
  \bibfield  {author} {\bibinfo {author} {\bibfnamefont {P.}~\bibnamefont
  {Rudra}}, \bibinfo {author} {\bibfnamefont {C.}~\bibnamefont {Ranjit}},\ and\
  \bibinfo {author} {\bibfnamefont {S.}~\bibnamefont {Kundu}},\ }\href
  {https://doi.org/10.1142/S0217732315501515} {\bibfield  {journal} {\bibinfo
  {journal} {Modern Physics Letters A}\ }\textbf {\bibinfo {volume} {30}},\
  \bibinfo {pages} {1550151} (\bibinfo {year} {2015})},\ \Eprint
  {https://arxiv.org/abs/https://doi.org/10.1142/S0217732315501515}
  {https://doi.org/10.1142/S0217732315501515} \BibitemShut {NoStop}%
\bibitem [{\citenamefont {García-Aspeitia}\ \emph {et~al.}(2018)\citenamefont
  {García-Aspeitia}, \citenamefont {Hernandez-Almada}, \citenamefont
  {Magaña}, \citenamefont {Amante}, \citenamefont {Motta},\ and\ \citenamefont
  {Martínez-Robles}}]{GarciaAspeitia_2018}%
  \BibitemOpen
  \bibfield  {author} {\bibinfo {author} {\bibfnamefont {M.~A.}\ \bibnamefont
  {García-Aspeitia}}, \bibinfo {author} {\bibfnamefont {A.}~\bibnamefont
  {Hernandez-Almada}}, \bibinfo {author} {\bibfnamefont {J.}~\bibnamefont
  {Magaña}}, \bibinfo {author} {\bibfnamefont {M.~H.}\ \bibnamefont {Amante}},
  \bibinfo {author} {\bibfnamefont {V.}~\bibnamefont {Motta}},\ and\ \bibinfo
  {author} {\bibfnamefont {C.}~\bibnamefont {Martínez-Robles}},\ }\bibfield
  {journal} {\bibinfo  {journal} {Physical Review D}\ }\textbf {\bibinfo
  {volume} {97}},\ \href {https://doi.org/10.1103/physrevd.97.101301}
  {10.1103/physrevd.97.101301} (\bibinfo {year} {2018})\BibitemShut {NoStop}%
\end{thebibliography}%

\end{document}